\documentclass[journal]{IEEEtran}
\ifCLASSINFOpdf
\else
\fi

\usepackage{graphics} 
\usepackage{epsfig} 
\usepackage{amsmath} 
\usepackage{amssymb}  
 \usepackage{csquotes}
\usepackage{bm}
\usepackage{amsmath}
\usepackage{url}
\usepackage{gensymb}
\usepackage[utf8]{inputenc}
\usepackage[T1]{fontenc}
\usepackage{textcomp}
\usepackage{gensymb}
\usepackage{ragged2e}
\usepackage{algorithm}
\usepackage{algpseudocode}
\usepackage{makecell}
\usepackage[noadjust]{cite}
\usepackage{multirow,tabularx}
\newcolumntype{Y}{>{\centering\arraybackslash}X}
\newenvironment{psmallmatrix}
  {\left(\begin{smallmatrix}}
  {\end{smallmatrix}\right)}
\begin{document}

\title{Modern WLAN Fingerprinting Indoor Positioning Methods and Deployment Challenges}
\author{Ali~Khalajmehrabadi,~\IEEEmembership{Student~Member,~IEEE,}
        Nikolaos~Gatsis,~\IEEEmembership{Member,~IEEE,}\\
       
        and~David~Akopian,~\IEEEmembership{Senior~Member,~IEEE}
\IEEEcompsocitemizethanks{\IEEEcompsocthanksitem Manuscript received on Aug, 6, 2016. The authors are with the Department of Electrical and Computer Engineering, The University of Texas at San Antonio. 
E-mails: ali.khalajmehrabadi@utsa.edu, nikolaos.gatsis@utsa.edu, david.akopian@utsa.edu.
}
\thanks{}}
\markboth{IEEE communications surveys \& tutorials (SUBMITTED)}
{}

%

\maketitle

\begin{abstract}
Wireless Local Area Network (WLAN) has become a promising choice for indoor positioning as the only existing and established infrastructure, to localize the mobile and stationary users indoors. However, since WLAN has been initially designed for wireless networking and not positioning, the localization task based on WLAN signals has several challenges. Amongst the WLAN positioning methods, WLAN fingerprinting localization has recently achieved great attention due to its promising results.  WLAN fingerprinting faces several challenges and hence, in this paper, our goal is to overview these challenges and the state-of-the-art solutions. This paper consists of three main parts: 1) Conventional localization schemes; 2) State-of-the-art approaches; 3) Practical deployment challenges. Since all the proposed methods in WLAN literature have been conducted and tested in different settings, the reported results are not equally comparable. So, we compare some of the main localization schemes in a single real environment and assess their localization accuracy, positioning error statistics, and complexity. Our results depict illustrative evaluation of WLAN localization systems and guide to future improvement opportunities.
\end{abstract}


%
\IEEEpeerreviewmaketitle

\section{Introduction}
\begin{figure}[t!]
      \includegraphics[scale=0.55]{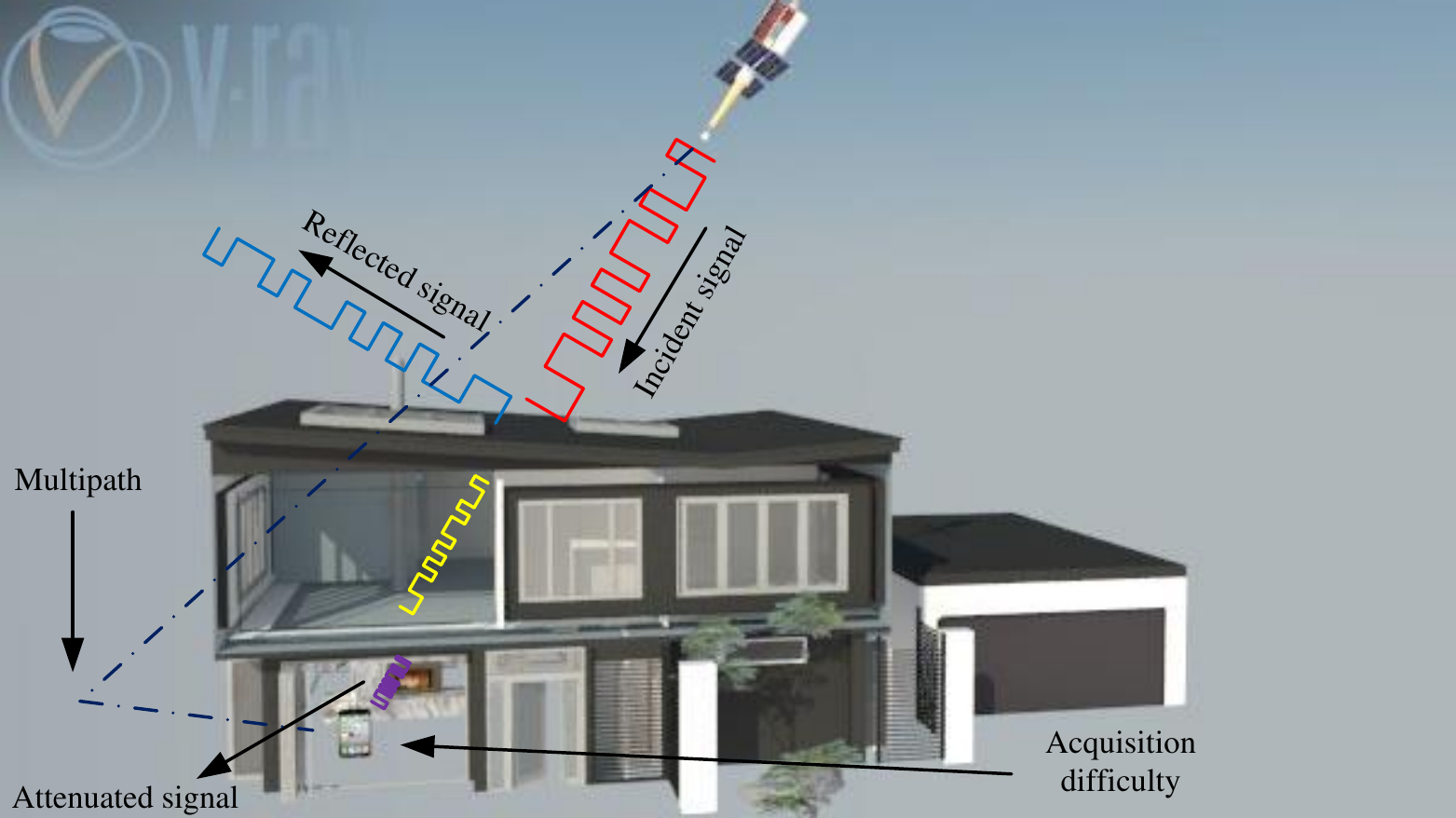}
           \centering

           \caption{The GPS needs direct line of sight between the satellite and the user and does not work indoors.}
           \label{fig6}
        \end{figure}

\IEEEPARstart{L}{ocation}-based services (LBSs) are currently in high demand and strongly drive the development of location-computing technologies \cite{r158}. In particular, indoor LBS will significantly improve network management and security \cite{r55,r56}, emergency personnel navigation \cite{r60,r123}, healthcare monitoring \cite{r57}, personalized information delivery \cite{r58}, context awareness  \cite{r59} and enable other applications. While US Global Positioning System (GPS) and other similar global navigation satellite systems (GNSS) provided good quality for outdoor positioning \cite{r1,r145,r146}, robust indoor positioning is still an open problem. The GPS and similar localization networks do not work indoors as they need direct Line-of-Sight (LOS) between the satellites and user which is not a case indoors as shown in Fig. \ref{fig6}.
\par Various techniques have been proposed for indoor positioning. From signaling perspective these approaches can be divided into two categories \cite{r90,r91}: (1) radio-based positioning such as radio frequency (RF) proximity sensors \cite{r15,r83,r163,r164,r165}, also called radio-frequency identification (RFID),   Ultra Wide Band (UWB) methods \cite{r92,r93},  Bluetooth-based methods \cite{r5,r160,r207}, ZigBee-based methods \cite{r188,r222}, Frequency Modulation (FM) methods \cite{r161,r162}, and IEEE 802.11 Wireless Local Area Networks (WLAN) based methods; and (2) non-radio-based positioning methods which utilize infrared (IR) \cite{r221},  ultrasonic and sound techniques \cite{r4, r61,r166,r176,r206,r221, r194}, visible light \cite{r167,r168}, inertial systems \cite{r142,r192} and magnetic field exploitation \cite{r169,r170}.
\par Many of the proposed technologies including RFID assume massive transceiver and infrastructure deployments and incur high maintenance costs. However, IEEE 802.11 WLANs are  alreadly broadly deployed to render a ubiquitous and continuous wireless network coverage which is exploited for localization purposes as well.  These networks operate in the several unlicensed bands such as 5-GHz (IEEE 802.11a) and 2.4-GHz (IEEE 802.11b/g) and others. Since these bands are unlicensed, several networks may transmit simultaneously and coexist with some interference distortions  \cite{r86,r87}.

\subsection{Indoor Localization Approaches }
\par
\begin{figure}[t!]
      \includegraphics[scale=0.5]{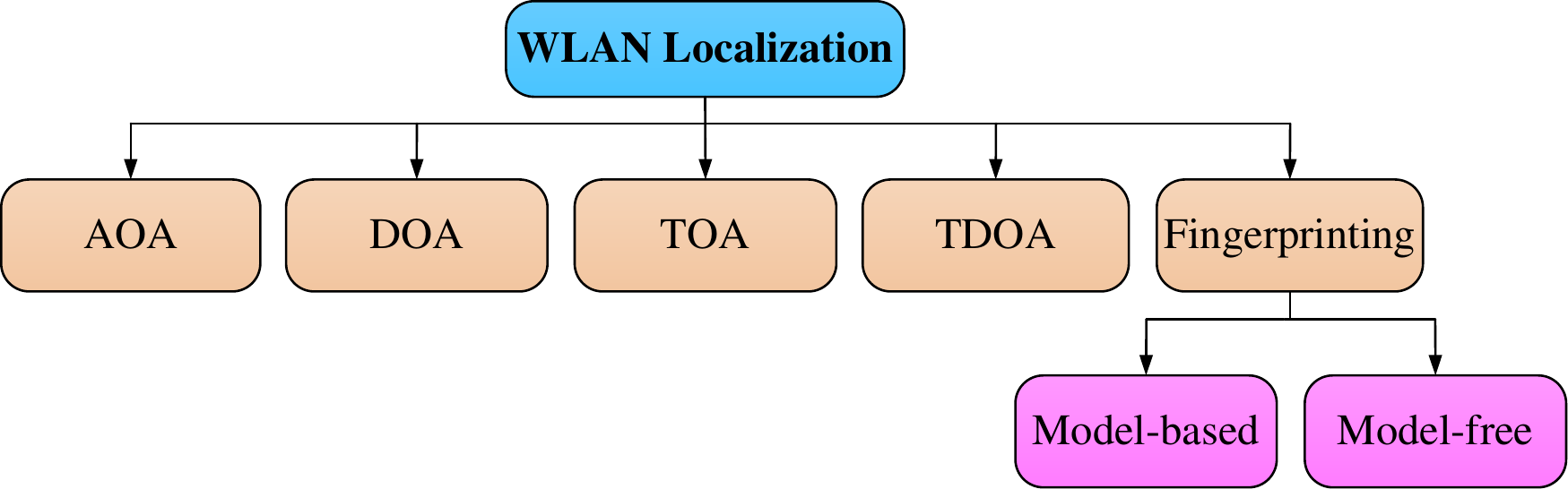}
           \centering

           \caption{WLAN localization schemes.}
           \label{fig5}
        \end{figure}

 Historically, from position computation perspective of radio-based signaling systems, the known approaches for WLAN positioning are of three main categories: (1) Angle of Arrival (AOA) and related Direction of Arrival (DOA) methods; (2) Time of Arrival (TOA) and related Time Difference of Arrival (TDOA) techniques; and (3) RSS exploitation methods (fingerprinting). These methods are shown in Fig. \ref{fig5} and will be reviewed next.
\par In AOA, the angle between the incident wave and a reference direction, known as orientation, is measured from at least two APs. The APs are equipped with an antenna array to be capable of determining the angle of the received signal. The intersection of the two virtual lines heading in the direction of the angles defines the user position \cite{r6,r7,r8}.
\par
\begin{figure}[t!]
      \includegraphics[scale=0.5]{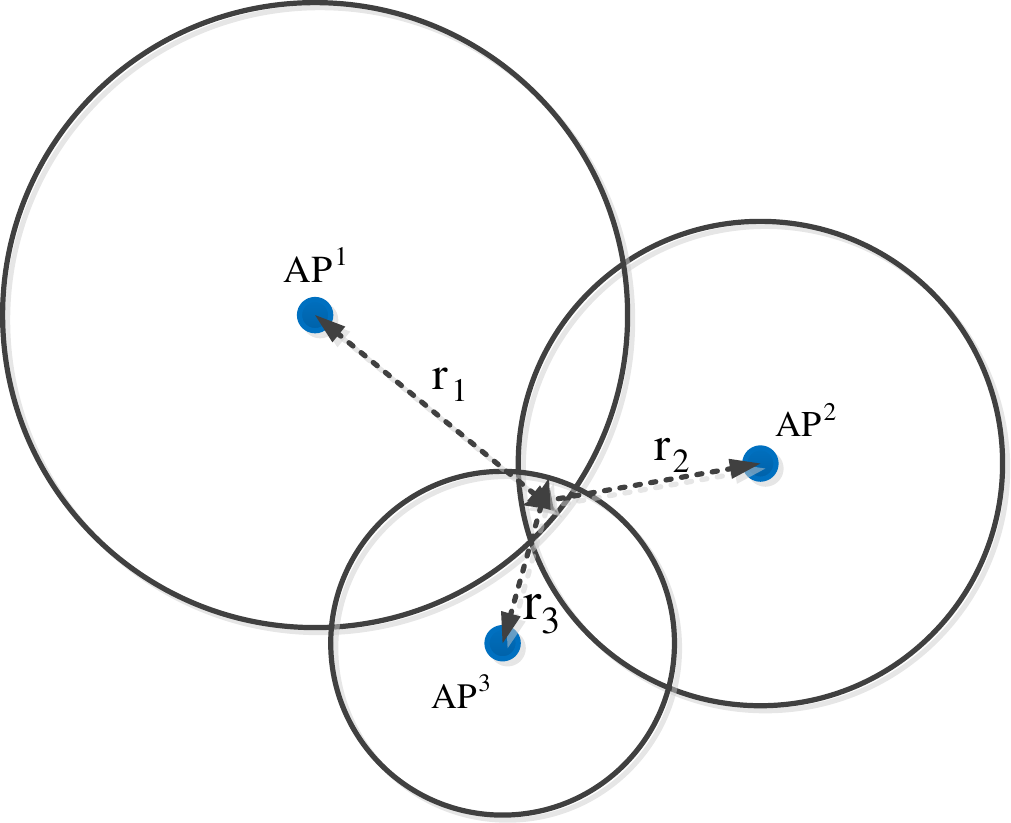}
           \centering

           \caption{Localization based on TOA.}
           \label{fig1}
        \end{figure}

 TOA techniques use the travel time that a wave takes from the transmitter to the receiver and transform it to range distance. At least three APs measure the TOA from a mobile device. For this positioning technique, normally, trilateration is applied \cite{r10,r142}. In trilateration technique, the APs coordinates are known.  Considering an AP as the locus, the range distance defines a circle of certain radius. The intersection of these circles associated with several loci allows to estimate the user’s position. However, there is a great probability that the circles do not intersect precisely at a point due to noisy measurements and the position is estimated with a limited accuracy. The localization based on TOA is shown in Fig. \ref{fig1}. To find the user's location, the following non-linear system of equations should be solved
\begin{equation}
\label{eq1}
\begin{split}
r_{i}=d_{i}+\epsilon_{i}&=\sqrt{(x_{i}-x)^{2}+(y_{i}-y)^{2}}+\epsilon_{i}  \\
 i&=1,\dots,N 
\end{split}
\end{equation}
where $r_{i}$ is the range distance computed from TOA, $\epsilon_{i}$ is the corresponding noise, $\mathbf{p}=(x,y)$ is the user's location to be estimated, and $\mathbf{p}_{i}=(x_{i},y_{i})$ is the $i$-th AP location. As illustrated in Fig. \ref{fig1}, due to the range measurement noise, the location of the user cannot be computed exactly and more sophisticated algorithms that minimize the mean square error (MSE) of the noise, such as least squares (LS), are applied. 

\par TDOA is a variation of TOA, in which a source signal is selected and the time difference of arrival between several spatially distributed APs are measured with respect to the source signal. Since the signal is received from several APs, the system locates the sending device on a hyperboloid \cite{r11,r12}.
\par The above approaches need direct AP-user LOS. Although some enhancements have been proposed for Non-LOS (NLOS) conditions \cite{r143,r144}, the localization errors are high \cite{r10}. In addition, the location of the APs should also be known which is a non-realistic assumptions as the location of the APs are generally unknown and is subject to change regularly for the purpose of providing maximum network coverage.  
\par WLAN fingerprinting methods which use Received Signal Strength (RSS), i.e. the power of received signals from WLAN Access Points (APs), have recently captured a lot of attentions. The reason is twofold: 1) WLANs are widely deployed in offices, business buildings, shopping malls, airports, home environments, etc. and provide ubiquitous coverage for the area. 2) The mobile and wireless receivers all contain the NICs to provide the RSS measurements, and thus, there is no need to install any additional hardware, leading to a reduction in infrastructure installation, equipment and labor costs. Usually, NICs are able to capture distinct RSS magnitudes at a rate of either 0.5 or 1 samples per second.  
\par In general, the RSS exploitation approaches are divided into two broad categories: \textit{model-based (path loss)}  and \textit{model-free (radio map)} approaches.
\par The model based approaches use the collected RSS fingerprints to train the parameters for the predefined propagation models \cite{r73,r15,r206,r228}. These techniques assume a prior path loss model for the indoor propagation which is a logarithmic decay function of the distance from the APs as \cite{r111}
\begin{equation}
\label{eq80}
PL=PL_{0}+10\gamma \text{log}_{10}\frac{d}{d_{0}}
\end{equation}
where $PL$ is the path loss measured in dB, $d$ is the length of the path, $d_{0}$ is the reference distance, and $\gamma$ is the path loss parameter. Using the collected RSS, the distance $d$ between the AP and the user is computed and the location of the user is estimated using the trilateration which incurs knowing the AP locations. To render a more accurate modeling and decrease the discrepancy between the RSS measurements and the model, a random component is added to the model to compensate for the RSS variations \cite{r74}. However, the underlying assumption of symmetric signal power decaying in indoor environments is questionable as the RSS attenuations decay at different rates in different directions due to asymmetric indoor structure.
\par
\begin{figure}[t!]
               \includegraphics[scale=0.33]{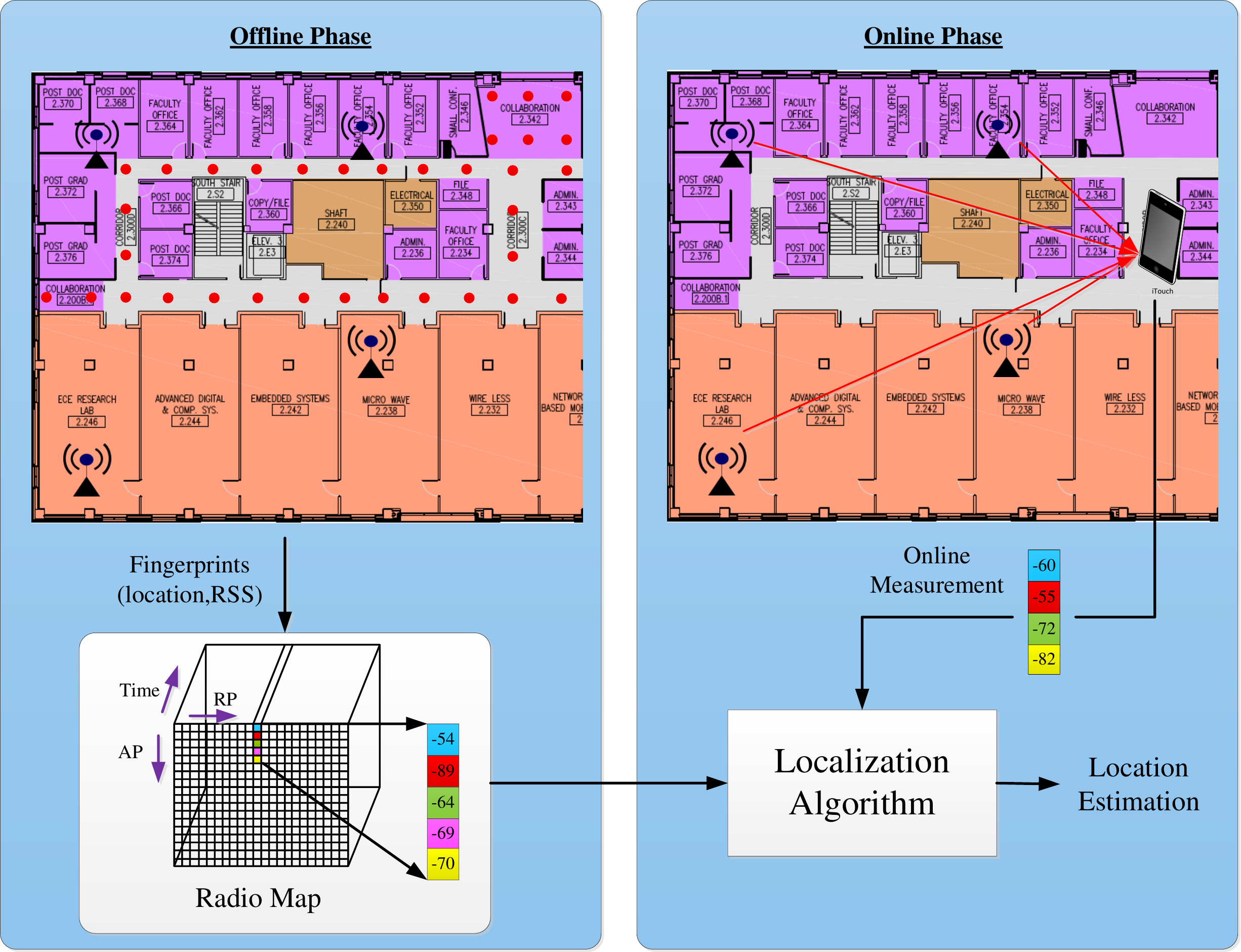}
                    \centering

                    \caption{The typical illustration of indoor WLAN fingerprinting localization system.}
                    \label{fig17}
                 \end{figure}

 The radio map based techniques, also called \textit{fingerprinting techniques}, make the use of dense AP deployments in indoor areas. A set of RSS  or other measurements serve as a fingerprint which  should be more or less unique for each location. In most cases, WLAN fingerprinting consists of \textit{offline} and \textit{online} phases. A schematic of typical WLAN fingerprinting localization is depicted in Fig. \ref{fig17}.  First of all, a set of predefined points, referred to as \textit{Reference Points} (RPs), also called landmarks, grid or survey points, are selected. So, these terms may be used interchangeably throughout this paper. During an offline phase, a survey is conducted and multiple copies of RSS measurements are read at each RP from available APs throughout a time interval. The database of fingerprints for all RPs makes a \textit{radio map} for the whole area. Then during online phase, user observes RSS measurements at his location and applies algorithms to associate these measurements to the radio map entries finding similar fingerprints, and using associated RP locations for estimating the user’s position. A combination of TOA and trilateration has also been introduced in which the contour of the signal strengths are a selection of RPs in the area rather than a circle around the AP \cite{r189}.
 \par Fingerprinting emerges as a straightforward and plausible alternative  offering both accuracy and ubiquity (all modern smartphones come with Wi-Fi capabilities).
\subsection{Criteria for the Categorization of Fingerprinting Methods }
\par WLAN fingerprinting methods differ in computational requirements. Since the computational complexity of the localization systems are high, some approaches assign the localization task on high-power servers, and hence, called \textit{server-based localization}. Quite the opposite, \textit{client-based approaches} minimize the computational complexity of the localization procedure and the positioning computations are performed in resource-limited hand-held wireless devices. Client-based methods are sometimes considered more preferable for users as privacy issues are typically associated with the server-based techniques. Without loss of generality, client-based approaches are considered in the following.
\par
\begin{figure}[t!]
      \includegraphics[scale=0.53]{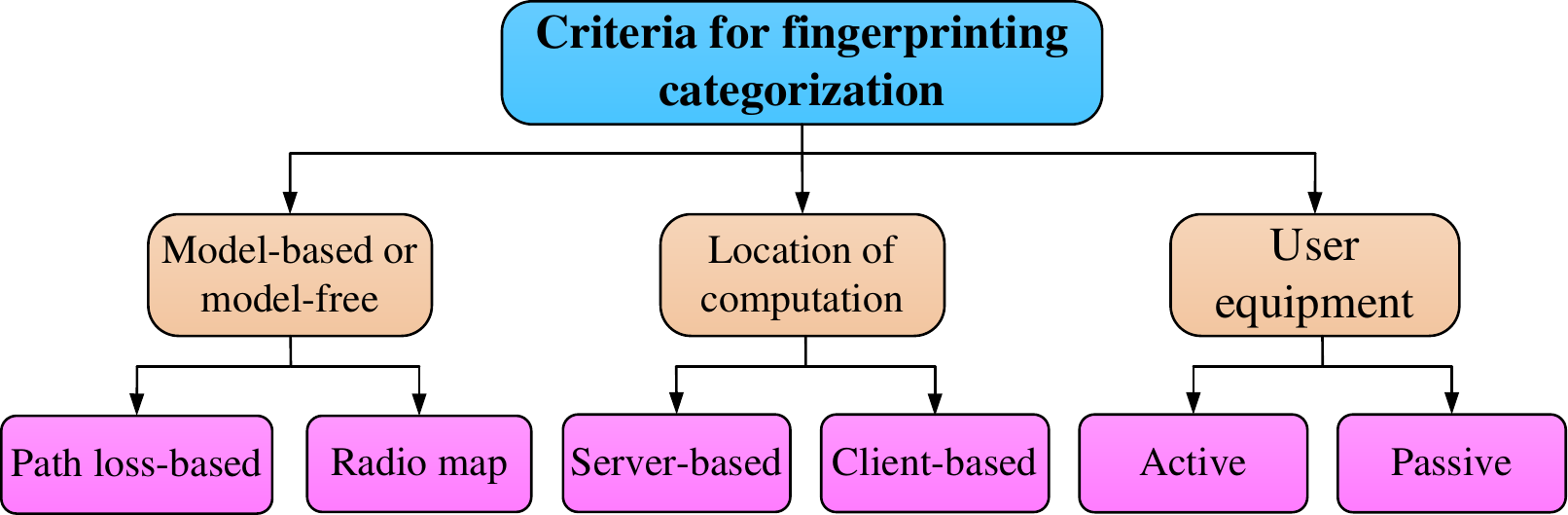}
           \centering

           \caption{Different criteria for the categorization of fingerprinting localization approaches.}
           \label{fig8}
        \end{figure}

 The user is required to carry a wireless device such as laptop, tablet, and smart phone. These devices are required to capture the RSS measurements for the localizations. However, some methods have been recently proposed that does not require the user to carry any device, known as \textit{passive (device-free) localization} \cite{r125, r126, r129,r131}. In passive localization, RSS measurements are taken from wireless devices available in the area which basically measure the changes in RSS profile in the presence of the user at different positions. The passive methods are not discussed in this paper. A summary of above-mentioned methods is  provided in Fig. \ref{fig8}.
                       
                       \begin{figure}[t!]
                             \includegraphics[scale=0.5]{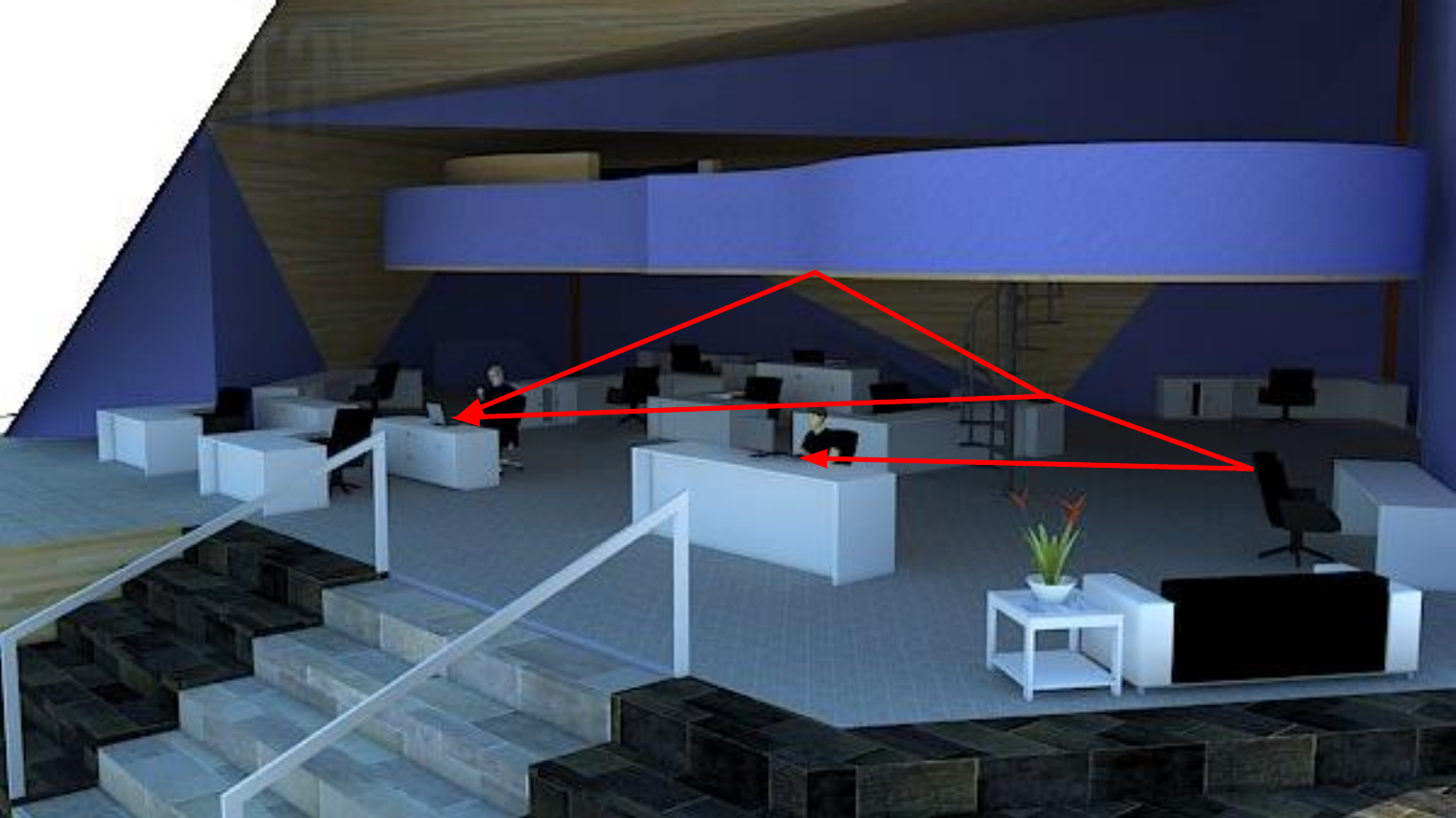}
                                  \centering

                                  \caption{The multipath profile of the WLAN signals is a major problem for localization.}
                                  \label{fig7}
                               \end{figure}

                               \begin{table*}
                               \label{Category}                                
                               \caption{Paper Organization and Contents}
                              \begin{center}
                                                             \begin{tabular}{ |l|l|c| } 
                                                             
                                                             \hline
                                                             \multirow{1}{*}{\textbf{Part I}} &  \begin{tabular}[x]{@{}l@{}} \textbf{Section II:} WLAN Fingerprinting Localization: \\  Problem Formulation and Conventional Approaches \end{tabular} & \begin{tabular}[x]{@{}c@{}}Discusses the fundamental concepts \\  and the early fingerprinting  approaches\end{tabular}     \\ 
                                       
                                                             \hline
                                                             \multirow{9}{*}{\textbf{Part II}} & \textbf{Section \ref{RP Clustering}:} RP Clustering& \begin{tabular}[x]{@{}c@{}}Restricts the localization into a  sub-region\end{tabular}   \\
                                                                                            \cline{2-3}
                                                                                            &  \textbf{Section \ref{Exploitation of APs for Localization}:} Exploitation of APs& Utilizes a suitable subset of APs\\
                                                                                            \cline{2-3}
                                                                                            &  \textbf{Section \ref{Advanced Density, weight estimation, and Data Metric}:} Advanced Density and Weight Estimation Methods& \begin{tabular}[x]{@{}c@{}}Exploits representative features   of the data\end{tabular}  \\
                                                                                            \cline{2-3}
                                                                                            &  \textbf{Section \ref{Sparsity-based Localization}: }  Sparsity-based Localization& \begin{tabular}[x]{@{}c@{}} Reformulates the WLAN localization  \\ to a sparse recovery problem \end{tabular}\\
                                                                                            \cline{2-3}
                                                                                            &  \textbf{Section \ref{Assisted Localization}:}   Assisted Localization& \begin{tabular}[x]{@{}c@{}} Employs available sensors in the \\ environment and mounted sensors on the device \end{tabular}\\
                                                                                            \cline{1-3}
                                                              \hline
                                                      \multirow{5}{*}{\textbf{Part III}} & \textbf{Section \ref{Radio Map Construction}:} Radio Map Construction& \begin{tabular}[x]{@{}c@{}} Discusses different methods of \\ collecting fingerprints\end{tabular}\\
                                                                                         \cline{2-3}
                                                                                      &  \textbf{Section \ref{Outlier Detection}:} Outlier Detection& \begin{tabular}[x]{@{}c@{}} Accounts for the possible outliers in \\ the online measurements \end{tabular}\\
                                                                                          \cline{2-3}
                                                                                       &  \textbf{Section \ref{Heterogeneous Devices}: } Heterogeneous Devices& \begin{tabular}[x]{@{}c@{}} Considers the differences between \\ fingerprinting and localization devices\end{tabular}\\
                                                                                \hline       
                                                             \end{tabular}
                                                             \end{center}
                               \end{table*}

\subsection{Fingerprinting Localization Challenges}
\par Several challenges face the WLAN fingerprinting localization schemes. RSS measurements are distorted by shadowing and NLOS propagation due to the presence of walls, doors, furniture, objects, human, \cite{r69,r70,r76,r107}. Fig. \ref{fig7} shows a typical office environment and a wireless router signals which travel different paths to the wireless devices.  So, the propagated signal encounters with severe frequency selective multipath fluctuations and hence cannot be considered wide sense stationary (non-WSS) \cite{r72}. Moreover, WLAN operates on unlicensed frequencies of 2.4GHz and 5GHz, open to cordless phones, microwaves and the resonance frequency of water. These lead to interference from such devices and signal absorption by the human body. These phenomena make RSS densities non-Gaussian and time varying. In addition, there are various AP networks in a typical area which add extra interference. Also, it is possible that the wireless network coverage degrades due to AP failures \cite{r94,r99}. 
\par There are also logistical problems in fingerprinting WLAN localization. First of all, the surveying stage is very time consuming as the surveyor needs to carry the recording device to each RP and record the RSS for a time period. Furthermore,  pre-processing techniques are usually exploited to reduce search area of the user location to a smaller region rather than the whole area. The smaller region is usually selected by clustering the area. This eliminates the need for a comparison of the online measured RSS with all the RPs fingerprints and hence, the computation time decreases significantly. In the online phase, a subset of RSS measurements should be selected as not all measurements provide beneficial information. Normally, an assessment of measurement sanity is conducted and a subset of APs are selected for positioning. An elaborate discussion on these methods are provided in this paper.

\subsection{What This  Paper Brings to the Scene}
Wireless indoor localization has been previously reviewed \cite{r180,r207,r105,r181,r182,r183,r184,r185,r106,r172,r212}. Though impressive, most of the previous surveys did not comprehensively cover all stages of a complete fingerprinting localization system. More importantly, these surveys have discussed the current localization approaches generally and did not dig into technical aspects.
\par This paper provides a technical overview of the state-of-the-art WLAN fingerprinting positioning approaches and practical implementation issues. The discussion of this paper is divided into three main parts which are overviewed next. Table I lists the related sections and contents of each part:

\begin{enumerate}
\item Problem Formulation and Conventional Approaches (Section \ref{WLAN Fingerprinting Localization: Problem Formulation and Conventional Approaches}): In this part, we discuss the fundamental fingerprinting concepts and unify the different notations used in the literature. Then, the conventional approaches, which have been proposed in the early stages of Wi-Fi based localization, are organized in three general categories.
\item State-of-the-Art Approaches: The wide variety of recent trends toward Wi-Fi- based localization can be organized along the following paths:

\begin{itemize}
\item \textit{Refinements of the Conventional Localization: } Since the conventional methods cannot achieve the necessary localization accuracy and the online running time cannot pace with the user's motion, refinements have been introduced. They have focused on RP clustering (Section \ref{RP Clustering}), exploitation of APs (Section \ref{Exploitation of APs for Localization}), and advanced density and weight estimation methods (Section \ref{Advanced Density, weight estimation, and Data Metric}).  These techniques are direct modifications of the conventional approaches. 
\item \textit{Sparsity-based Localization (Section \ref{Sparsity-based Localization}):} A reformulation of WLAN localization has been recently introduced which exploits sparse recovery methods.
\item \textit{Assisted Localization (Section \ref{Assisted Localization}):} Aside from merely utilizing WLAN fingerprints for localization, some methods gain assistance from available resources in the environment and user's device to achieve superior localization accuracy. These methods may integrate sensory signatures built  in the modern wireless devices, track the user's motion, exploit the available environment landmarks, or utilize the peer-to-peer collaboration between devices, and collectively fall under assisted localization.  
\end{itemize}

\item Deployment Challenges:  Localization schemes face laborious deployment challenges which constraint their applicability as real positioning systems. Even with advanced localization techniques, practical systems should account for several challenges listed next.
 \begin{itemize}
 \item \textit{Radio-map Construction (Section \ref{Radio Map Construction}):} An existing problem with fingerprinting methods is the need for dense survey of the area. Previous works attempt to decrease the time and cost of fingerprinting tasks through crowd-sourcing, implicit or unlabeled data collection, and radio map interpolation.
 \item \textit{Outlier Detection (Section \ref{Outlier Detection}):}   APs are easily prone to infrastructure problems that render faulty readings. These faulty readings are called \textit{outliers}. Outliers can occur both on fingerprints during the survey process and more importantly during the online phase. The fingerprint outliers are  easier to detect.  The presence of outliers during the online  phase implies that the user's location should be estimated using faulty measurements. 
 \item \textit{Heterogeneous Devices (Section \ref{Heterogeneous Devices}):} Wireless devices obtain  RSS fingerprints through their Network Interface Cards (NICs). The sensitivity of the wireless devices differ as the NICs chipsets are different and the position of the antenna on the device affects the RSS readings.
 \end{itemize}
\end{enumerate}
\par After the theoretical discussion, we provide a numerical evaluation of the representative approaches based on  localization accuracy and positioning error statistics in Section \ref{Numerical Experiments and Comparisons}. The methods are tested on the same set of fingerprints collected at the University of Texas at San Antonio (UTSA).  These comparisons provide illustrative guidelines for future improvements. A critical summary and future directions are provided in Section \ref{Critical Summary and Outlook}. 
\section{WLAN Fingerprinting Localization: Problem Formulation and Conventional Approaches}
\label{WLAN Fingerprinting Localization: Problem Formulation and Conventional Approaches}
This section provides the definitions and formulation of the WLAN fingerprinting localization, and a description of the conventional localization methods comes in sequel.

\subsection{Problem Formulation}
In fingerprinting, the area is divided into a set of RPs   $ \mathcal{P}=\left\{\mathbf{p}_{j}=(x_{j},y_{j})| j=1,\ldots,N  \right\} $ where $  \mathcal{P} $ defines the set of RP Cartesian coordinates, which are not necessarily set apart in equal distances. The mobile device records RSS fingerprints at time instants $t_{m},\ m=1,\ldots,M$, with RSS magnitudes $ (r_{j}^{i}(t_{1}),\ldots, r_{j}^{i}(t_{M})) $ at each RP, where $ i $ indicates the AP index from the set of APs, $\mathcal{L}= \left\{AP^{1},\ldots,AP^{L}\right\} $. It is typical to take the same number of training samples, $ M $, at each RP. The RSS fingerprints from all APs at time $t_{m}$ at $\mathbf{p}_{j}$ are organized in a vector $ \mathbf{r}_{j}(t_{m})=[r_{j}^{1}(t_{m}),\ldots,r_{j}^{L}(t_{m})]^{T} $. The entire radio map at recording instant $ t_{m} $ is represented as
\begin{equation}
\label{eq2}
\begin{split}
 \mathbf{R}&(t_{m})=\left( \mathbf{r}_{1}(t_{m}),\ldots,\mathbf{r}_{N}(t_{m})\right)  =\\
 &\begin{pmatrix}
   r_{1}^{1}(t_{m}) & r_{2}^{1}(t_{m})  & \cdots & r_{N}^{1}(t_{m}) \\
   r_{1}^{2}(t_{m}) & r_{2}^{2}(t_{m})  & \cdots & r_{N}^{2}(t_{m}) \\
   \vdots  & \vdots  & \ddots & \vdots  \\
   r_{1}^{L}(t_{m}) & r_{2}^{L}(t_{m}) & \cdots & r_{N}^{L}(t_{m}) 
  \end{pmatrix},\\
  & \ \ \ \ \ \ \ \ \ \ \ \ \ \ \ \ \ \ \  m=1,\dots,M.
  \end{split}
\end{equation}

Let also $ \mathbf{r}_{j}^{i}=[r_{j}^{i}(t_{1}),\ldots,r_{j}^{i}(t_{M})]^{T} $, $ \mathbf{r}^{i}(t_{m})=[r_{1}^{i}(t_{m}),\ldots,r_{N}^{i}(t_{m})]^{T} $, and $ \mathbf{r}_{j}(t_{m})=[r_{j}^{1}(t_{m}),\ldots,r_{j}^{L}(t_{m})]^{T} $  indicate a vector of RSS fingerprints for different time instants, different RPs, and different APs, respectively. If the time sequence of radio maps, $  \mathbf{R}(t_{m}) $, is averaged over the recording time, the time averaged radio map is denoted as 
\begin{equation}
\label{eq3}
\begin{split}
 \mathbf{\Psi}=\left( \boldsymbol \psi_{1},\ldots,\boldsymbol \psi_{N}\right)  =
 \begin{pmatrix}
   \psi_{1}^{1} & \cdots & \psi_{N}^{1} \\
   \vdots  &  \ddots & \vdots  \\
   \psi_{1}^{L} & \cdots & \psi_{N}^{L} 
  \end{pmatrix}\\
  \end{split}
\end{equation}
where $ \boldsymbol \psi_{j}=[\psi_{j}^{1},\ldots ,\psi_{j}^{L}]^{T}$, and $\psi _{j}^{i}=\frac{1}{M}\sum_{m=1}^{M} r_{j}^{i}(t_{m})$. 
\par A subset of RPs with the most similarity to the online measurement is denoted by $\mathcal{K}$ where $\left| \mathcal{K}\right| =K$. This similarity is defined differently in each localization method and will be discussed in detail later.
\par In the online phase, the mobile user receives the online RSS measurements, $ \boldsymbol y=(y^{1},\ldots,y^{L})^{T} $.
The goal of a localization scheme is to find the user's location, $  \hat{\mathbf{p}}= (\hat{x},\hat{y})$, based on a rule that compares the received online measurements against radio map fingerprints as:
\begin{equation}
\label{eq5}
\hat{\mathbf{p}}=f(\mathbf{R}, \boldsymbol y).
\end{equation}
where $\mathbf{R}$ denotes  the collection of radio maps at all recording instances. Some techniques (especially the advanced probability methods in Section \ref{Advanced Density, weight estimation, and Data Metric}) need multiple online measurements which are indexed by time instants $t_{m'}$ as $ \boldsymbol y(t_{m'})=(y^{1}(t_{m'}),\ldots,y^{L}(t_{m'}))^{T} $. Next, three  conventional  localization approaches are discussed. 
\begin{figure}[t!]
      \includegraphics[scale=0.63]{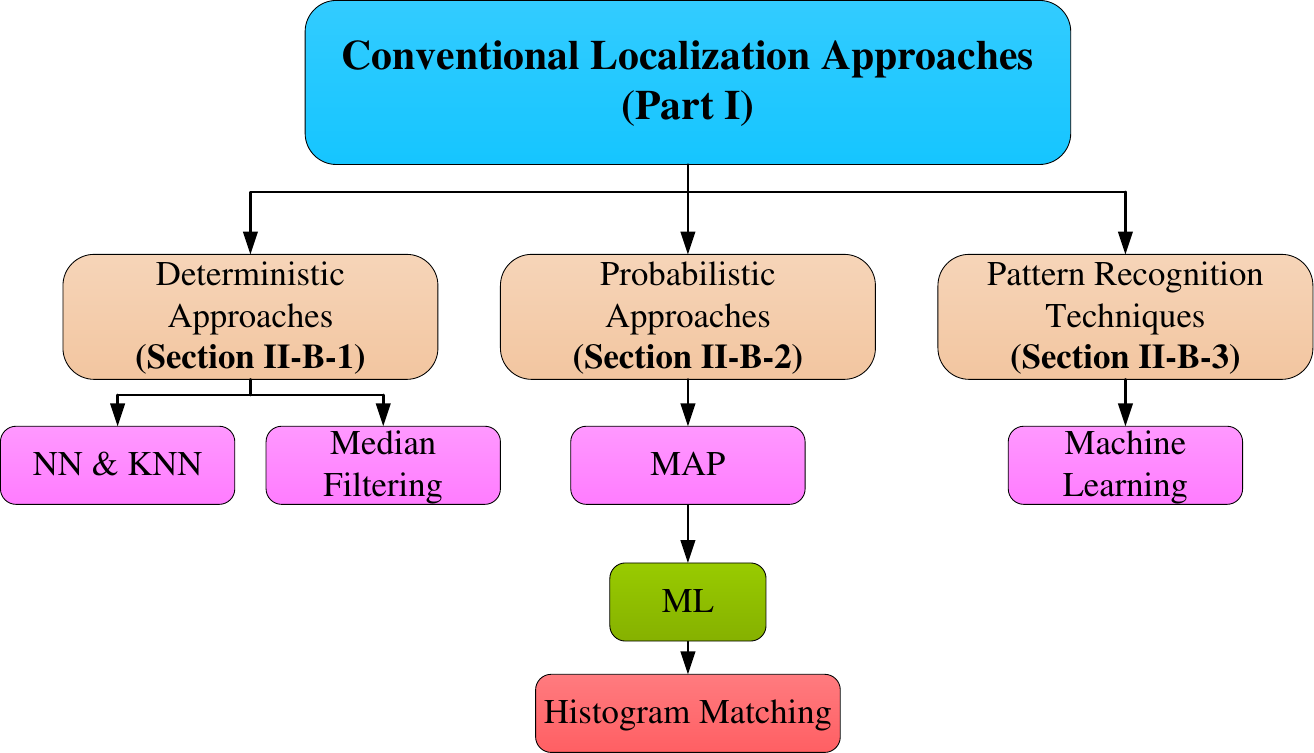}
           \centering

           \caption{Conventional localization approaches.}
           \label{fig2}
        \end{figure}
        \begin{table*}[t!]
                              \caption{Conventional Fingerprinting Localization Approaches and Representative References}
                               \label{table1}
                              \begin{center}
                              \begin{tabular}{ |c|c| } 
                                \hline
                             \textbf{Methods} &\textbf{ Related Works}\\
                              \hline
                                 \hline
                              Deterministic & KNN \cite{r15,r82,r14,r18,r79,r105,r106}, Median filtering \cite{r17,r105,r106}  \\ 
                                \hline
                            Probabilistic  & MAP \cite{r76,r66,r65}, ML \cite{r105}, Weighted probabilistic \cite{r21,r64,r232}\\ 
                                \hline
                              Pattern recognition & CCA \cite{r25}, SVM \cite{r26,r27}, Neural Networks \cite{r14}, Linear Discriminant Analysis \cite{r220} \\ 
                                \hline
                              \end{tabular}
                              \end{center}
                               \end{table*}
\subsection{Conventional Localization Approaches}
\label{Conventional Localization Approaches}
In this section we elaborate on the early WLAN fingerprinting localization approaches \cite{r175,r181,r182,r183}. A diagram summarizing these approaches is shown in Fig. \ref{fig2} along with a categorization of the related works in Table \ref{table1}.

\subsubsection{Deterministic Approaches}
In deterministic approaches, the general form of position estimation is achieved through selecting RPs whose fingerprints are the closest to the  online RSS measurements as
\begin{equation}
\label{eq6}
\hat{\mathbf{p}}=\underset{j=1,\dots,N}{\text{argmin}} d(\breve{\mathbf{r}}_{j},\boldsymbol y) 
\end{equation}
where $\breve{\mathbf{r}}_{j}$ is the representative fingerprint value at RP $j$  \cite{r15,r79} and $ d(\breve{\mathbf{r}}_{j},\boldsymbol y)$ defines a typical distance  metric \cite{r179}. In case of time-average, the representative value is  $\boldsymbol {\psi}_{j}$.  Euclidean distance is a well-known distance metric for \eqref{eq6} defined as
\begin{equation}
\label{eq7}
\begin{split}
  d(\breve{\mathbf{r}}_{j},\boldsymbol y)=\Arrowvert \boldsymbol y -\breve{\mathbf{r}}_{j}\Arrowvert_{2} \ \
 j=1,\dots, N. 
\end{split}
\end{equation}
A solution which finds the RP with the minimum Euclidean distance among measurements is known as the nearest neighbor (NN) method. 
\par  Median filtering has also been used  to improve the robustness of the KNN method to  unusual fingerprint readings \cite{r17}. In this case, the $i$-th entry of $\breve{\mathbf{r}}_{j}$ is $\breve{r}_{j}^{i}=\mathrm{med} \{r_{j}(t_{m}), \  m=1,\ldots,M \}$.
\par If instead of selecting a single RP with the least distance, a set of closest RPs are selected, the method is known as K-nearest neighborhood (KNN) \cite{r15,r82,r105, r106}. In KNN, the user position is usually the centroid of a set of $K$ RPs with the least distances $d(\breve{\mathbf{r}}_{j},\boldsymbol y)$ 
\begin{equation}
\label{eq8}
 \hat{\mathbf{p}}_{KNN} =\frac{1}{K}\sum_{j \in \mathcal{K}}^{} \mathbf{p}_{j}.
\end{equation}
The weighted KNN approach differentiates  RPs by assigning weights in \eqref{eq8} proportional to the inverse of their corresponding $d(\breve{\mathbf{r}}_{j},\boldsymbol y)$. So, RPs that are similar to the online measurement, receive higher weights. KNN weights can also be computed based on the inverse of the RSS variance at each RP \cite{r18}, or cosine similarity \cite{r233}. An RP can be excluded from engaging in positioning if its total RSS variation is above a predefined threshold as it is unreliable \cite{r74}.
  
\par Since the number and availability of APs varies across the localization area, \eqref{eq7} is typically estimated over the common visible APs and missing APs’ readings are replaced by a boundary number indicating weak signal (usually -95 dBm).

\subsubsection{Probabilistic Approaches}
 A single RSS fingerprint may not be a sufficient representation of the data because of the time-varying nature of indoor propagation. The performance of deterministic localization approaches can be improved if instead of a single representative RSS fingerprint, all  fingerprints are used. In probabilistic approaches, the whole ensemble of RSS fingerprints are utilized to provide statistical characteristics of the area. 
\par The underlying approach in probabilistic localization is the Maximum \textit{ A Posteriori }(MAP) estimation \cite{r153,r182}. The MAP estimates the location of the user based on maximizing the conditional probability of the location given the received online measurement
\begin{equation}
\label{eq10}
\hat{\mathbf{p}}= \underset{j=1,\dots, N}{\text{argmax} }\ f(\mathbf{p}_{j}|\boldsymbol y)
\end{equation}
where $f(\mathbf{p}_{j}|\boldsymbol y)$ is the conditional probability that the user is in $\mathbf{p}_{j}$  given the received online vector $\boldsymbol y$. The equivalent reformulation of \eqref{eq10} is achieved through the Bayes rule
\begin{equation}
\label{eq11}
\begin{split}
f(\mathbf{p}_{j}|\boldsymbol y)=\frac{f(\mathbf{p}_{j},\boldsymbol y)}{f(\boldsymbol y)}&=\frac{f(\boldsymbol y|\mathbf{p}_{j})f(\mathbf{p}_{j})}{\sum_{j=1}^{N}f(\boldsymbol y|\mathbf{p}_{j})f(\mathbf{p}_{j})}\\
 j&=1,\dots, N
\end{split}
\end{equation}
The probability $f(\mathbf{p}_{j})$ is the distribution of the user location over the area and is usually assumed to be uniform, i.e. $f(\mathbf{p}_{j})=\frac{1}{N}$ since there is no prior knowledge regarding the user location and all survey points are equally probable. Therefore,  $f(\mathbf{p}_{j})$ can be ignored in the maximization problem \eqref{eq10}. Likewise, the denominator in \eqref{eq11} is the same for all $j=1,\dots, N$. Therefore, the MAP estimation in \eqref{eq10} is equivalent to the following problem 
\begin{equation}
\label{eq13}
\hat{\mathbf{p}}= \underset{j=1,\dots, N}{\text{argmax} } \ f(\boldsymbol y|\mathbf{p}_{j})
\end{equation}
known as Maximum Likelihood (ML) estimation \cite{r154}. Another alternative to the ML estimate of \eqref{eq13} is to select three non-collinear RPs with the highest probability. The user's location can be estimated through an interpolation between these RPs by solving a system of two equations with two unknowns \cite{r104}.
\par ML estimate  picks the RP with the maximum statistical similarity to the online measurement, however, if the user is in between the RPs only a single RP is not a suitable location estimate of the user. To this end, the convex hull of the RPs that surround the user's location provides a suitable estimate. Therefore, \eqref{eq13} can be replaced by an estimate that utilizes all (or a subset of) RPs with corresponding weights as follows \cite{r155,r156}
\begin{equation}
\label{eq14}
\begin{split}
\hat{\mathbf{p}}=\sum_{j=1}^{N} w _{j} \mathbf{p}_{j}, \ w _{j}=\frac{f(\boldsymbol y|\mathbf{p}_{j})}{\sum_{j=1}^{N}f(\boldsymbol y|\mathbf{p}_{j})}.
\end{split}
\end{equation}
The ML estimate of \eqref{eq13} renders the most similar RP as the user's location. This leads to a high localization error if the user is not exactly at one RP. The supremacy of \eqref{eq14} over \eqref{eq13} is that it renders the user's location as the weighted convex combination of the RPs that own the most similar fingerprints to the online measurements.
\par The previous discussion reveals that the task of positioning relies on estimating the prior density $f(\boldsymbol y|\mathbf{p}_{j})$. There are two main approaches regarding fingerprint distribution estimation: parametric and non-parametric estimation. Parametric estimation methods try to map the data to known analytical distributions, e.g., Gaussian, to approximate temporal RSS characteristics \cite{r66,r64}. This assumption has been questioned in several works, e.g., \cite{r76}. Early approaches consider the RSS distribution as log-normal \cite{r20}. However, it is shown that the distribution is not typically log-normal but left skewed, stationary only over small time frames, and the user’s presence makes it multi-modal \cite{r21,r72,r198}. Moder density estimation methods use kernel functions, and and overview is provided in Section \ref{Advanced Density, weight estimation, and Data Metric}.
\par Non-parametric estimation methods do not assume any known distribution matches with the RSS fingerprints. Instead, the fingerprint distributions are generated using histogram matching of radio-map fingerprints \cite{r19,r65,r66}.  In histogram matching, the whole data is quantized into multiple levels and the frequency of each bin is calculated for the estimation of $f(\boldsymbol y|\mathbf{p}_{j})$. The histogram consists of the concatenation of these bins. However, a large number of time samples are needed at each RP to generate a histogram. Besides, the histogram is primarily dependent on bin width and the choice of origin \cite{r67,r68,r194}.

\begin{figure}[t!]
      \includegraphics[scale=0.37]{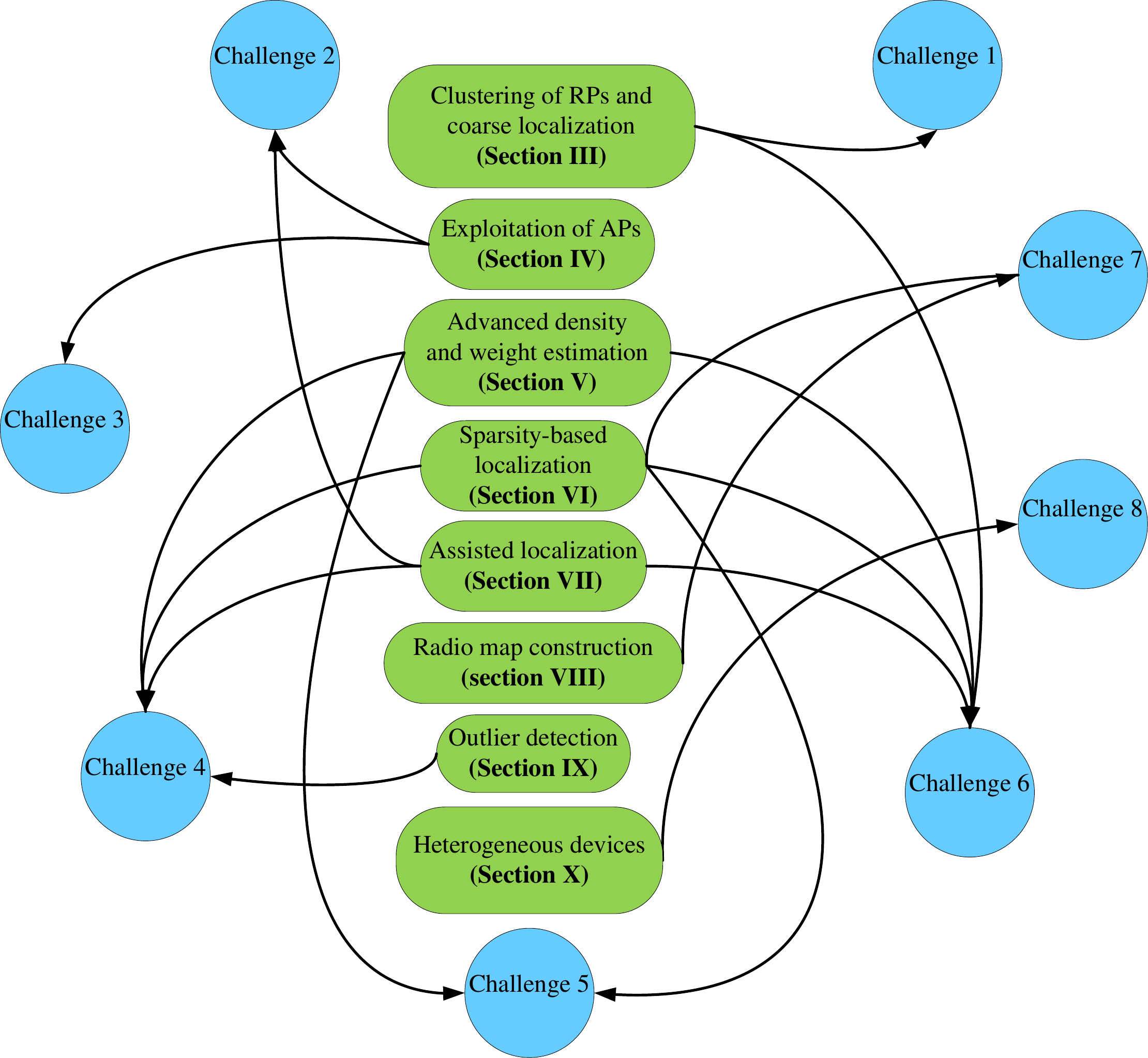}
           \centering

           \caption{Challenges of fingerprinting and modern solutions.}
           \label{fig18}
        \end{figure}
\subsubsection{Pattern Recognition Techniques}
The basic idea of pattern recognition methods is based on classifiers, that are trained using surveyed fingerprinting data and then used to discriminate unknown RSS measurements during the online phase. In the training phase, the system tunes the internal classifier model knowing a radio map database. In the testing phase, the received RSS data from unknown locations are processed by the classifier by estimating the most likely location. The difference between pattern recognition approaches is in their pattern-matching techniques. The outcome of the pattern recognition algorithm is typically a likelihood of various locations given observed measurements, which allows to estimate the centroid of the all candidate positions as the solution . Support Vector Machine (SVM), Canonical Correlation Analysis (CCA), Neural Network, and linear discriminant analysis   are examples of contemporary pattern recognition schemes \cite{r25,r26,r27,r220}.

\subsection{Inadequacies of Conventional Methods and Overview of Recent Works}
The conventional WLAN fingerprinting localization methods face several challenges which degrade the positioning accuracy and introduce biased estimations. These challenges motivate all remaining parts and are listed next in this paper and are shown in Fig. \ref{fig18} with corresponding state-of-the-art solutions detailed as follows:
\begin{itemize}
\item [c1)]The number of RPs increases with the area size, which increases the required memory needed to store the surveyed data and the computing resources. 
\item [c2)]APs do not necessarily provide independent information and the fingerprints can be correlated. 
\item [c3)]APs have limited coverage area and may not be accessible to all RPs in the surveyed area. Utilizing distant APs with weak signals at the user location degrades positioning accuracy.  
\item [c4)]Possible  faulty RSS measurements may incur biased position estimates.
\item [c5)] The distribution of RSS fingerprints is non-Gaussian, skewed, multimodal, and time-varying.
\item [c6)]Most proposals for conventional methods give low accuracy guarantees.
\item [c7)]The radio map construction is labor intensive and time consuming.
\item [c8)]The difference between surveying device and the user's device leads to heterogeneity of fingerprints readings which impose a great error on localization.
\end{itemize}  
\par Practical positioning schemes have attempted to address these issues \cite{r172,r173, r174}. Fig. \ref{fig18} maps the challenges with corresponding solutions. Note that one solution may address several challenges simultaneously.
\par To address challenges c1 and c6, offline RP clustering and online coarse localization have been proposed.  In RP clustering, the RPs are divided into groups (clusters) based on a similarity metric. Then, the localization coarsely estimates the user location in a subset of RPs and then the fine location of the user is estimated within this subset. The RP clustering reduces the computational burden, and guides the fine localization step.
\begin{figure*}[t!]
      \includegraphics[scale=0.52, angle=0]{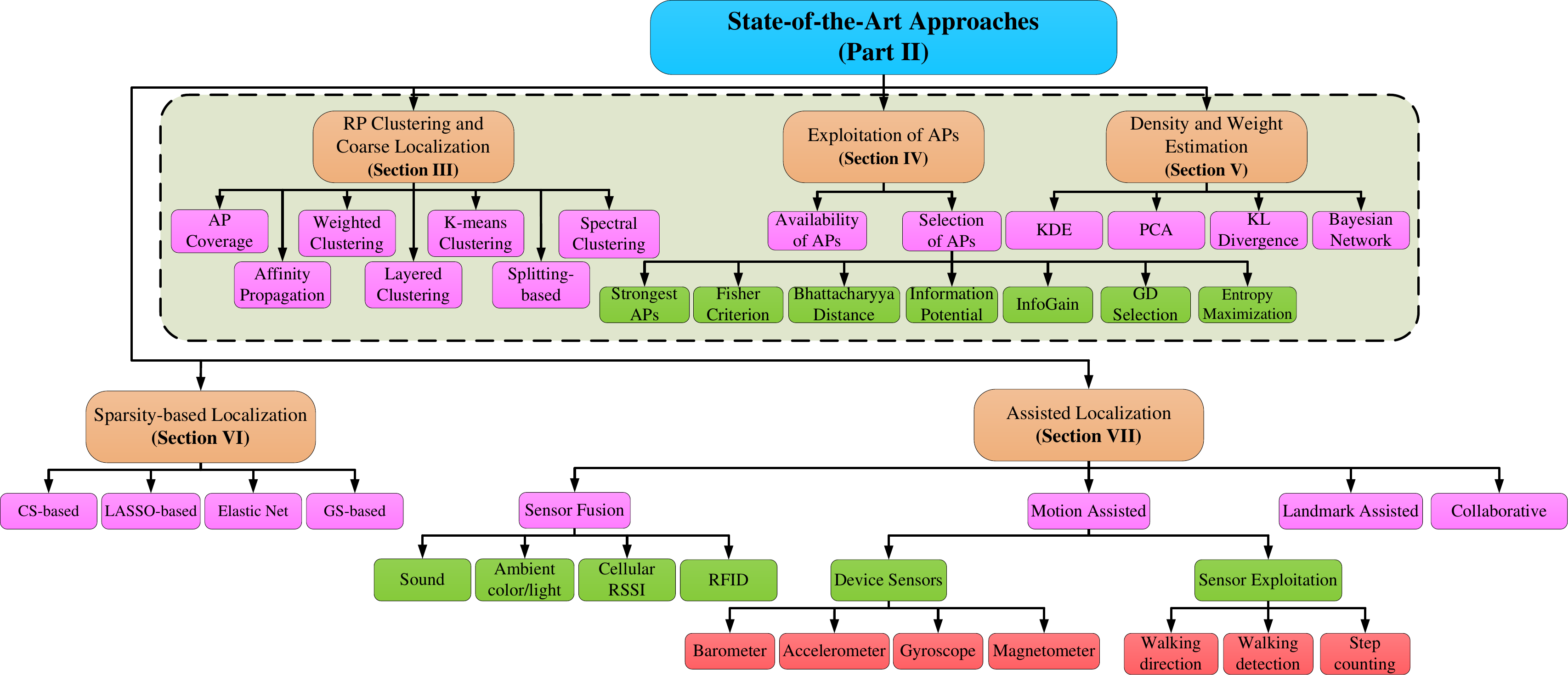}
           \centering

           \caption{State-of-the-art localization approaches.}
           \label{fig3}
        \end{figure*}

\par Challenges c2 and c3 are addressed through AP selection, in which an evaluation metric assigns scores to APs. Generally, the score defines the suitability of each AP for localization considering the online measurement of the user. Then, the best set of APs that can provide distinguishable information are used in localization. AP selection discards the APs that do not provide independent information, and biased location estimation due to distant APs. 
\par Challenges c4, c5, and c6 should be treated with accurate metrics that measure the distance between the fingerprints and online measurements. In recent probabilistic methods, the RSS fingerprints distributions are estimated through more sophisticated schemes that account for the multimodality of the distribution. Also, advanced techniques have been introduced for weight estimation in \eqref{eq14}. In addition, the Wi-Fi fingerprints can be integrated with additional environmental features, inertial device sensors, and collaboration between devices to use all the available information and deliver more accurate location estimations. Furthermore, recent approaches have introduced a new solution to  the WLAN fingerprinting problem via sparse recovery methods. Above all, inordinate readings in online measurements are treated with outlier detection methods. 
\par To tackle challenge c7, recent techniques propose to record the radio map with the help of users or at a coarser grid with subsequent interpolation in between RPs at a finer grid. 
\par Challenge c8 is treated with approaches that match the online measurements with the offline fingerprints. These methods are discussed in Section \ref{Heterogeneous Devices}.
\par Fig. \ref{fig3} categorizes the state-of-the-art solutions (part II) that come as refinements and enhancements to conventional approaches. The shaded box denotes the three tasks that a typical modern localization system performs. Sparsity-based localization and assisted localization may also be combined with these tasks to improve the localization accuracy. The corresponding literature is summarized in Table \ref{table2}. 
\par Fig. \ref{fig10} categorizes the deployment challenges (part III) with the localization methods. These challenges and the methods to address them are the contents of part III of the paper. Related works are listed in Table \ref{table4}.

\section{RP Clustering and Coarse Localization}
\label{RP Clustering}
In WLAN positioning, the characteristics of RSS fingerprints highly depend on environmental features and available APs. This motivated the recent works to constraint the positioning algorithm to a subset of RPs that show similar characteristics \cite{r28}. In other words, a coarse localization stage reduces the search space of the user location to a smaller number of RPs, which is followed by a finer search on the refined set of RPs \cite{r205}.  This procedure is typically called \textit{radio map clustering} or \textit{spatial filtering}. These terms are used interchangeably in this context. The clustering is an offline process where the members of a cluster are grouped together based on a similarity metric. A representative value of fingerprints shows the characteristics of  each cluster and is used for coarse localization. Specific clustering methods are surveyed next.
\begin{table}[t!]
                      \caption{State-of-the-Art Approaches and Related Literature }
                       \label{table2}
                      \begin{center}
                      \begin{tabular}{ |c|c| } 
                        \hline
                         
                      \textbf{Methods} & \textbf{Related Works}\\
                      \hline
                      \hline
                      RP clustering & \cite{r28,r21,r31,r139,r140,r141,r35, r80}  \\ 
                        \hline
                    Exploitation of APs  & \cite{r84,r85,r19,r35,r80,r30,r88,r27,r130} \\ 
                        \hline
                      Advanced density and weight estimation & \cite{r102,r21,r103} \\ 
                        \hline
                     Sparsity-based localization & \cite{r30,r121,r139,r141} \\ 
                                                \hline
                     Assisted localization & \cite{r205,r208,r215,r216,r217,r224,r225,r219,r179} \\ 
                          \hline
                      \end{tabular}
                      \end{center}
                       \end{table}
\subsection{Clustering Using AP Coverage} 
One spatial filtering method is based on the assumption that neighbor RPs receive similar RSS fingerprints \cite{r21, r31}. The intuition is that neighboring RPs should receive RSS readings from the same set of APs. The scheme relies upon defining the \textit{continuous coverage} of an AP over a subset of RPs.  First,  the set of time slots for which the fingerprints corresponding to each AP are above a threshold $\gamma$ is computed:
 \begin{equation}
 \label{eq16}
 \begin{split}
 \mathcal{T}_{j}^{i}&=\left\lbrace m \in \left\lbrace 1, \dots , M\right\rbrace \arrowvert r_{j}^{i}(t_{m}) \ge \gamma \right\rbrace,\\
  i&=1,\ldots,L, \ \  j=1,\ldots,N.
 \end{split} 
   \end{equation} 
An AP is considered reliable for RP $j $ if its RSS fingerprints are above a threshold \enquote{most of the time.} The indicator $ I_{j}^{i}$ denotes the APs whose readings satisfy \eqref{eq16} for, e.g., 90\% of the time during the fingerprint phase:
   \begin{equation}
   \label{eq17}
 I_{j}^{i}= \begin{cases}
          1 & \ \lvert \mathcal{T}_{j}^{i} \rvert \ge 0.9M\\
          0 & \text{otherwise}
      \end{cases} \ \ \ 
      i=1,\ldots,L, \ \ j=1,\ldots,N
       \end{equation} 
where $\left| \ . \ \right| $ denotes the cardinality and $I_{j}^{i}$ is called the \textit{coverage indicator} of AP $i$ at RP $\mathbf{p}_{j} $. Next, let $\mathcal{L}_{j}$ be the set of APs that satisfy \eqref{eq17} for RP $\mathbf{p}_{j} $, i.e.
        \begin{equation}
        \label{eq18}
      \mathcal{L}_{j}=\left\lbrace i \in \left\lbrace 1, \dots ,L \right\rbrace  \big | I_{j}^{i}=1 \right\rbrace, \ j=1\ldots,N.
 \end{equation} 
A binary coverage vector, $\mathbf{I}_{ j}=[I_{j}^{1},\ldots,I_{ j}^{L}]$, is assigned to each RP as an indication of the difference between RPs. Likewise a coverage vector $\mathbf{I}_{\boldsymbol y}=[I_{\boldsymbol y}^{1},\ldots,I_{\boldsymbol y}^{L}]$ is defined for online measurement $\boldsymbol y$ where the $i$-th entry is given by
\begin{equation}
\label{eq19}
I_{\boldsymbol y}^{i}=
\begin{cases}
         1 & \text{if } y^{i}\ge \gamma\\
         0              & \text{otherwise}
     \end{cases}
     \ \ i=1,\ldots,L
\end{equation}
\par The coarse localization is performed by selecting a subset of $ \tilde{\mathcal{P} } \subseteq \mathcal{P} $ RPs whose coverage vector $\mathbf{I}_{ j}$'s has distance from $\mathbf{I}_{\boldsymbol y}$ below a threshold as follows
\begin{equation}
\tilde{\mathcal{P}}=\left\lbrace j \in \{1,\ldots,N\} \big |d_{H}(\mathbf{I}_{\boldsymbol y}, \mathbf{I}_{j}  ) \le \eta \right\rbrace
\end{equation}
where the Hamming distance between $\mathbf{I}_{\boldsymbol y}$ and $\mathbf{I}_{j}$ is defined as
\begin{equation}
\label{eq21}
\begin{split}
d_{H}(\mathbf{I}_{\boldsymbol y}, \mathbf{I}_{j}  )= \sum _{i=1}^{L}|I_{\boldsymbol y}^{i}-I_{j}^{i}|, \ j\in \{1,\dots,N\}.
\end{split}
 \end{equation} 

\subsection{Affinity Propagation} 
The affinity propagation considers that the set of all RPs are the nodes $\mathcal{V}$ of a graph $\mathcal{G}=(\mathcal{V},\mathcal{E})$. The set of edges consists of all pairs $(j,j'), \ j,j'=1,\ldots,N $.  This method is based on an iterative message exchange between the nodes to find clusters and an exemplar (cluster head) for each cluster  \cite{r30}. The messages are simply the negative of the Euclidean distance  between the fingerprints $\boldsymbol \psi_{j}^{i}$ and $\boldsymbol \psi_{j'}^{i}$ at two different RPs ${j} $ and ${j'} $. Message passing between nodes reaches to a decision convergence in which the exemplars and corresponding clusters are defined. 
\par In the online phase, the online measurement is compared against the cluster heads and a set of clusters whose cluster heads have the least distances are selected as the coarse location of the user.
 \begin{figure*}[t!]
         \includegraphics[scale=0.7]{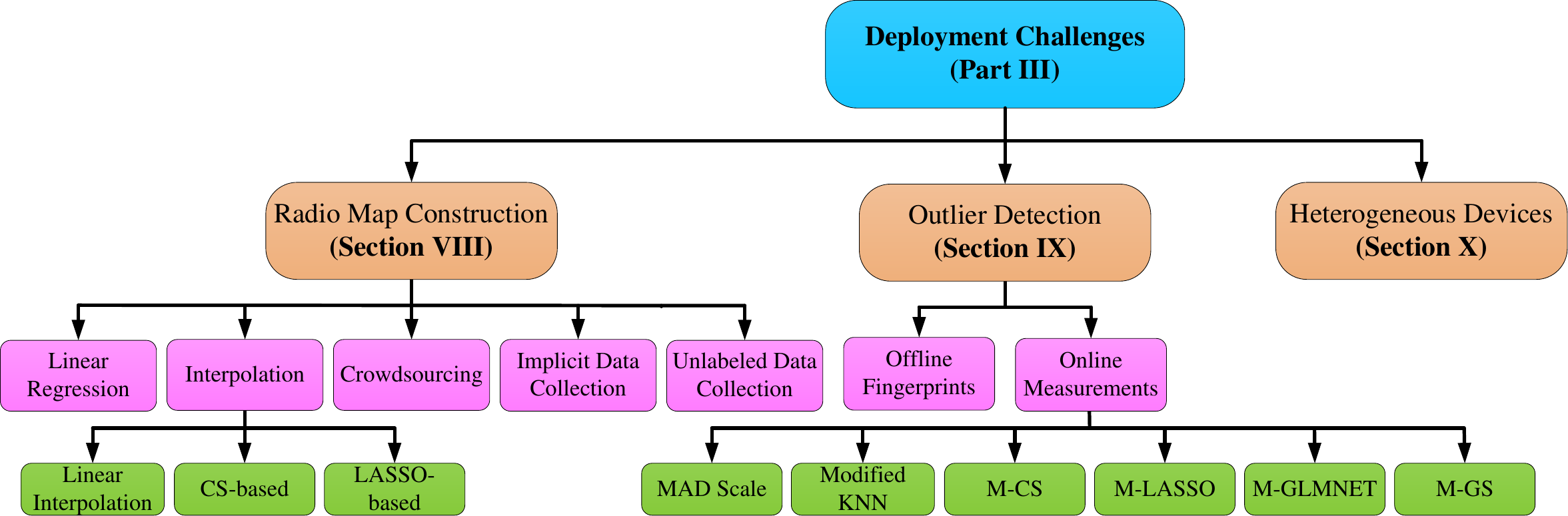}
              \centering

              \caption{Deployment challenges.}
              \label{fig10}
           \end{figure*}  \begin{table}[t!]
                                           \caption{Deployment Methods and Representative References}
                                            \label{table4}
                                           \begin{center}
                                           \begin{tabular}{ |c|c| } 
                                             \hline
                                          \textbf{Methods} &\textbf{ Related Works}\\
                                           \hline
                                              \hline
                                           Radio Map Construction &\cite{r108, r109, r135,r171,r189,r193,r226,r110, r112,r113,  r133, r16, r136, r132, r134, r30, r139}  \\ 
                                             \hline
                                         Outlier Detection  & \cite{r94, r204, r40, r95,r96,r97,r98, r36, r101, r151, r152, r139, r140}\\ 
                                             \hline
                                           Heterogeneous Devices & \cite{r227,r229,r186,r190} \\ 
                                             \hline
                                           \end{tabular}
                                           \end{center}
                                            \end{table} 
\subsection{K-means Clustering} 
K-means clustering also finds the clusters and their associated cluster centroids iteratively, however, the centroid of each cluster is updated at each iteration \cite{r35, r80}. In this method, the number of clusters should be defined a priori through a training set. Block-based weighted clustering is a further evolution of K-means clustering proposed in \cite{r137}, where a weighted least squares objective function is used. The distance between an RP and a centroid is minimized and the weights are obtained through a polynomial function of this distance. 
\subsection{Splitting-based Clustering} 
Unlike the conventional clustering schemes where a similarity measure groups RPs into a cluster, splitting-based clustering starts from the whole area and at each iteration (level) splits the area into four clusters. Then, the mean and variance of each cluster  defines a score, which signifies the distinction among the current level subclusters for AP $i$ \cite{r130}:
  \begin{equation}
  \label{eq20}
   \begin{split}
   \xi^{i}_{k}=\frac{\sum_{k,k'\in \mathcal{C}}^{}( \rho_{k}^{i}- \rho^{i}_{k'})^{2}}{\sum_{k=1}^{|\mathcal{C}|}\sigma_{k}^{i}}, \ i=1,\ldots,L\ 
    \end{split}
  \end{equation}
 where $ \mathcal{C}$ is the set of clusters, $k,k' \in \mathcal{C}$ are two distinct clusters, and $\rho_{k}^{i}$ and $\sigma_{k}^{i}$ are the mean and variance of fingerprints $ \psi_{j}^{i}$ of AP $i$ in cluster $k$, respectively. Each subcluster is labeled with a subset of $\mathcal{L'}_{k}\subseteq \mathcal{L} $ of APs that provide the largest score in \eqref{eq20}.  The  $\rho_{k}^{i}, \ i \in \mathcal{L'}_{k} $ is stored for subcluster $k$.  If $\xi^{i}_{k}$ is above a threshold for AP $i$, the cluster is divided into sub-clusters again. The clustering process ends when no subcluster satisfies this criterion.
 \par In coarse localization, the comparison with the online measurement $ \boldsymbol y $ is started from the first level of clusters. An Euclidean metric is computed between the online measurements and  each subcluster's mean, where the difference is computed only on the labeled APs for sub-cluster $k$.
\subsection{Weighted clustering} 
The weighted connection (edge) between two nodes is regulated by a similarity measure between the nodes \cite{r139}. This similarity is based on the fact that spatially close RPs should receive similar readings from the same set of APs. The similarity metric that reveals this feature is 
\par  The similarity $s(j,j')$ between RPs ${j}$ and ${j'}$ is defined as
  \begin{equation}
  \label{eq22}
 \begin{split}
   s(j,j')&=\begin{cases}
             \frac{1}{d_{H}(\mathbf{I}_{\boldsymbol j}, \mathbf{I}_{j'}  ) }, & \text{if} \ d_{H}(\mathbf{I}_{\boldsymbol j}, \mathbf{I}_{j'}  ) \ne 0\\
             \varLambda & \text{otherwise}
         \end{cases} \ \ \ 
    \\
   \forall j,j'&=1,\dots, N \ , \ j\neq j' 
   \end{split}
  \end{equation}
  which is proportional to the inverse of Hamming distance between two different RPs, and $\varLambda$ is a sufficiently large number.
   \par The trust on an RP includes the stability of the fingerprint readings  through the fingerprinting time. Hence, the variance of readings for all RPs is also computed as
   \begin{equation}
   \label{eq23}
   \begin{split}
    \Delta _{j}^{i}&=\frac{1}{M-1}\sum_{n=1}^{M} (r_{j}^{i}(t_{m})-\psi_{j}^{i})^{2} \\
   i&=1,\dots, L , \ j=1,\dots, N.
   \end{split}
   \end{equation} 
   The variance of ${j}$ is the average of variances in the set of APs that obey \eqref{eq18}, i.e. $\mathcal{L}_{j}$:
   \begin{equation}
   \label{eq24}
   \begin{split}
   \Delta _{j}=\frac{1}{\left| \mathcal{L}_{j} \right| }\sum_{l\in  \mathcal{L}_{j}}^{}\Delta _{j}^{l}, \ j=&1,\dots, N.
   \end{split}
   \end{equation}
\par The cluster, $\mathcal{C}(k)$, is the set comprising the cluster head, $\mathrm{CH}(k)$, and its followers, $\mathcal{FL}(k)$: 
      \begin{equation}
      \label{eq27}
      \mathcal{C}(k)=\left\lbrace  \mathrm{CH}(k) \right\rbrace \cup \mathcal{FL}(k).
      \end{equation}
\par  An RP is randomly selected as the cluster head $\text{CH}(k)$. The criteria for RP ${j}$ to be in its cluster, i.e. being assigned as $\mathcal{FL}(k)$, is that the similarity between RP $j$ and $\text{CH}(k)$ should be greater than a predefined value as
  \begin{equation}
  \label{eq26}
  \begin{split}
  j\in \mathcal{FL}(k) \ \ \ \text{if} \ \ \ s(j,\text{CH}(k))\ge \eta, \ k=1,\dots,K.
  \end{split}
  \end{equation}
Since the above criterion for each node may be satisfied for more than one cluster, each node has a table of CHs it belongs to. So, a node might be follower to more than one CH. 
\par The initial set for the potential cluster heads is the entire RPs. Once one cluster head and its followers are selected, the next cluster head is chosen randomly from the remaining set of RPs. This  process continues till the set of RPs that do not belong to any cluster is exhausted. 

\par Once all cluster members are defined, a representativity test is conducted within each cluster to select the best node as the representative node of that cluster, CH. This may lead to switching the CH of that cluster. The test measures the suitability of the CH to represent the characteristics of its followers.
    A node is selected as the CH when it has the least variance of fingerprints amongst all the cluster members:
  \begin{equation}
  \label{eq28}
  \begin{split}
  \text{CH}(k)&=\left\lbrace j \in \mathcal{C}(k)  | \Delta_{j}=\text{min} \ \left\lbrace \Delta_{l}\right\rbrace , l \in \mathcal{C}(k) \rvert \right\rbrace \\
 k&=1,\dots,K.
  \end{split}
  \end{equation}
\par The coarse localization is performed by selecting the cluster whose CH has the least distance from the online measurement $\boldsymbol{y}$. If the cluster with the minimum distance has RPs common with other clusters, then the neighbor cluster RPs are also included.  
\subsection{Layered Clustering}
 Another method for clustering RPs using the AP coverage vector is proposed in \cite{r140,r141}. This method is a layered clustering of RPs based on their similarity to the online reading. So, unlike the previous methods, the clustering is performed in the online phase. After defining the AP coverage vector as in \eqref{eq17} and \eqref{eq19} for the offline fingerprints and online vector $\boldsymbol y$, the Hamming distance $d_{H}(\mathbf{I}_{\boldsymbol y}, \mathbf{I}_{j}) $ between the online measurement coverage vector and that of each RP is computed.
\par The minimum and maximum of the Hamming distance over the area is defined as
  \begin{equation}
  \label{eq30}
  \begin{split}
  d_{H}^{min}&= \underset{j=1,\dots,N}{\text{min}} \ d_{H}(\mathbf{I}_{\boldsymbol y}, \mathbf{I}_{j})\\
  d_{H}^{max}&= \underset{j=1,\dots,N}{\text{max}} \ d_{H}(\mathbf{I}_{\boldsymbol y},  \mathbf{I}_{j}).
  \end{split}
  \end{equation}
Then, the group Hamming range is defined, as follows
  \begin{equation}
  \label{eq31}
  \begin{split}
  r&=\frac{d_{H}^{max}-d_{H}^{min}}{K}
  \end{split}
  \end{equation}
where $K$ is the number of groups (clusters) and is defined experimentally or from a training set. RPs are clustered with respect to their Hamming distances to the online measurement. Specifically, the distance range $\left[ d_{H}^{min}, d_{H}^{max} \right] $ is partitioned in $K$ groups collected in set $\mathcal{D}$ 
  \begin{equation}
  \label{eq32}
  \mathcal{D}=\left\lbrace  \left[ d_{k-1},d_{k}\right]  \big | d_{k}=d_{H}^{min}+kr, \ k=1,\dots,K\right\rbrace
  \end{equation}
where $d_{0}=d_{H}^{min}$. Then, ${j}$ is assigned to group $k$ if and only if
  \begin{equation}
  \label{eq33}
  d_{k-1}\le d_{H}(\mathbf{I}_{\boldsymbol y}, \mathbf{I}_{j}) \le d_{k}.
  \end{equation}
It could happen that $d_{H}(\mathbf{I}_{\boldsymbol y} , \mathbf{I}_{j})=d_{k} $, so, ${j}$ may belong to groups $k$ and $k+1$. In this case, ${j}$ is randomly assigned to one of these groups. The corresponding weight for each group is the inverse of the average of group Hamming distance
  \begin{equation}
  \label{eq34}
  w_{k}=\frac{2}{d_{k-1}+d_{k} } \ \ \ \forall k=1,\dots, K.
  \end{equation}
\par During the fine localization, all groups accompany in localization through corresponding weights. This clustering scheme is not for coarse localization, but is used (together with the weights) for the group sparsity based localization (Section \ref{Sparsity-based Localization}) that combines coarse and fine localization in a single step.
 \subsection{Spectral Clustering}
The similarity measure in spectral clustering is the pairwise cosine similarity between two RPs $j, j'$ as
\begin{equation}
s(j,j')=\frac{\langle \boldsymbol \psi_{j}, \boldsymbol \psi_{j'} \rangle}{ \Arrowvert \boldsymbol \psi_{j}\Arrowvert  \Arrowvert \boldsymbol \psi_{j'}\Arrowvert}
\end{equation}
RPs are grouped into a predefined number of clusters, $K$, so that the similarity of RSS vectors within the cluster is maximized \cite{r224}, i.e.,
\begin{equation}
\mathrm{max}\sum_{k=1}^{K}\sum_{j \in \mathcal{C}(k)}^{} s(\boldsymbol \psi_{j}, \boldsymbol \rho_{k})
\end{equation}
where  $\boldsymbol \rho_{k}$ is the average of fingerprint vectors $\boldsymbol \psi_{j}$ within cluster $k$. 
\section{Exploitation of APs for Localization} 
\label{Exploitation of APs for Localization}
The complexity of the indoor propagation environment causes several challenges associated with APs. We first elaborate on these challenges and then discuss approaches to address them. 
\subsection{Challenges Related to APs}
The challenges with APs can be generally divided into three main categories: 1) the unavailability of APs; 2) large set of available APs, from which a subset of  APs should be selected; and 3) faulty APs. The first two issues are elaborated in this section and the last one is discussed  Section \ref{Outlier Detection}.
\par
\begin{figure}[t!]
             \includegraphics[scale=0.5]{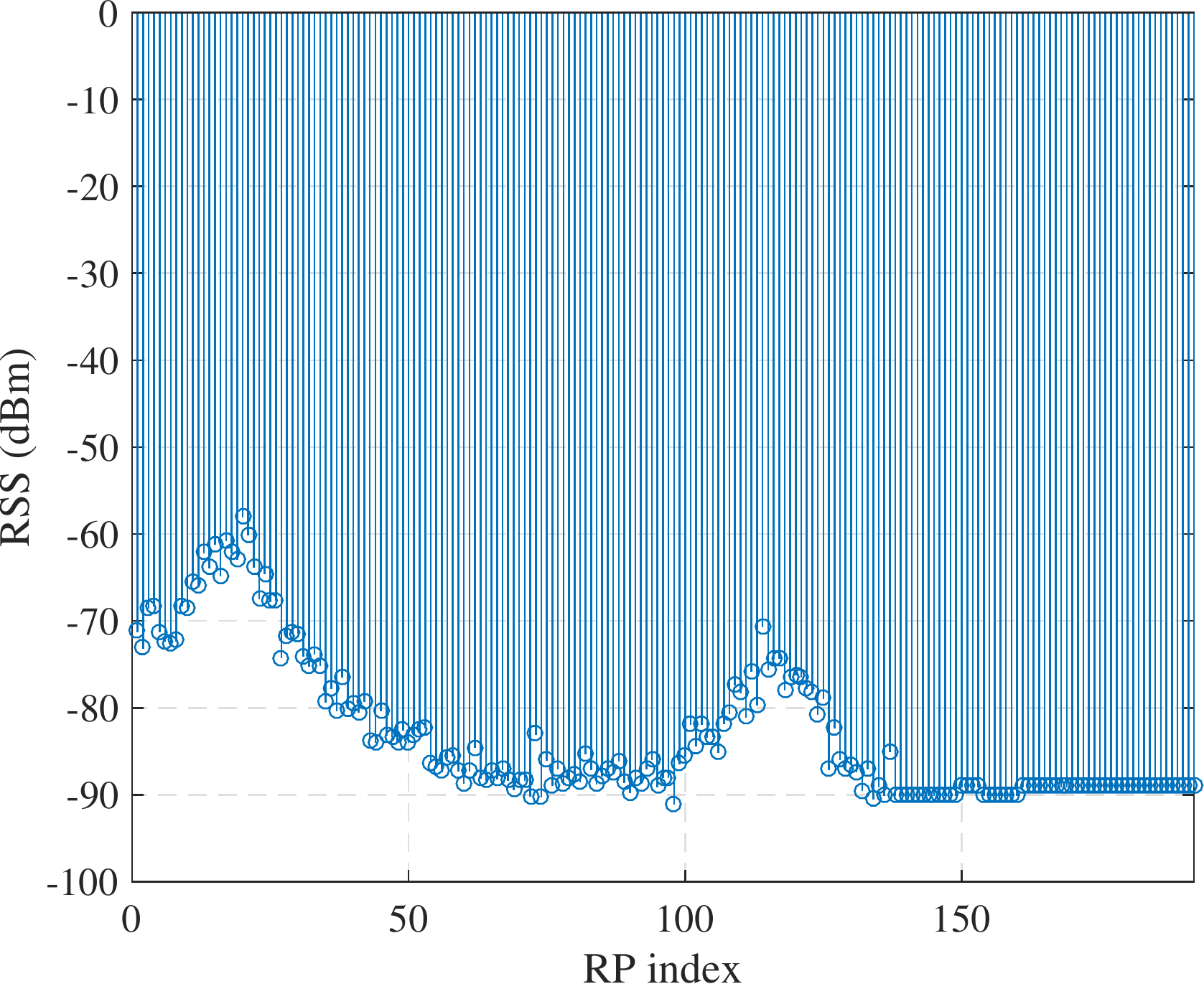}
             \centering
             \caption{The RSS profile for a single AP.}
              \label{fig4}    
      \end{figure}

 The main reason for unavailability of APs is the range limitation. For instance, a typical IEEE 802.11b AP provides a coverage of less than 100m at 5.5 Mbps. To provide a ubiquitous network coverage, multiple APs are installed in buildings. So, not all APs can provide RSS signatures for a single RP. In large areas, a subset of APs is visible on the user's device, however, if the user moves far from the previous location, another subset of APs become visible. For instance, Fig. \ref{fig4} shows the RSS profile for a single AP in a real environment. The RPs are numbered as specified by the horizontal axis index. This figure indicates that the device cannot receive signals from the AP when located at the RPs beyond 140.
\par In addition, due to the wide deployment of APs in indoor settings, including all APs in positioning is not recommended due to the following issues:
\begin{itemize}
\item  The number of available APs is usually more than the minimum  needed (3 APs for 2D localization and 4 APs for 3D localization) for positioning.
\item Advanced APs can transmit in different channels with  different Media Address Control (MAC) settings. Including all MAC addresses for one AP does not add information to the system. For example, during an experiment in a typical real office environment, we found a total of 268 MAC addresses. 
\item  APs usually provide correlated readings. This correlation may occur in three ways: 1) Neighbor RPs may receive correlated fingerprints from a specific AP because the RSS fingerprints are obtained from the same signal received in close locations. This prevents the distinguishability between RPs. 2) A pair of APs may provide correlated fingerprints for an RP.  This issue occur when the APs are located close to each other but belong to different networks. So, APs from different networks produce similar measurements and  engaging all  may introduce biased position estimates, incur overfitting, and impose time and computational complexity \cite{r85}.  3) Fingerprints at one RP may be correlated during the fingerprinting time \cite{r232, r84}, which incurs a large difference between the fingerprints and online measurements.
\end{itemize}
\par To mitigate the above effects two major tasks are usually performed, namely, 1) feature selection that  maps the information of APs to other domains to obtain more distinguishable representation; and 2) AP selection, whereby a subset of APs that better represents the characteristics of the environment is selected. The focus of this paper is on  AP selection.

\subsection{AP Selection Methods}
\label{Selection of APs}
\par AP selection can be performed in both offline and online phases. If the AP selection is performed in the offline phase, a subset of APs is selected using only the radio map regardless of the online measurements. However, if the characteristics of the environment are different from the online localization phase, this selection mechanism fails to choose a suitable subset of APs. Hence, the RSS readings from specific APs are selected in the online phase utilizing the online measurements explicitly or implicitly. In explicit utilization, a subset of the APs are selected considering only the online measurement. In implicit utilization, the selection of APs is performed exploiting both online measurements and radio map. One method is to select  a subset of APs from the online measurements and apply the offline AP selection techniques on this subset. Another method is to apply the offline AP selection methods on the radio map using only the RPs that have been selected at the coarse localization stage. This way, the fingerprints of the RPs that are the most similar to the online measurements assist in AP selection.
\par To formulate the process, define an AP selection matrix $\mathbf{\Phi}$ which selects a subset of APs $ \mathcal{L'} \subseteq \mathcal{L} $. Let $L' \le L$ be the cardinality of $\mathcal{L'}$.  The $i$-th row of $\mathbf{\Phi}$, i.e., $ \mathbf{\Phi}^{i} $, is a $ 1\times L $ vector that defines the selected AP through zeroing out all indices except the selected AP index as
  \begin{equation}
  \mathbf{ \Phi}^{i}=[\ldots,\underbrace{1}_\text{Index of selected AP},\ldots], \  i=1,\ldots, L'.
  \end{equation}
Hence, the modified localization problem of   \eqref{eq5} is
\begin{equation}
\hat{\mathbf{p}}=f(\mathbf{\Phi}\mathbf{R}, \mathbf{\Phi} \boldsymbol y).
\end{equation}
and the localization methods of Section \ref{WLAN Fingerprinting Localization: Problem Formulation and Conventional Approaches} are performed on $\mathbf{y}=\mathbf{\Phi} \boldsymbol y$ instead of $\boldsymbol y$.
\par Although a plethora of methods have been already introduced in \cite{r80}, the following provides a summary of recently introduced AP selection methods:
 
\subsubsection{Strongest APs (MaxMean)} The early studies  advocate to select APs based on their signal strengths in the online phase and select the same set of APs from the radio map fingerprints \cite{r19}. The intuition is that the strongest APs provide most coverage time and render more accurate measurements. Different set of APs  are selected if the user travels into different locations. The strongest AP selection scheme, however, may not always render a suitable criterion \cite{r35}.
\subsubsection{Fisher Criterion} The Fisher criterion is a metric that quantifies the discrimination ability of each AP across RPs and takes into account the stability of AP fingerprints. This metric uses the statistical properties of the radio map fingerprints and selects APs based on their performance during the offline fingerprinting period. A score is assigned to each AP separately as
  \begin{equation}
  \label{eq35}
    \begin{split}
   \zeta^{i}&=\frac{\sum_{j=1}^{N}( \psi_{j}^{i}-\bar{ \psi}^{i})^{2}}{\frac{1}{M-1}\sum_{m=1}^{M}\sum_{j=1}^{N}(r_{j}^{i}(t_{m})- \psi_{j}^{i})^2},\\
   &\ \ \ \ \ \ \ \ \ \ \ \ \ i=1,\dots,L
    \end{split}
    \end{equation}
    where 
    \begin{equation}
    \label{eq36}
    \bar{ \psi}^{i}= \frac{1}{N}\sum_{j=1}^{N} {\psi}_{j}^{i} .
    \end{equation} 
This criterion is based on the fact that APs with higher variance should receive smaller scores as they are less reliable. This score is sorted decreasingly for all APs and a number of APs with the highest scores are selected \cite{r21,r80,r156,r30}. However, the Fisher discriminant analysis for AP selection considers the offline fingerprints only. If the APs are not available in the online phase or provide faulty online measurements, then this criterion is not a suitable one. This issue is discussed in Section \ref{Outlier Detection}.
\subsubsection{Bhattacharyya distance} This AP selection method is better suited to methods that utilize statistical properties of the radio map as it measures the distance between the probability densities of the fingerprints from two APs $i,\ i'$ at RP $j$:
  \begin{equation}
  \label{eq37}
  \begin{split}
    d_{B}(r_{j}^{i},r_{j}^{i'})=-\mathrm{ln}\left( \int_{}^{} \sqrt{f_{j}^{i}(r_{j}^{i})f_{j}^{i'}( r_{j}^{i'})dr_{j}^{i}dr_{j}^{i'}} \right) 
    \end{split}
    \end{equation}
    where $f_{j}^{i}(r_{j}^{i})$ and $f_{j}^{i'}( r_{j}^{i'})$ are the fingerprint distributions at APs $ i$ and $ i'$. Although the fingerprints are not Gaussian distributed, computing \eqref{eq37} under the Gaussian assumption provides an acceptable distance measure \cite{r21}. This measure gives a score to each pair of APs. Unlike the previously discussed methods, this method selects  a subset $\mathcal{L'}$ of APs by choosing for each RP $j$ the pairs of APs with the highest scores.  To this end, it needs an exhaustive search over  
    $\begin{psmallmatrix}L \\    
                  L'\end{psmallmatrix}$ $\begin{psmallmatrix} L'  \\             
                                      2\end{psmallmatrix}$ pairs to find the ones with the smallest distance according to \eqref{eq37}.
\subsubsection{Information Potential (IP)} This AP selection method also measures the distance between the RSS fingerprints \cite{r21}
    \begin{equation}
    \label{eq38}
    \begin{split}
      d_{I}(r_{j}^{i},r_{j}^{i'})=-\mathrm{ln}\left( \frac{1}{M^{2}}\sum_{m=1}^{M} \sum_{m'=1}^{M} k(r_{j}^{i}(t_{m}),r_{j}^{i'}(t_{m'})) \right) 
      \end{split}
      \end{equation}
  where $k(\cdot,\cdot)$ is a kernel function of the distance between each single fingerprint at APs $i, \ i'$  and RP $ j$.  The selection of APs is similar to the one under the Bhattacharyya distance. This criterion also selects pairs of APs and hence, suffers from the exhaustive search set.
\subsubsection{Information Gain (InfoGain)}  This offline criterion  selects the  APs with the highest discriminative power. The discriminative power of AP $i$ is measured through the mutual information between two random variables \cite{r35} as follows
  \begin{equation}
    \label{eq39}
    \begin{split}
      \text{InfoGain}(r^{i}_{j})=H(\mathbf{p})-H(\mathbf{p}|r^{i}_{j})
      \end{split}
      \end{equation}      
where $H(\mathbf{p})=-\sum_{j=1}^{N}f(\mathbf{p}_{j})\mathrm{log}f(\mathbf{p}_{j})$,  $H(\mathbf{p}|r^{i}_{j})=-\sum_{j=1}^{N}f(\mathbf{p}_{j})\sum_{v}^{}f_{j}^{i}(r_{j}^{i}=v|\mathbf{p}_{j})\mathrm{log}\frac{f_{j}^{i}(r_{j}^{i}=v|\mathbf{p}_{j})f(\mathbf{p}_{j})}{f_{j}^{i}(r_{j}^{i}=v)}$, and $v$ is one possible value of signal strength for AP $i$. The distributions $f_{j}^{i}(r_{j}^{i}=v|\mathbf{p}_{j})$ and $f_{j}^{i}(r_{j}^{i}=v)$ are estimated analytically, through histograms, or using kernels. A modification that ranks APs jointly by InfoGain and mutual correlation has also  been introduced \cite{r88}.
\subsubsection{Entropy Maximization}
The entropy maximization for AP $i$ discretizes the RSS range into $u \in \{1, \dots, U\}$ levels. The probability of occurrence for level $u$ is
\begin{equation}
f^{i}(u)=\frac{N_{u}^{i}}{N^{i}}
\end{equation}
where $N_{u}^{i}$ is the number of RPs whose RSS is in level $u$ and $N^{i}$ is the number of RPs that detect AP $i$. The entropy of AP $i$ is given by
\begin{equation}
H^{i}=-\sum_{u=1}^{U}f^{i}(u)\mathrm{log}_{2}f^{i}(u).
\end{equation}
A subset of the APs with the maximum entropy are selected for localization.
\subsubsection{Group Discrimination (GD)} The idea  is that a group of APs that provides the maximum discrimination is selected, rather than choosing the APs independently.  This method finds the best subset $\mathcal{L'} \subset \mathcal{L}$ of APs that produce the least score for the subset. The score is defined as follows  \cite{r27}
  \begin{equation}
    \label{eq40}
    \begin{split}
      & \xi_{j,j'}=\sum_{m=1}^{M}\sum_{m'=1}^{M} \alpha_{j}\alpha_{j'}k( r^{i}_{j}(t_{m}), r^{i}_{j'}(t_{m'})), \ i \in  \mathcal{L'} \\
      &\text{Score}_{\mathcal{L'}}=\sum_{j,j' \in \mathcal{P}}^{} \xi_{j,j'}
      \end{split}
      \end{equation}      
where $\alpha_{j}$ and $\alpha_{j'}$ are the coefficients for RPs $j$ and $j'$ and $k(\cdot,\cdot)$ is an exponential kernel function. There is a total of $\frac{L!}{L'!(L-L')!}$ combinations that needs to be searched, and hence, this method needs an exhaustive search over the set of APs.
  
\subsubsection{Joint Selection} This method is similar to  Fisher criterion. However, instead of computing the differentiability of RPs with respect to the mean RSS value $\bar{ \psi}^{i}$, the mutual differentiability between RPs is computed \cite{r130,r88}:
  \begin{equation}
  \label{eq41}
  \begin{split}
  \xi^{i}=\frac{\underset{1<j<j'<N}{\sum \sum}(\psi^{i}_{j}-\psi^{i}_{j^{\prime}})^{2}}{\frac{1}{M-1}\sum_{m=1}^{M}\sum_{j=1}^{N}(r_{j}^{i}(t_{m})- \psi_{j}^{i})^2}.
  \end{split}
  \end{equation}
A subset of $L'$ APs with the least scores is selected. 
\section{Advanced Density And Weight Estimation Methods}
\label{Advanced Density, weight estimation, and Data Metric}
The fine localization accuracy highly depends on the distance between online measurements and RSS radio map fingerprints. An incorrect metric may not lead to a representative difference and can cause a biased estimation towards specific RPs. To better exploit the features in the offline fingerprints and provide a more accurate comparison with the online measurements, more advanced techniques have been proposed for fingerprint density estimation---as a modification to \eqref{eq11} from which the weight for each RP is computed--or for directly computing weight for each RP--- as a modification to \eqref{eq14}. In this section, we elaborate on these methods.

\subsection{Kernel Density Estimation (KDE) Method} 

\par  As discussed previously, one of the approaches is the non-parametric estimation of the RSS prior distribution. Since the usual analytical assumptions on the prior probability, such as Gaussianity, do not necessarily hold, the parametric estimation cannot exactly capture the empirical characteristics of the fingerprints \cite{r23}. An alternative approach is to estimate the empirical fingerprint distributions non-parametrically. 
\par An approach to estimate the empirical distributions is to use kernel density estimation (KDE) as follows \cite{r21, r29}:
  \begin{equation}
  \label{eq42}
  \hat{f}(\boldsymbol y|\mathbf{p}_{j})=\frac{\sigma^{-L'}}{M}\sum_{t=1}^{M}k\left( \frac{\boldsymbol y-\mathbf{r}_{j}(t_{m})}{\sigma}\right) 
  \end{equation}
where $k(\cdot)$ is the kernel function, $\sigma$ is the kernel width estimated through either training sequence or analytical solutions provided for Gaussian kernels \cite{r102}, and $L'$ is a the number of the APs used for localization. The KDE is based on a superposition of kernel functions centered around the fingerprints.
\par The kernel functions can also be used for weight computations. Consider first that the weights are obtained through an average normalized inner product between the fingerprints and online measurements
  \begin{equation}
  \label{eq43}
  w_{j}=\frac{1}{M}\sum_{t=1}^{M}\frac{\langle \boldsymbol y,\mathbf{r}_{j}(t_{m})\rangle}{\Arrowvert \boldsymbol y \Arrowvert\Arrowvert  \mathbf{r}_{j}(t_{m}) \Arrowvert}
  \end{equation}
where $\langle\cdot\rangle$ denotes the inner product. This metric basically measures the angles between the online measurements and radio map fingerprints. As discussed earlier at the beginning of this section, if the AP readings are correlated, this angle is small, and hence, not a representative metric. 
\par For the sake of better differentiability between APs, an alternative approach is to map the data to another space where the difference between these angles becomes larger and more distinguishable. Let $\varphi$ be a nonlinear mapping such that $\varphi:\mathbf{x}\longmapsto\varphi(\mathbf{x})$. The transformed weights of \eqref{eq43} take the form
  \begin{equation}
  \label{eq44}
  \begin{split}
  w_{j}&=\frac{1}{M}\sum_{t=1}^{M}\frac{\langle \varphi (\boldsymbol y), \varphi(\mathbf{r}_{j}(t_{m}))\rangle}{\Arrowvert \varphi( \boldsymbol y) \Arrowvert\Arrowvert  \varphi(\mathbf{r}_{j}(t_{m})) \Arrowvert}\\
  &=\frac{1}{M}\sum_{t=1}^{M}\frac{ k (\boldsymbol y, \mathbf{r}_{j}(t_{m}))}{\sqrt{k( \boldsymbol y, \boldsymbol y)   k( \mathbf{r}_{j}(t_{m}),\mathbf{r}_{j}(t_{m}))}   }.
  \end{split}
  \end{equation}
One should note that in \eqref{eq44} the kernel function computes the inner product between the mapped online measurement and fingerprints and thus, the specific definition of the nonlinear mapping is circumvented. 
\subsection{Principal Component Analysis (PCA) Method}
An alternative approach for computing $\hat{f}(\boldsymbol y|\mathbf{p}_{j})$ is to map the online measurements to the domain of its principal components (PCs)  \cite{r103}. First, the sample covariance of the fingerprints at RP $j$ is computed as
\begin{equation}
  \label{eq48}
  \begin{split}
  \mathbf{C}_{\boldsymbol y|\mathbf{p}_{j}}(i,i')&=\frac{1}{M}\sum_{m=1}^{M}(r_{j}^{i}(t_{m})- \psi_{j}^{i})(r_{j}^{i'}(t_{m})- \psi_{j}^{i'})^{T}\\
  & i,i'=1,\ldots,L, \ j=1,\ldots,N
  \end{split}
  \end{equation}
and the global covariance matrix has entries
\begin{equation}
  \label{eq483}
  \begin{split}
  \mathbf{C}_{\boldsymbol y}(i,i')&=\frac{1}{MN}\sum_{j=1}^{N}\sum_{m=1}^{M}(r_{j}^{i}(t_{m})- \psi_{j}^{i})(r_{j}^{i'}(t_{m})- \psi_{j}^{i'})^{T}\\
  i,i'&=1,\ldots,L.
  \end{split}
  \end{equation}
The eigenvectors of the global covariance matrix are defined as follows: 
\begin{equation}
  \label{eq482}
  \mathbf{C_{\boldsymbol y}} \cdot \mathbf{v}_{\ell}= \lambda_{\ell} \cdot \mathbf{v}_{\ell}, \ \ell=1,\ldots, L.
  \end{equation}
A transformation of the data to its PCs is achieved through concatenating the eigenvectors corresponding to the eigenvalues sorted decreasingly as
\begin{equation}
  \label{eq47}
  \begin{split}
  \mathbf{A}=[\mathbf{v}_{1},\ldots,\mathbf{v}_{L}],\  \lambda_{1} \ge  \lambda_{2} \ge \ldots,\lambda_{L}.\\
  \end{split}
  \end{equation}
The fingerprints and the online measurements should be transformed to the PC domain. To this end, the online measurement, the radio map, and the covariance matrix are mapped to the PC domain through multiplication with the transformation matrix $\mathbf{A}$ as
  \begin{equation}
  \label{eq45}
 \mathbf{q}=\mathbf{A} \boldsymbol y, \ \bm{\mu}_{\mathbf{q}|\mathbf{p}_j}=\mathbf{A} \bm{\psi}_j, \ \mathbf{C}_{\mathbf{q}|\mathbf{p}_{j}}=\mathbf{A} \mathbf{C}_{\boldsymbol y|\mathbf{p}_{j}}\mathbf{A}^{T}, \ j=1,\ldots\,N.
  \end{equation}
The posterior probability of \eqref{eq11} is computed using only the first $L' \le L$  PCs as
  \begin{equation}
  \label{eq46}
  \hat{f}( \mathbf{q'}|\mathbf{p}_{j})=\prod_{i=1}^{L'}\frac{1}{\sqrt{2\pi \mathbf{C}_{\mathbf{q}|\mathbf{p}_{j}}(i,i')}} \exp \left( -\frac{1}{2}\frac{(q^{i}- \mu^{i}_{\mathbf{q}|\mathbf{p}_j})^{2}}{\mathbf{C}_{\mathbf{q}|\mathbf{p}_{j}}(i,i')}\right) 
  \end{equation}
where $\mathbf{q'}$ is the first $L'$ entries of $\mathbf{q}$ as $\mathbf{q}=[q^{1},\ldots,q^{L'}]^{T}$. 
  
\subsection{KL-Divergence Method} 
The Kullback-Leibler (KL) divergence is fundamentally a distance between two probability density functions---namely of the online measurements $f^{i}(y^{i})$ and RSS fingerprints $f_{j}^{i}(r_{j}^{i})$---written as a kernel function \cite{r229,r230}. To obtain the probability density function of the online measurement,  the user should remain at his/her location to get multiple online measurements. The symmetrized KL divergence between two probability functions is computed as
\begin{equation}
D(f^{i}(y^{i}),f_{j}^{i}(r_{j}^{i}))=\mathrm{KL}(f^{i}(y^{i})|| f_{j}^{i}(r_{j}^{i}))+\mathrm{KL}(f_{j}^{i}(r_{j}^{i})||f^{i}(y^{i}))
\end{equation}
where $\mathrm{KL}(\cdot,\cdot)$ is the KL divergence
\begin{equation}
\mathrm{KL}(X||Y)=\sum_{v}^{}f(X=v)\mathrm{log}\left( \frac{f(X=v)}{f(Y=v)}\right) .
\end{equation}
The KL divergence is combined with a kernel function in order to yield the weights for location estimation: 
\begin{equation}
w_{j}=\mathrm{exp}\left( -\alpha\sum_{i=1}^{L}D(f^{i}(y^{i}),f_{j}^{i}(r_{j}^{i}))\right).
\end{equation}

\subsection{Geometry-based Localization}
Tilejunction \cite{r234}, Sectjunction \cite{r235}, and Contour-based trilateration \cite{r237} are recent methods that exploit the geometry of the area to come up with weights which are found by solving a convex optimization problem with environmental constraints such as presence of walls. Define
\begin{equation}
\Gamma^{i}_{j}= (y^{i}-\psi_{j}^{i})^{2}+(\sigma^{i})^{2}+(\Delta_{j}^{i})^{2}
\end{equation} 
where $\sigma^{i}$  is the variance of   $ \psi_{1}^{i},\ldots ,\psi_{N}^{i}$ and $\Delta_{j}^{i}$ was defined in \eqref{eq23}. The difference between the offline fingerprints and the online measurements are computed through the expected signal difference as 
\begin{equation}
\Gamma_{j}= \frac{1}{N}\sum_{i=1}^{L}\Gamma^{i}_{j}
\end{equation}
The user's location is estimated as $\hat{\mathbf{p}} =\sum_{j=1}^{N}w_{j}\mathbf{p}_{j}$, where the weights are computed from the following linear program:
\begin{equation}
\begin{split}
\underset{\{w_{j}\}_{j=1}^{N}}{\mathrm{argmin}}& \sum_{j=1}^{N}w_{j} \Gamma_{j}\\
\mathrm{s.t.} \ \ &\text{Environment constraints}\\
& \sum_{j=1}^{N}w_{j}=1,\ w_{j}\ge 0.
\end{split}
\end{equation}
\section{Sparsity-based Localization}
\label{Sparsity-based Localization}
As the computational complexity of the probabilistic approaches is high and the localization accuracy of the deterministic approaches is low, a new reformulation of the WLAN localization problem has been proposed. This section elaborates on the sparse reformulation of the WLAN localization problem and introduces the methods that solve the sparsity-based localization problems.
\subsection{Measurement Model Enabling Sparse Recovery}
The localization problem can be  interpreted as finding only one location among  all RPs, which is the closest  to the user position. The localization can be transformed into a sparse recovery problem with only one selection out of many options \cite{r30,r121}. Let the location vector be recast as a sparse vector as
  \begin{equation}
  \label{eq49}
  \boldsymbol \theta = [ 0,\ldots,0,1,0,\ldots,0 ]^{T}
  \end{equation}
where all entries of $\boldsymbol \theta$  correspond to radio-map RPs and 1 corresponding to the index of the RP to which the user is the closest. The equivalent measurement model that enables sparse recovery is
  \begin{equation}
  \label{eq50}
  \mathbf{y}=\mathbf{\Phi \Psi \boldsymbol\theta +\boldsymbol\epsilon}
  \end{equation}
where $  \boldsymbol \Phi  $ is the AP selection matrix, i.e. the matrix that selects certain elements of $\mathbf{\Psi}$ corresponding to selected APs (Section \ref{Selection of APs}), $ \boldsymbol \Psi $ is the modified radio map matrix,$ \ \boldsymbol \epsilon $ is the error vector, and $ \mathbf{ y }$ is the online  captured RSS vector from specific APs as
  \begin{equation}
  \label{eq51}
  \mathbf{y}=\boldsymbol \Phi \boldsymbol y.
  \end{equation}
Since the dimension of $ \boldsymbol y $ is less than that of $ \boldsymbol \theta $,  \eqref{eq50} is an under-determined problem. next, we discuss the sparsity-based localization methods that solve this problem.
\subsection{CS-based Localization}
Although  under-determined problems may have infinite solutions,  the location vector $ \boldsymbol \theta $ in  \eqref{eq50} is sparse as the user can only be in one of the RP locations. This type of problems can be addressed through Compressive Sensing (CS), and may have unique solutions if certain conditions are satisfied. The CS problem can be solved via the convex optimization
  \begin{equation}
  \begin{split}
  \label{eq52}
   \hat{\boldsymbol\theta} =\underset{\boldsymbol\theta}{\text{argmin}} \Arrowvert \boldsymbol\theta \Arrowvert_{1} \\
    s.t.\  \mathbf{y}=\mathbf{\Phi \Psi \boldsymbol\theta}
  \end{split}
  \end{equation}
where $ \Arrowvert \boldsymbol \theta \Arrowvert _{1} $ is the $\ell_{1}$-norm of $ \boldsymbol\theta $. Using the $ \ell_{1}$-norm, the CS renders a sparse vector. Under certain conditions listed shortly, problem \eqref{eq52} has a unique solution. Several algorithms have been proposed to solve this problem, e.g. greedy algorithms  \cite{r50}, iteratively re-weighted linear least-squares (IRLS)  \cite{r52}, and  basis pursuit \cite{r120}. 
  \par The  CS  formulation faces several challenges. We enumerate these challenges next and provide improvements in the ensuing subsections.
  \begin{enumerate}
  \item In order to obtain a unique sparse solution in CS formulation, the sensing matrix $\mathbf{\Phi}$  and the basis matrix  $\mathbf{\Psi}$ should obey two criteria \cite{r157}:
    \begin{itemize}
    \item \textit{Restricted Isometry Property (RIP)}: This property states that the multiplication of the sensing and the basis matrix, i.e. $\mathbf{\Phi \Psi}$, should approximately preserve the Euclidean norm of the positioning vector \cite{r157}. Mathematically, this condition is expressed as
     \begin{equation}
         (1-\delta_{s})\Arrowvert\boldsymbol\theta \Arrowvert_{2}^{2} \le \Arrowvert \mathbf{\Phi \Psi}\boldsymbol\theta \Arrowvert_{2}^{2} \le  (1+\delta_{s})\Arrowvert\boldsymbol\theta \Arrowvert_{2}^{2}.
          \end{equation}
    where $\delta_{s}$ is a small positive number. The above condition for example would be satisfied if the matrix $ \mathbf{\Phi \Psi}$ were orthonormal.
    \item \textit{Mutual Incoherence:} This requires that the rows of $\mathbf{\Phi}$ cannot sparsely represent the columns of  $\mathbf{\Psi}$ and vice versa. Smaller coherence leads to a better chance to reconstruct a unique and optimum sparse solution \cite{r115}.
    \end{itemize}
    To induce the above conditions, an orthonormalization procedure on the radio map is applied \cite{r30}. Nonetheless, this procedure does not make $ \boldsymbol\Phi\boldsymbol\Psi $ completely orthonormal, as it is not square.
    
  \item The computational complexity of the optimization algorithm increases with the size of area and hence makes the positioning impractical in small hand-held devices. Hence, a preprocessing step is required to reduce the searching area and this is done through the radio map clustering algorithms (Section \ref{RP Clustering}).
    
    \item The CS  optimization formulation assumes that the model \eqref{eq50} does not contain the measurement error $\boldsymbol \epsilon$ and attempts to find the RPs whose fingerprints match the online measurements exactly. 
  \end{enumerate}

\subsection{LASSO-based Localization}  
\par The shortcomings of the CS localization are overcome by recent sparse recovery methods which do not need the orthogonalization step, and not rely on special properties of the matrix $ \boldsymbol\Phi\boldsymbol\Psi $, which may not be valid in practice. In addition to recovering a sparse vector, the proposed localization methods use \eqref{eq50} as the model, and thus, work better with noisy measurements.
\par The localization accuracy can be improved if the sparse recovery problem also suppresses the error between the online measurement vector and radio map fingerprints. The LASSO localization minimizes the $\ell_{1} $-norm of the location vector and the $\ell_{2} $-norm of the residuals \cite{r139}. The convex optimization problem for localization is reformulated as
    \begin{equation}
    \label{eq53}
   \hat{\boldsymbol \theta} =\underset{\boldsymbol \theta}{\text{argmin}} \ \left[ \frac{1}{L'}\Arrowvert \mathbf{y-\mathbf{H} \boldsymbol \theta}\Arrowvert_{2}^{2} + \lambda \Arrowvert\boldsymbol \theta\Arrowvert_{1}\right]
    \end{equation}
where $ \mathbf{H}=\mathbf{\Phi \Psi }$ and $ \lambda \ge 0$ is a tuning parameter. This problem is also known as $\ell_{1} $-penalized least squares, which incorporates feature and model selection into the optimization \cite{r147}.  The first component seeks coefficients that minimize the residuals, and the second one promotes a sparse $\boldsymbol \theta$.  LASSO has been shown to be more indifferent to  correlated RSS fingerprints. The parameter $ \lambda$ is a tuning parameter that regularizes between minimizing the residuals and the sparse vector solution. This parameter can be tuned experimentally or using cross validation (CV) \cite{r228}.
\subsection{GLMNET-based Localization}
\par Suppose there are correlated predictors in the modified radio map. If the user is exactly at an RP, the online measurement is supposed to be very similar to the fingerprints of that RP. Another possible case is when the user is between two RPs with similar environmental features. The location estimation problem in both cases is expected to assign higher coefficients to the points with correlated fingerprints. Hence, the correlated predictors should be allowed to jointly borrow strength from each other. GLMNET-based localization incorporates the above features  as follows \cite{r139}:
    \begin{equation}
    \label{eq54}
    \begin{split}
  \hat{\boldsymbol \theta} &=\underset{\boldsymbol \theta}{\text{argmin}} \ \left[ \frac{1}{L'}\Arrowvert \mathbf{y-\mathbf{H }\boldsymbol \theta}\Arrowvert_{2}^{2} +  P_{\alpha}  \right]\\
  P_{\alpha}&=\lambda \bigl((1-\alpha)\Arrowvert \boldsymbol \theta \Arrowvert _{2}^{2}+\alpha \Arrowvert \boldsymbol \theta \Arrowvert_{1})\bigr)
    \end{split}
    \end{equation}
where $ \lambda \ge 0$ is a tuning parameter and $0\le\alpha\le 1 $ is a compromise between ridge regression and LASSO. Ridge regression promotes the shrinkage of the coefficients of correlated radio map columns towards each other and is expressed by the $\Arrowvert \boldsymbol \theta \Arrowvert _{2}^{2}$ objective. Hence, we take advantage of the correlation between the radio map readings. If $\alpha=0 $, the above formulation amounts to the ridge regression. As  $\alpha$ increases from $0$ to $1$ for a given $\lambda$ the sparsity of the solution increases monotonically from $0$ to the sparsity of the LASSO solution. This formulation therefore jointly considers the correlated predictors and finds a sparse solution for the user's pose. This estimator is known as GLMNET \cite{r148}. 
\par The computational complexity of the above optimization problem grows with the number of predictors (size of radio map). Therefore, the previously mentioned coarse localization schemes in Section \ref{RP Clustering} reduce the size of the area that the optimization problems  seek for the solution and hence reduce the computation time. This allows these procedures to be executed on resource-limited devices.  
\subsection{Group Sparsity (GS)-based Localization} 
\par Since there is no guarantee that the cluster within which the solution is searched is the correct cluster, Group Sparsity (GS)-based localization is proposed which utilizes all the clusters, each with a different weight in the following optimization \cite{r140}
    \begin{equation}
      \label{eq55}
     \hat{\boldsymbol \theta} =\underset{\boldsymbol \theta}{\text{argmin} }\ \left[ \frac{1}{L'}\Arrowvert \mathbf{y-\mathbf{H} \boldsymbol \theta}\Arrowvert_{2}^{2} + \lambda_{1} \Arrowvert\boldsymbol \theta\Arrowvert_{1}+\lambda_{2}\sum_{k=1}^{K}w_{k}\Arrowvert\boldsymbol \theta_{k}\Arrowvert_{2}\right]
      \end{equation}
where  $\boldsymbol \theta_{k}$ is a segment of position vector corresponding to group $k$, $w_{k}$ is the weight assigned to group $k$, $K$ is the total number of groups,  and $ \lambda_{1}, \lambda_{2} \ge 0$ are tuning parameters. The weights $w_{k}$ can be obtained from any of the coarse localization methods discussed in Section \ref{RP Clustering}. The first component minimizes the  impact of online measurement noise considering that the RSS fingerprint noises have already been minimized through time-averaging of the fingerprints. The second component promotes sparsity in the position vector $\boldsymbol \theta$. The last term provides the sparsity among the groups (clusters) so that the recovered vector's nonzero elements are concentrated within a single group. This term basically plays the role of coarse localization. This minimization is known as Sparse Group Lasso (SGL) \cite{r149,r150}.

\section{Assisted Localization}
\label{Assisted Localization}
In this Section, different techniques that employ additional information from environmental and common wireless device sensors to assist the Wi-Fi fingerprinting localization are detailed.
 \subsection{Sensor Fusion Assistance for Localization}
\label{Sensor Fusion Localization}
\par Although Wi-Fi signals provide redundant wireless RSS data for fingerprinting, these networks are not originally designed for localization and the measurements captured by wireless devices can be distorted due to various phenomena. wireless devices are  untenable  to reduce the unobtrusiveness of cues or increase comprehension. Most of wireless devices such as smartphones encompass other sensors that can provide additional assistance to RSS-based fingerprinting. Fig. shows some of the additional sensor data which can be extracted from smart mobile devices and fused with RSS-based localization. Several examples are listed in the following. 
\subsubsection{Sound  or Ambient color/light}
 Nearly all of the wireless devices contain speakers and most of them accommodate microphones. The ambient sound renders coarse location specific features if a dataset of sound fingerprints for different places are available. For instance, the ambient sound of a restaurant is different from that of shopping malls. Hence, the ambient sound can help in defining the area that the user is and so, prevent from large localization errors \cite{r205}. 
 \par For a case example, consider shopping malls where ambient light and color conditions are brand-specific. So, these thematic colors along with the lighting styles may provide location-specific signatures which can be fused with the Wi-Fi RSS fingerprints \cite{r205}. However, the lights and lightening styles are subject to frequent changes.
\par While useful, such light/color based fingerprints may often change, and it was shown that floor imagery provides more reliable measurements.    
\subsubsection{RSSI from cellular base stations  \cite{r208}} Although the use of Wi-Fi fingerprints helps to achieve finer localization accuracy due to dense deployments, RSSI measurements from other networks, such as cellular, can serve as additional fingerprints, especially in areas with low density Wi-Fi deployments, in weak signal conditions and when Wi-Fi fingerprinting cannot resolve location ambiguities.
\subsubsection{RFID \cite{r165}} Different from Wi-Fi signals, the RFID tags should be installed and thus require infrastructure upgrades. However, they provide independent location estimation with the RSS fingerprints. The question then arises on how optimally integrate the location estimations from two different sources. Let $\hat{\mathbf{p}}=\{\hat{\mathbf{p}}_{1},\ldots,\hat{\mathbf{p}}_n\}$ be the estimated locations of the user from $n$ different localization procedures. The final user's location, $\hat{\mathbf{p}}=\sum_{j=1}^{N}\beta _{j}\hat{\mathbf{p}}_{j}$, can be estimated through a weighted combination of the individually estimated locations where the weights should be assigned so that the variance of the final estimated location is minimized. This variance is
\begin{equation}
\mathrm{Var}(\hat{\mathbf{p}})=\mathrm{Var}(\sum_{j=1}^{n}\beta _{j}\hat{\mathbf{p}}_{j})=\boldsymbol \beta^{T} \mathrm{diag}(\sigma_{1}^{2},\ldots,\sigma_{n}^{2})\boldsymbol \beta
\end{equation}
where $\boldsymbol \beta=(\beta_{1},\ldots,\beta_{n})$ aggregates weights corresponding to location estimation procedures, and $\mathrm{diag}(\sigma_{1}^{2},\ldots,\sigma_{n}^{2})$  is an $\mathrm{n} \times \mathrm{n}$ with diagonal matrix. The optimal $\beta$ can be obtained through the following minimization problem:
\begin{equation}
\begin{split}
\boldsymbol \beta=&\underset{\boldsymbol \beta}{\mathrm{argmin}} \ \  \boldsymbol \beta^{T} \mathrm{diag}(\sigma_{j}^{2})\boldsymbol \beta\\
& \mathrm{s.t.} \ \ \Arrowvert \boldsymbol \beta \Arrowvert_{1}=1,\\
& \ \ \ \ \ \ \ \ \ \ \boldsymbol \beta \succeq 0
\end{split}
\end{equation}
and the closed form optimal solution is
\begin{equation}
\beta_{j}^{*}=\left( \sigma_{j}^{2}\sum_{j=1}^{n}\frac{1}{\sigma_{j}^{2}}\right) ^{-1}.
\end{equation} 
 \subsection{Motion Assisted Localization}
 \label{Motion Assisted Localization}
Through the widespread deployment of  Micro Electro-Mechanical System (MEMS)  sensors in the smart wireless devices, the Pedestrian Dead Reckoning (PDR) is proliferating as a feasible option for indoor tracking. The set of the sensors that are being used for indoor tracking are called the Inertial Measurement Units (IMUs). These sensors underpin the localization through providing additional details regarding the user's motion such as counting user's steps, inertial navigations, and heading directions \cite{r215}. The Dead-reckoning systems use these sensors and estimate the change of the position of the user with respect to his past location rather than delivering absolute location. In this section, we first introduce the available devices and then discuss the possible ways to utilize these sensors.
\subsubsection{Available Motion Sensors}
\par The sensors that are available in smart devices which can support the localization are:
\begin{itemize}
\item \textit{Barometer:} Measures the atmospheric pressure. The readings of the atmospheric pressure may indicate a special location.
\item \textit{Accelerometer:}  Shows the 3-D acceleration of the user while carrying the device. When the user lifts her foot the acceleration increases and when the foot is planted the acceleration decreases, all leading to a cyclic peak-valley motion pattern. 
\item \textit{Gyroscope:}  Measures the angular velocity of the device and show the orientation of the user.
\item \textit{Magnetometer:} Provides the strength and direction of the earth magnetic field in the environment through which we can know the heading direction of the user.
\end{itemize}
\subsubsection{Sensor Exploitation}
The sensors in smart mobile devices are used to collect  various user's motion patterns \cite{r215}. Through the detection of motion patterns, the following information can be obtained:
\begin{itemize}
\item \textit{Walking direction \cite{r216}:} It is needed to compute location in the first place which leverages application-specific opportunities such as crowd-sourcing of the Wi-Fi data and knowing the user's facing direction. 
\item \textit{Walking detection \cite{r217}:} Although the motion sensors can be exploited to deliver user's motion, utilizing these sensors for long time  consumes a great portion of battery. A smarter localization procedure is to turn on the signals only when the user moves.
\item \textit{Step counting \cite{r217}:} The most common user's location detection though the motion sensors is to estimate the user's location through the distance that the user has passed from a starting point. 
\par Accelerometers provide the 3-D acceleration of the wireless device, and although the obtained data depends on the position and orientation of device with respect to the user, it provides useful information on the user's step length. The accelerometers are triggered based on the lifting and planting of the user's foot. The passed distance of the user is detected from counting the strides along with the stride length. Several methods have been proposed to detect the number of passed strides such as peak detection, zero-crossing, cycle detection, correlation analysis, and fast Fourier transform. 
\end{itemize}
\par These techniques help to estimate the vicinity of the user's location through techniques such as Kalman Filters, Information Potential (IP) \cite{r177,r195,r197}, Conditional Random Fields (CRF) \cite{r231, r226}, nonlinear filters \cite{r38}, etc..

 \subsection{Land-mark Assisted Localization}
 \label{Land-mark Assisted Localization}
The landmark assisted localization helps to harness certain locations in indoor environment which represent identifiable signatures of their surrounding area. These landmarks trace to two types of assistance: 1) calibrating the dead-reckoning schemes, thereby curbing the error growth; 2)  assistance in coarse localization, which warrants more attainable precision. 
\par Landmarks provide specific features to the user depending on the sensors that the user is using for localization. Basically, there are two different landmarks in a typical environment:
\begin{itemize}
\item \textit{Seed landmarks (SLMs):} These are the physical landmarks which can be associated with their actual locations, such as elevators, escalators, and stairs.  For instance, using a camera, the user can match the images of the environment with a database of the available environment images.
\item \textit{Organic landmarks (OLMs):} These are the landmarks that are associated with detecting sensory signatures that are area-specific and confined to a small area.  For instance, an elevator affects the z-dimensional pattern of the phone' accelerometer. 
\end{itemize}
\par As an example, UnLoc \cite{r224} looks for certain structure in the buildings- stairs, elevators, escalators, entrances- that force the user to have predictable motion patterns. For instance, the method checks the confidence level of the GPS as an indicator of user entrance from outdoors to indoors. SemanticSLAM \cite{r225} also checks the gyroscope angular readings  readings to recognize the turns at the end of corridors, classrooms, etc. 
\subsection{Collaborative Localization}
\label{Collaborative Localization}
The idea of collaborative localization is to exploitation the sensors in mobile phones to find the distance between wireless devices, through which the relative locations between neighboring devices are obtained. These relative locations serve as additional constraints in location estimation and improves the localization accuracy \cite{r165,r219,r176}. The range (distance) between wireless devices can be obtained through the following sensors:
\begin{itemize}
\item \textit{Acoustic ranging:} If the wireless devices contain the speaker-microphone, the distance between the devices can be obtained through transmitting acoustic signals and use the TOA to compute the ranges (distances) between the wireless devices \cite{r213}. The ranges help in reducing the search space of the user's location and solve a system of equations to find relative users locations in the collaborative constellation. 
\item \textit{Bluetooth:}  The efficacy of the bluetooth for proximity estimation has been shown in \cite{r159} for collaborative localization and offers accuracies up to 1.5 m.  The fingerprints from the bluetooth of wireless devices are collected in a data base which train the coefficients of a RSSI-distance model.
\end{itemize}

 \section{Radio Map Construction}
 \label{Radio Map Construction}
 This section starts part III of the paper, whose structure is provided in Fig. \ref{fig10} along with the related works in Table \ref{table4}.
 \par A major problem in WLAN positioning systems is the surveying scale in terms of collecting RSS data at large number of RPs for high accuracy positioning. With large scale deployments, the  upfront cost of the deployment effort becomes tremendous.  Furthermore, the radio-map changes over the time and should be periodically calibrated. The size of this dataset is increasing with the size of the area, granularity of the RPs, the number of APs, and the recording length. As this process is labor intensive, some works have focused on reducing the efforts of data collection such as model-based map generation \cite{r108}, Simultaneous
 Localization And Mapping (SLAM) techniques \cite{r109}, and dynamic radio map construction \cite{r135}. 
 \par Crowdsourcing approaches introduce the participatory role of the user during localization \cite{r171,r189,r193,r134}. A dedicated surveyor does not  collect the fingerprints, but the users help to update the radio map if they volunteer. The tedious task of fingerprinting is split between involved users. However, the accuracy  of the data decreases as the fingerprinting time is short and the location of the fingerprints cannot be guaranteed. 
 \par Another simplified data collection tasks resides on implicit data collection, in which the users help with collecting the data through their  daily life routines. For instance, mobile devices can be configured to implicitly collect surveying data without direct involvement of the users.  If part of the data is labeled with its corresponding locations, users can also collect some data without any location association (label), called unlabeled data collection. Then the unlabeled data can be associated with locations through some algorithms such as Hybrid Generative/Discriminative Learning \cite{r226}.
\par AP power profiling has been addressed in \cite{r110}. In this approach, the fingerprints (location, RSS) are considered as Gaussian Processes (GP) and a  model is used to define the relation between the locations and the fingerprints. The coefficient matrix of the regression model is estimated using different learning methods such as linear regression, nonlinear GP, Gaussian Kernel Learning, and augmented path-loss model. Once the coefficient matrix has been estimated using a training set, the RSS values of an unknown location is estimated using a zero-mean GP regression  \cite{r112,r113}. 
  \par Linear interpolation has also been used for interpolation of RSS measurements between RPs \cite{r132, r133}. With the assumption that three non-colinear RPs $j_{1},j_{2},j_{3}$ have been chosen, RSS values for an RP that is inside the convex hull of these RPs is computed as
  \begin{equation}
  \label{eq70}
  r_{j}^{i}(t)=\lambda_{1}r_{j_{1}}^{i}(t)+\lambda_{1}r_{j_{2}}^{i}(t)+\lambda_{1}r_{j_{3}}^{i}(t)
  \end{equation}
  where $\lambda_{1}+\lambda_{2}+\lambda_{3}=1$ \cite{r132}. 
  \par Other interpolation methods that use the minimum and mean of the RSS values of the three non-colinear RPs $j_{1},j_{2},j_{3}$ have also been introduced \cite{r132}. The RSS fingerprint of the nearest RP may also be used as the RSS of the virtual RP \cite{r134}. A more sophisticated method is to use a weighted average of the close RPs \cite{r134}.  
 \par Sparse recovery methods can also be  used in the offline phase to reconstruct the radio map from a lower number of RSS fingerprints. Let $\mathbf{F}$ be the $N\times N$  Fourier transform matrix that linearly transforms the vector of radio map fingerprints to its equivalent representation in the frequency domain as
  \begin{equation}
  \label{eq71}
  \boldsymbol \psi^{i} _{f}=\mathbf{F}\boldsymbol \psi^{i}, \   i=1,\dots,L.
  \end{equation}
  The vector $\boldsymbol \psi^{i} _{f}$ is sparse; that is, most of the frequency components are zero; see e.g., \cite{r27}. This observation helps to reconstruct the radio map in the subsequent discussions. Then, consider a matrix that defines the relation between all RPs and those over which fingerprints have been taken. To this end, we define an $S\times N$ matrix $\mathbf{A}$ whose rows are 1-sparse vectors $\mathbf{a}^{i}=[0,\cdots,1,\cdots,0]$ denoting the index of the RP that is measured during radio map fingerprinting. Let $S<N$ be the total number of RPs where fingerprints are recorded. In essence, $\mathbf{A}$ selects the RPs in which  actual fingerprints are recorded.
  \par The model for the offline radio map interpolation corresponding to AP $i$  can be represented as
  \begin{equation}
  \label{eq72}
  \begin{split}
  \mathbf{b}^{i}=\mathbf{A}\boldsymbol \psi^{i}=\mathbf{A}\mathbf{F}^{-1}\boldsymbol \psi^{i} _{f} \ \ \ \forall i=1,\dots,L.
  \end{split}
  \end{equation}
  The model in \eqref{eq72} is an under-determined system of equations because $S<N$. However, since $\boldsymbol \psi^{i} _{f}$ is sparse, a unique solution exists for it. Two methods have been proposed to find the unique solution for \eqref{eq72}. The CS theory has been used for the interpolation \cite{r30} as

  \begin{equation}
  \label{eq73}
  \begin{split}
  \hat{\boldsymbol \psi}^{i} _{f} =\underset{\boldsymbol \psi^{i} _{f}}{\text{argmin}} \Arrowvert \boldsymbol \psi^{i} _{f} \Arrowvert_{1} \\
  s.t. \ \mathbf{b}^{i}=\mathbf{A}\mathbf{F}^{-1}\boldsymbol \psi^{i} _{f}.
  \end{split}
  \end{equation}
The LASSO has also been used for radio map interpolation \cite{r139} as
  \begin{equation}
    \label{eq74}
   \hat{\boldsymbol \psi}^{i} _{f} =\underset{\boldsymbol \psi^{i} _{f}}{\text{argmin} }\ \left[ \frac{1}{2}\Arrowvert \mathbf{b}^{i}-\mathbf{A}\mathbf{F}^{-1}\boldsymbol \psi^{i} _{f}\Arrowvert_{2}^{2} + \lambda_{1} \Arrowvert \boldsymbol \psi^{i} _{f}\Arrowvert_{1}\right]
    \end{equation}
which has the form of the group sparse recovery \eqref{eq3} with $\lambda_{2}=0$. The above formulation minimizes the error between the measured RSS fingerprints and the interpolated fingerprints, while the second term promotes sparsity of  the RSS fingerprints in the Fourier domain.
\par The previous optimizations \eqref{eq73} or \eqref{eq74} are solved for all APs. The reconstructed radio map rows are computed as
    \begin{equation}
    \label{eq75}
    \hat{\boldsymbol \psi}^{i}=\mathbf{F}^{-1}\hat{\boldsymbol \psi}^{i}_{f}.
    \end{equation}
Using \eqref{eq75}, RSS fingerprints can be measured on a smaller number of RPs, and the radio map is interpolated in between RPs at a finer granularity.  
\section{Outlier Detection}
\label{Outlier Detection}
In this section we first discuss the possible causes of outliers and then an overview of outlier detection and mitigation methods is provided.
APs may experience faults during their operations due to the following causes:
\begin{itemize}
\item Some APs become intermittently unavailable or provide erroneous RSS measurements due to unexpected failures, jamming, power outages, or intentional adversary attacks that may weaken or strengthen the AP signals.
\item The indoor obstacles introduce a multipath profile to the traveling signals.
\item There is no guarantee that the APs that have been visible during the fingerprinting time are visible during the online localization phase.
\item Modern APs are able to adapt their transmit power based on the traffic.
\end{itemize}
Due to the previous reasons, the AP characteristics in the fingerprinting phase may not match those in the online phase. In such cases, online readings of APs are not trustable. These inordinate online measurements are called \textit{outliers}.
\par An outlier occurs when the online measurement from an AP is significantly different than any fingerprint in the area.  This hurdle has surprisingly received little attention in the  literature. Note that existing AP selection schemes select the APs based on the AP performance during the fingerprinting period, and are therefore not well-suited to mitigate outliers which occur in the online phase. 
\par Outliers may also occur during the fingerprinting period. However, some post-sanitary measures such as authentication of beacon nodes, radio map collection over various periods, validation, and attack detection help to remedy any impersonation and data corruption \cite{r127,r128}. 
\par Next, an overview of the schemes for the detection of outliers in the online measurements is provided. Some approaches focus on outlier detection and improve the localization performance of conventional methods \cite{r94, r204, r40}. A categorization of the recently proposed outlier detection schemes for WLAN localization is depicted in Fig. \ref{fig10}.
\subsection{Hampel Filter} 
Hampel filter has been extensively used for outlier detection in statistical data \cite{r95,r96,r97,r98} and has been introduced as an offline and online outlier detection procedure in \cite{r99}. It replaces the outlier-sensitive mean and standard deviation estimates with the outlier-resistant median and median absolute deviation from the median (MAD). The latter is defined as 
        \begin{equation}
        \label{eq56}
        R_{j}^{i}=1.4826\times \mathrm{median} \left\lbrace \left|r_{j}^{i}(t_{m})- \mathrm{median}(\mathbf{r}_{j}^{i})\right| \right\rbrace 
        \end{equation}
The factor $1.4826$ was chosen so that the expected value of $\mathrm{R}_{j}^{i}$ is equal to the standard deviation for normally distributed data. The MAD-scale substitute of the data is
        \begin{equation}
        \label{eq57}
        MAD_{j}^{i}(t_{m})=\frac{\left|r_{j}^{i}(t_{m})- \mathrm{median}(\mathbf{r}_{j}^{i})\right|}{R_{j}^{i}}.
        \end{equation}
\subsection{Modified Distance-based Outlier Detection}
\par A modified KNN method has been proposed as an alternative fault tolerant localization method \cite{r36}. The Euclidean distance between the online measurements and fingerprints over a modified subset of APs is defined  as
        \begin{equation}
        \label{eq58}
        \begin{split}
          d_{\mathrm{Euc}}(\boldsymbol \psi_{j},\boldsymbol y) 
          & =\sqrt{\sum_{i\in \mathcal{A'}\cap \mathcal{A'}_{y}}^{}\left(  y^{i} -\psi_{j}^{i}\right) ^{2}+\sum_{i\in \mathcal{A'}_{y} \setminus \mathcal{A'}}^{}\left( y^{i} -\psi_{j}^{i}\right) ^{2}}, \\
           j&=1,\dots, N 
        \end{split}
        \end{equation}
where $\mathcal{A'}$ and $ \mathcal{A'}_{y}$ are respectively the subsets of APs available during fingerprinting at $\mathbf{p}_{j}$ and in $\boldsymbol y$. The first summation component is on a subset of APs that are available in both fingerprinting and online phase while the second term sums over the APs that are available only in the online period and not in the fingerprinting period.
    \par A more comprehensive model of outliers has been proposed \cite{r101}, which considers different causes of outliers as 
        \begin{equation}
        \label{eq59}
         y^{i}=b_{1}y^{i}+b_{2}( y^{i}+n(i))+b_{3} y^{i}_{bog}+b_{4}c_{NaN}
        \end{equation}
where $b_{k}\in\left\lbrace 0,1 \right\rbrace, k=1,\ldots,4$, and $\sum_{k=1}^{4}b_{k}=1$, which means only one of the components is active at a time. The second term models the extra noise due to jammed APs, where $n(i)\sim \mathcal{N}(0,\sigma^{2})$, $y^{i}_{bog}$ models the bogus APs that imitate an actual AP, and $c_{NaN}$ models the unavailability. The localization procedure contains a modified distance which switches between the Euclidean and median distances as
        \begin{equation}
        \label{eq60}
        d_{\mathrm{mod}}(\breve{\mathbf{r}}_{j},\boldsymbol y)=\mathrm{min}\left( d_{\mathrm{Euc}}(\breve{\mathbf{r}}_{j},\boldsymbol y),d_{\mathrm{med}}(\breve{\mathbf{r}}_{j},\boldsymbol y)  \right) 
        \end{equation}
where $d_{Euc}$ and $d_{med}$ are given by \eqref{eq7} and $\breve{r}_j$ is replaced by the average or median fingerprint, as explained in Section II-B.
\subsection{Sparsity-based Outlier Detection}
\par Localization in the presence of outliers via sparse recovery methods has also considered. The main idea is that outliers are modeled exactly by augmenting \eqref{eq50}. Specifically, with $\bm \kappa$ denoting the outlier vector, the online measurements adhere to the following model:
        \begin{equation}
        \label{eq63}
        \mathbf{y}=\mathbf{\Phi \Psi \boldsymbol\theta +\boldsymbol \kappa+\boldsymbol\epsilon}.
        \end{equation}
The advantage of the previous model is that the outliers vector $\boldsymbol \kappa$ will be sparse as long as the number of corrupted APs is small, and can therefore be estimated jointly with the position indicator vector $\boldsymbol \theta$ via $\ell_{1}$-minimization. The premise of explicitly modeling the outliers for robust regression in a general statistical setting has been previously analyzed in \cite{r151} and \cite{r152}. In what follows, the CS, LASSO, and GLMNET approaches are modified so that the outlier vector $\bm\kappa$ can be estimated alongside the user position vector $\bm\theta$.
        \par The modified CS (M-CS) approach minimizes the weighted combination of the $\ell_{1}$ norms of $\bm\theta$ and $\bm\kappa$ \cite{r139}
        \begin{equation}
        \label{eq64}
        \begin{split}
         (\hat{\boldsymbol\theta}, \hat{\boldsymbol\kappa})&=\underset{\boldsymbol\theta, \boldsymbol\kappa}{\text{argmin}} \Arrowvert \boldsymbol\theta \Arrowvert_{1} +\mu \Arrowvert \boldsymbol \kappa \Arrowvert_{1} \\
          & s.t.\  \mathbf{y}=\mathbf{\Phi \Psi \boldsymbol\theta+\boldsymbol \kappa}.
        \end{split}
        \end{equation}
        \par The modified LASSO (M-LASSO)  minimizes the squared residuals, in addition to the  $\ell_{1}$  norms of the sparse vectors:
          \begin{equation}
          \label{eq65}
         (\hat{\boldsymbol\theta}, \hat{\boldsymbol\kappa}) =\underset{\boldsymbol\theta, \boldsymbol\kappa}{\text{argmin}} \ \left[ \frac{1}{|  \tilde{\mathcal{L}} |}\Arrowvert \mathbf{y-\mathbf{H} \boldsymbol \theta}-\boldsymbol \kappa\Arrowvert_{2}^{2} + \lambda \Arrowvert\boldsymbol \theta\Arrowvert_{1}+\mu \Arrowvert \boldsymbol \kappa \Arrowvert _{1}\right]
          \end{equation}
          where $ \mu >0 $ is a tuning parameter. 
          \par The modified GLMNET (M-GLMNET) amounts to the following optimization problem:
          \begin{equation}
          \label{eq66}
          \begin{split}
          (\hat{\boldsymbol\theta}, \hat{\boldsymbol\kappa}) =\underset{\boldsymbol\theta, \boldsymbol\kappa}{\text{argmin}} \ \left[ \frac{1}{|  \tilde{\mathcal{L}} |}\Arrowvert \mathbf{y-\mathbf{H} \boldsymbol \theta -\boldsymbol \kappa}\Arrowvert_{2}^{2} + P_{\alpha} \right]\\
           P_{\alpha}=\lambda \bigl[(1-\alpha)\Arrowvert \boldsymbol \theta \Arrowvert _{2}^{2}+\alpha \Arrowvert \boldsymbol \theta \Arrowvert_{1})\bigr]+\mu \Arrowvert \boldsymbol \kappa \Arrowvert _{1}.
          \end{split}
          \end{equation}

\par Finally, The modified Group-Sparsity (MGS)-based regression is formulated as \cite{r140}
      \begin{equation}
      \label{eq67}
      \begin{split}
        (\hat{\boldsymbol\theta}, \hat{\boldsymbol\kappa}) & =\underset{\boldsymbol \theta}{\text{argmin} }\ \left[ \frac{1}{|  \tilde{\mathcal{L}} |}\Arrowvert \mathbf{y-\mathbf{H}\boldsymbol \theta-\boldsymbol \kappa }\Arrowvert_{2}^{2} +P_{\alpha}\right]\\
       P_{\alpha}&=\lambda_{1} \Arrowvert\boldsymbol \theta\Arrowvert_{1}+\lambda_{2}\sum_{k=1}^{K}w_{k}\Arrowvert\boldsymbol \theta_{k}\Arrowvert_{2}+\mu \Arrowvert \boldsymbol \kappa \Arrowvert_{1}
      \end{split}  
        \end{equation}
\par In the previous joint localization and outlier detection formulations, the outlier vector, $\boldsymbol \kappa$, enables the optimization algorithms to discard the outliers in the online measurement vector. The terms promoting sparsity of the user's location vector and the outlier indicator vector have the weights $ \lambda $ and $ \mu $, respectively. Optimization problems \eqref{eq64}--\eqref{eq67} are convex problems which can be efficiently solved \cite{cvx,gb08}.

     \begin{figure*}[t!]
                         \includegraphics[scale=0.945,angle =90]{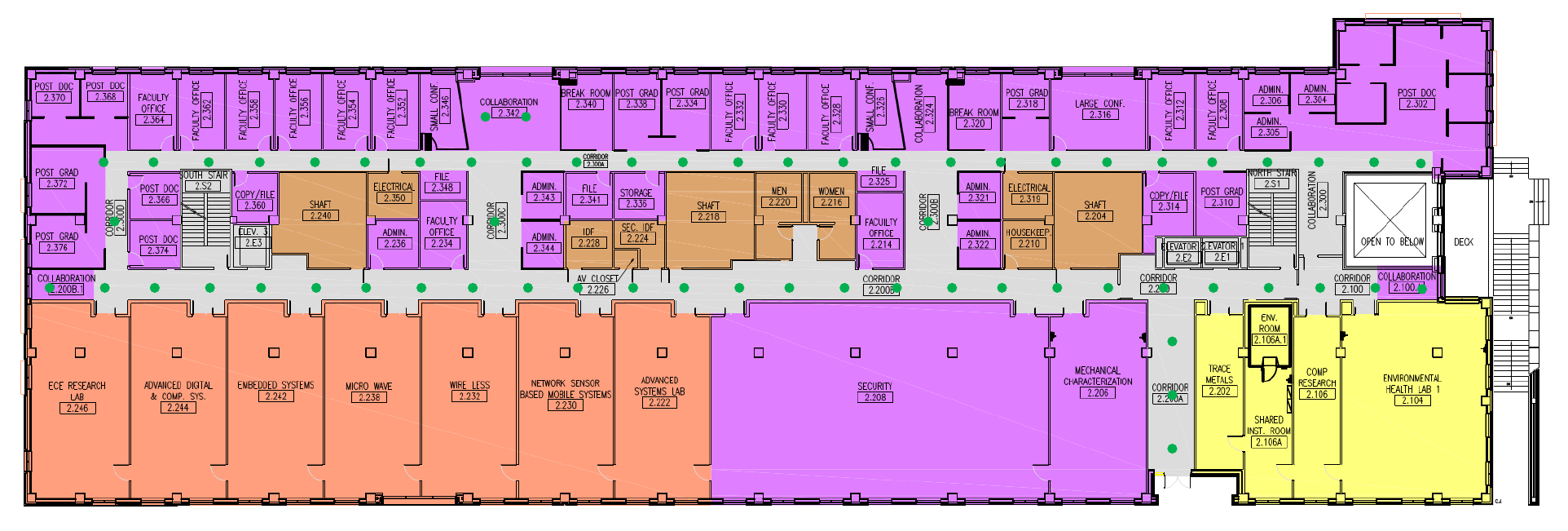}
                              \centering
                              \caption{The map of experimental environment. The green dots indicate the RP locations.}
                              \label{fig14}
                           \end{figure*}   

\section{Heterogeneous Devices}  
\label{Heterogeneous Devices} 
One of the issues related to the deployment of fingerprinting approaches is that wireless devices do not read equal RSS measurements if they are located in the same position, primarily, due to heterogeneous reception characteristics of embedded NICs.  Rapid growth of wireless devices from different manufacturers caused hardware variations amongst devices or even across models (same manufacturer), such as the receiving antenna gain, position of the antenna on the device, sensitivity, and Operating System (OS) characteristics. 
\par Hardware variation can significantly degrade the positional accuracy of RSS-based WiFi localization systems. RSS data (fingerprints and online measurements) can be transformed using linear regression, expectation maximization, and neural networks \cite{r227}. The Pearson correlation coefficient has also been used to find the similarity between RSS fingerprints and online measurements \cite{r227}. Device-invariant fingerprints can be derived from RSS measurements by proper normalization such as using signal strength ratios between pairs of APs instead of absolute RSS values. The rank-ordering of APs can also serve as device invariant measure \cite{r229}.
\par Some works have also used the Signal Strength Difference (SSD) instead of dealing with RSS fingerprints directly to compensate for different devices' hardware readings of RSS signals \cite{r186,r190}.
             
\section{Experimental Evaluation and Comparisons}
\label{Numerical Experiments and Comparisons}
In this section, we provide an illustration of the localization performance for some the approaches in the previous sections on a real indoor environment. The results render beneficial insights  as all the localization approaches are compared within a single environment.
\par The results are based on data collected at the second floor of the Applied Engineering and Technology (AET) building at the University of Texas at San Antonio which has an area of $576 \mathrm{ft }\times 35 \mathrm{ft}$. The map of the surveying area is provided in Fig. \ref{fig14}. The area represents a typical office environment as it includes several research labs, offices, library, study area, and break rooms.
\par The localization approaches have been assessed through their localization accuracy. Let $N_{t}$ be the number of the test points (online measurements taken at different positions). The Mean Absolute Error (MAE) is a measure of the localization accuracy defined as \cite{r121,r30,r21,r27}
\begin{equation}
   \text{MAE}=\frac{1}{N_{t}}\sum_{n=1}^{N_{t}}\sqrt{ \left( \hat{\mathbf{p}}(n)-\mathbf{p}(n)\right) ^{T}\left( \hat{\mathbf{p}}(n)-\mathbf{p}(n)\right) } .
   \end{equation} 
where $\mathbf{p}(n)$ and $\hat{\mathbf{p}}(n)$ are the true and estimated positions, respectively. To define the spread of the localization errors, the cumulative distribution function (CDF) of the localization errors is also evaluated. 

\par First, we assess the performance of the localization approaches without clustering and coarse localization.  The performance of localization methods is then evaluated together with one of the coarse localization techniques of Section \ref{RP Clustering}.
\par The localization approaches that have been selected are as follows: KNN, KDE, CS, LASSO, GLMNET, GS, and Contour-based localization. Table \ref{table3} shows the formula based on which the user's location is estimated. 

\begin{table}[t!]
                      \caption{Numerical Test Methods with Corresponding Formula}
                       \label{table3}
                      \begin{center}
                      \begin{tabular}{ |c|c| } 
                        \hline
                      \textbf{Methods} & \textbf{Related Computing Formula}\\
                      \hline
                         \hline
                      KNN & \eqref{eq8}  \\ 
                        \hline
                    KDE  & \eqref{eq42} and \eqref{eq14}\\ 
                        \hline
                     CS & \eqref{eq52} \\ 
                        \hline
                        GLMNET & \eqref{eq54} \\ 
                                             \hline
                         LASSO & \eqref{eq53} \\ 
                         \hline
                     GS & \eqref{eq55} \\ 
                     \hline
                      Contour-based & \cite{r189} \\ 
                                          \hline
                      \end{tabular}
                      \end{center}
                       \end{table}

\subsection{Localization Error Without  Coarse Localization} 
\label{no clustering}
The  methods in this subsection have been implemented without utilizing any coarse localization. However, for reducing the number of APs, the Fisher criterion \eqref{eq35} has been applied. 
\par
\begin{figure}[t!]
        \includegraphics[scale=0.51]{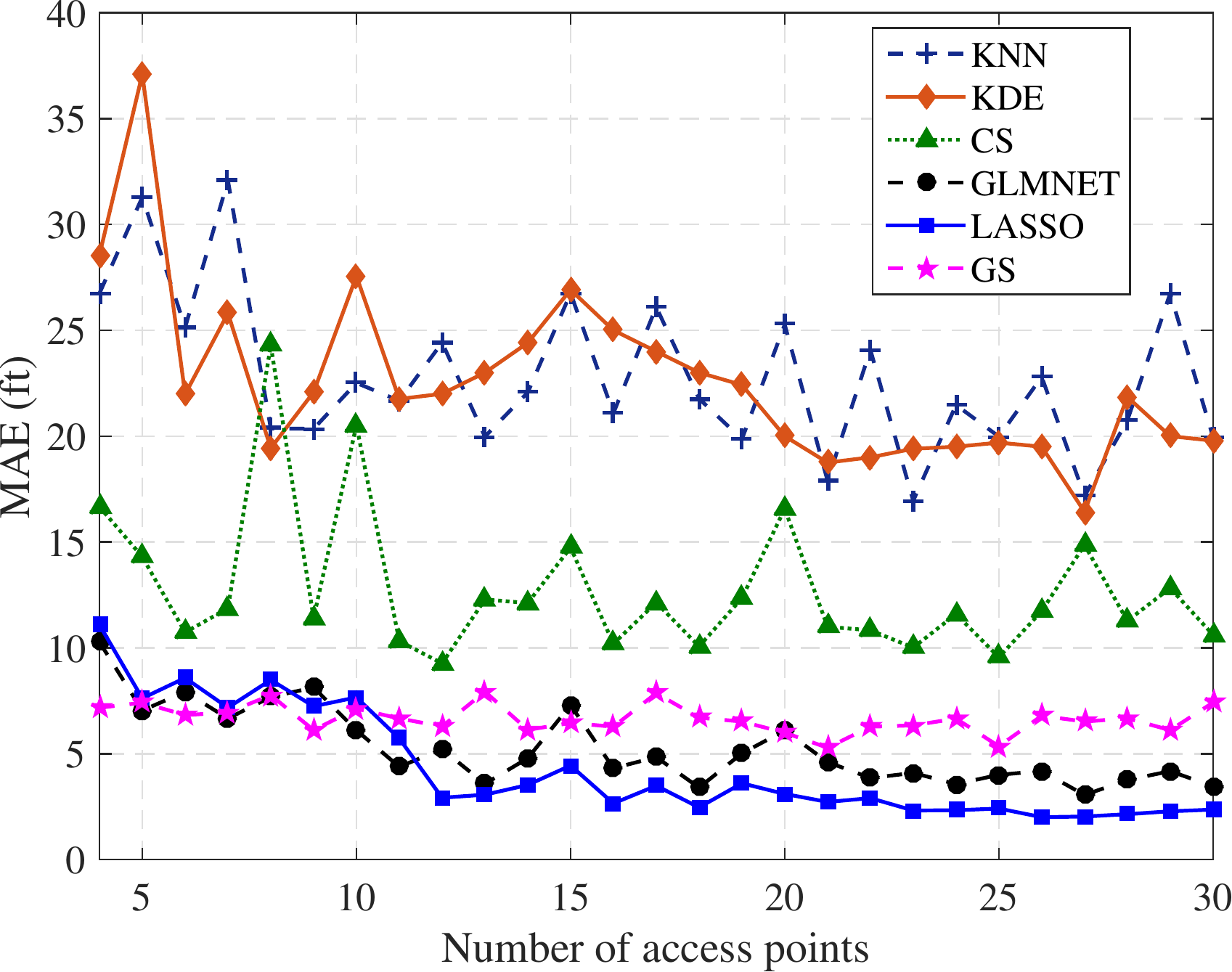}
             \centering

             \caption{Localization error comparison for  different number of APs without clustering.}
             \label{fig13}
          \end{figure}

 Fig. \ref{fig13} illustrates the localization error versus an increasing number of APs. For the KNN method, $K=10$ RPs have been selected. The  kernel widths for KDE approach have been computed through the recommendations given in \cite{r21}. The probability density  of the RSS fingerprints had to be estimated in the online phase because the APs  engaging in the localization should be known for the KDE approach. The GS approach needs the corresponding weight for each cluster which is computed through the layered clustering method ($K=10$).  The results show  high localization errors for all approaches although the errors decrease as the number of APs increases. However, the sparse recovery methods show higher accuracy, among which  the GS-based localization shows the highest localization accuracy if less than 10 AP are used. The GS accuracy slightly improves if more APs are used. Overall, LASSO-based localization shows the least localization error if more APs are used.

 \par
\begin{figure}[t!]
                      \includegraphics[scale=0.505]{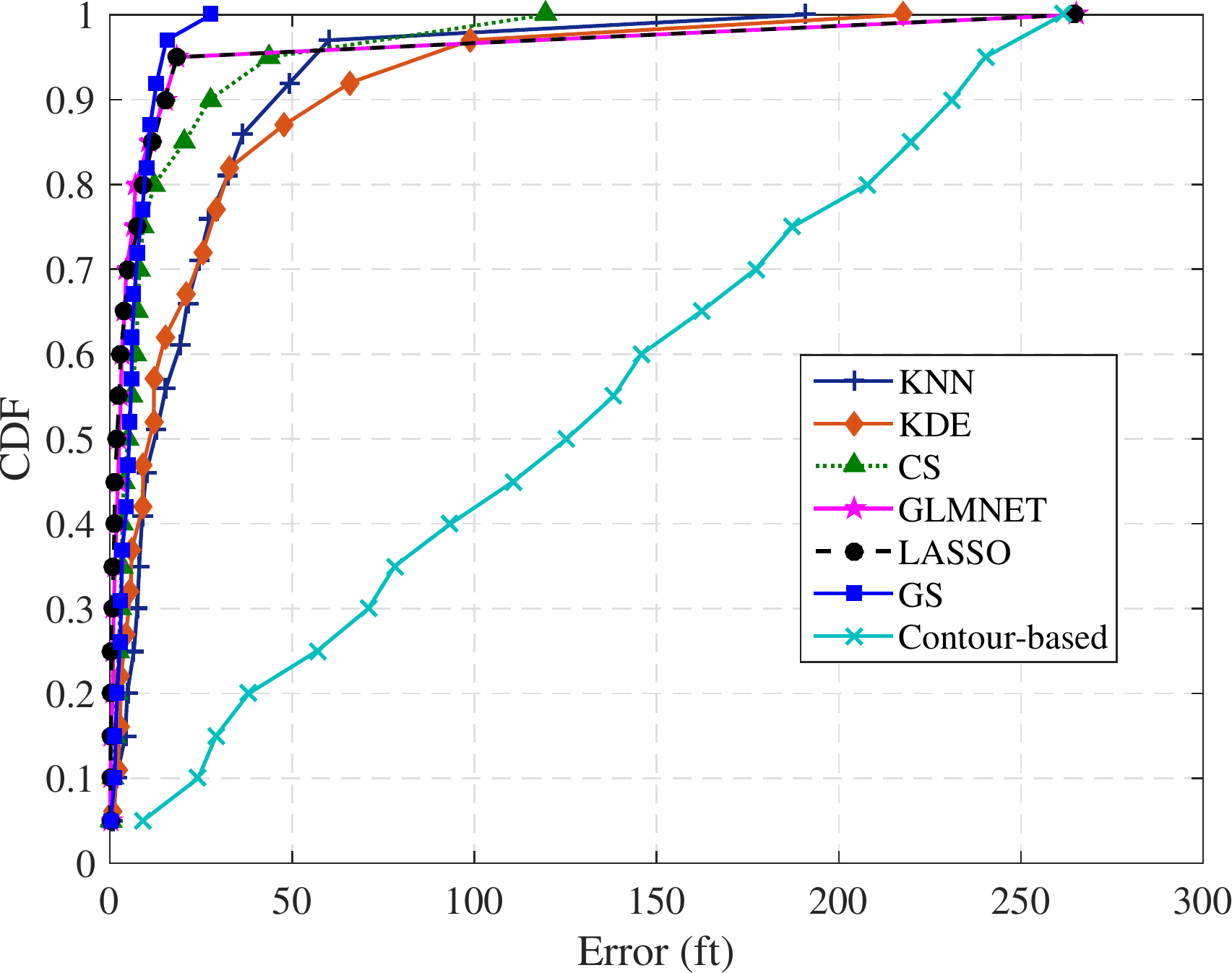}
                           \centering

                           \caption{The CDF of the localization error for 10 APs without  clustering.}
                           \label{fig15}
                        \end{figure}

 The localization error distribution is shown in Fig. \ref{fig15} when 10 APs have been used for localization.  The contour-based approach introduces the largest errors because it needs an estimation of the path loss parameters. These parameters are assumed uniform for an AP along all directions which is not a suitable assumption in complex indoor environments. The KNN and KDE techniques do not  show satisfactory performance either.

\subsection{Localization Error With Coarse Localization} 
\label{clustering}
As shown in the previous subsection, the localization accuracy is low without coarse localization in large surveying areas. To enhance the  performance, the user's location is first estimated in the coarse localization stage, and the fine localization step is applied afterwards. To show that the localization accuracy is enhanced with coarse localization, the clustering using the AP coverage vector has been utilized for the KNN and KDE approaches as in \cite{r21}, weighted clustering has been used for CS, LASSO, and GLMNET, and layered clustering has been used for GS. 
\par
\begin{figure}[t!]
          \includegraphics[scale=0.52]{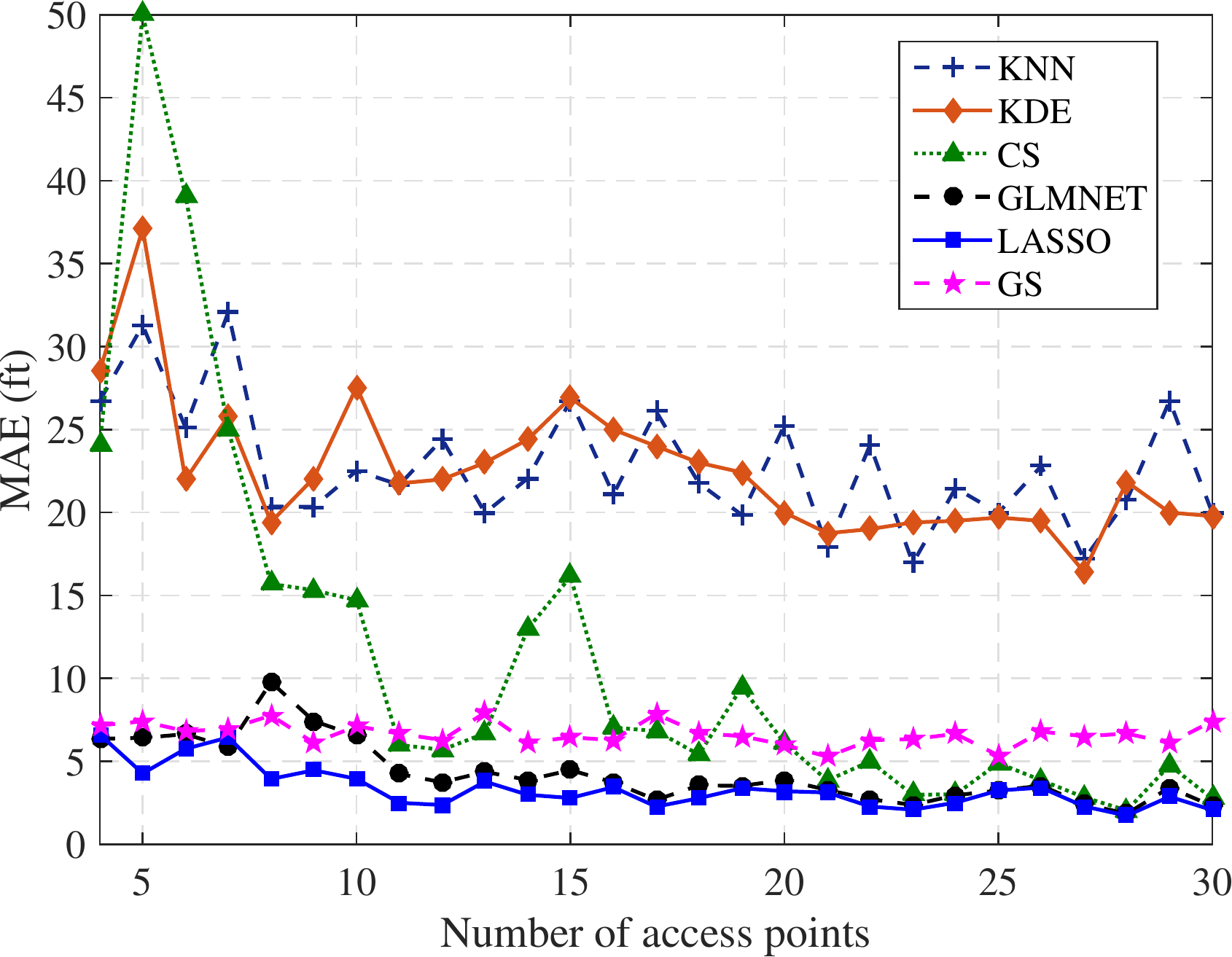}
               \centering

               \caption{ Localization error comparison for  different number of APs with clustering.}
               \label{fig12}
            \end{figure}

 Fig. \ref{fig12} shows the average localization error for an increasing number of APs. Increasing the number of AP  slightly improves the  KNN, KDE and GS approaches, however, the localization error decreases from 10 ft to 2 ft for LASSO and GLMNET  if the number of engaged APs is increases from 4 and 29. However, it is evident that the localization error for CS, LASSO, and GLMNET  has overall been decreased dramatically compared to when no coarse localization was used.
\par
\begin{figure}[t!]
              \includegraphics[scale=0.5]{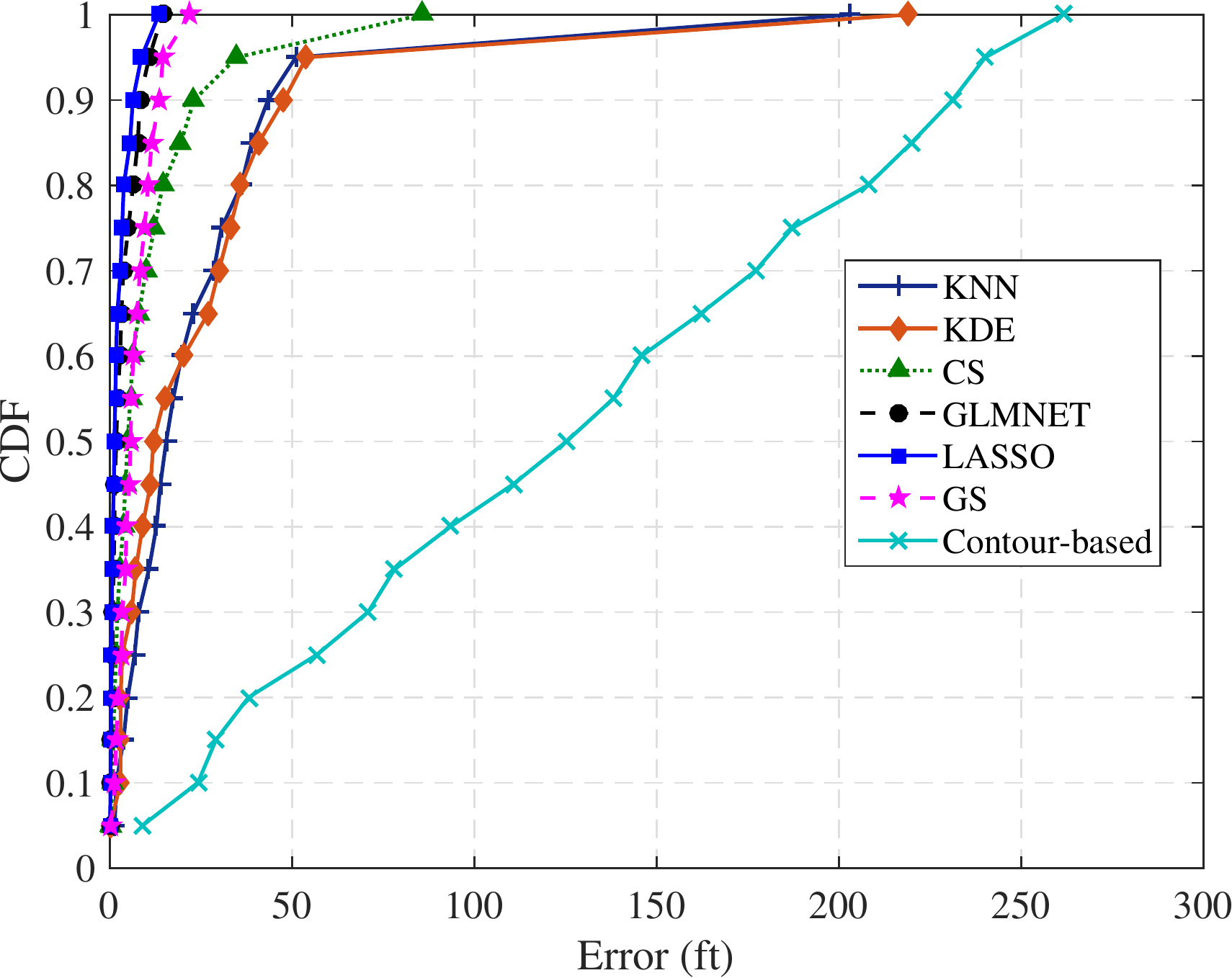}
                   \centering

                   \caption{The CDF of the localization error with 10 APs and clustering.}
                   \label{fig16}
                \end{figure}

 The distribution of the localization error is depicted in Fig. \ref{fig16} when only 10 APs are utilized in localization. Comparing  Figs. \ref{fig16} and \ref{fig15} reveals that the errors of CS, LASSO, and GLMNET are  greatly decreased and the 80\% of the errors are less than 20 ft. However, the KNN and KDE methods render unacceptably high localization errors. 

  \begin{table*}
                             \label{Comparetable}                                
                            \caption{Comparison of Representative Fingerprinting Localization Approaches and Technical Details }
                              \begin{center}
                                 \begin{tabular}{ |c|c|c|c|c|c|c| }                                                  
                                 \hline
                                 \textbf{Scheme} & \begin{tabular}[x]{@{}c@{}}\textbf{RP Selection} \\ \textbf{Technique}\end{tabular}  & \begin{tabular}[x]{@{}c@{}}\textbf{AP Selection} \\ \textbf{Technique} \end{tabular}& \begin{tabular}[x]{@{}c@{}} \textbf{Fine Localization}\\ \textbf{ Technique} \end{tabular} & \begin{tabular}[x]{@{}c@{}} \textbf{Reported }\\ \textbf{Accuracy at } \\ \textbf{50 \%} \end{tabular}  & \begin{tabular}[x]{@{}c@{}}\textbf{Testbed}  \\  \textbf{ Information} \end{tabular} \\
                                 \hline
                                 RADAR \cite{r15} & -- & Strongest AP & KNN  & $\sim 8.16\mathrm{m}$ & \begin{tabular}[x]{@{}c@{}}70 RPs, N/A \\ > 20 samples/RP \\area: $43.5\mathrm{m} \times 22.5\mathrm{m}$ \end{tabular}  \\
                                 \hline 
                                \begin{tabular}[x]{@{}c@{}} Cosine \\ Similarity \cite{r233}  \end{tabular}    & -- & -- & WKNN  & $\sim 3.5\mathrm{m}$ & \begin{tabular}[x]{@{}c@{}} 213 RPs, $0.5\mathrm{m}$ apart\\ 100 samples/RP\\ area: N/A \end{tabular} \\
                                 \hline
                                Horus \cite{r232} & \begin{tabular}[x]{@{}c@{}}Incremental\\ Triangulation  \end{tabular} & Strongest AP & Weighted probabilistic  & $\sim 0.6\mathrm{m}$ & \begin{tabular}[x]{@{}c@{}}110 RPs, $2.13\mathrm{m}$  \\ 100 samples/RP  \\ area: $35.9\mathrm{m} \times 11.8\mathrm{m}$ \end{tabular} \\
                                  \hline  
                                Tilejunction \cite{r232} &  \begin{tabular}[x]{@{}c@{}}Entropy \\Maximization  \end{tabular} & Spectral Clustering & Linear programming & $\sim 6\mathrm{m}$ & \begin{tabular}[x]{@{}c@{}} $183$ RPs, $5\mathrm{m}$ apart \\ $15$ samples/RP \\  area: $2000{\mathrm{m}}^{2}$ \end{tabular} \\
                                                                \hline     
                                PCA \cite{r103} & -- & -- & Weighted probabilistic  & $\sim 1.6\mathrm{m}$ & \begin{tabular}[x]{@{}c@{}}$45$ RPs, $2\mathrm{m}$ apart,\\ $100$ samples/RP \\   area: $24.6\mathrm{m} \times 17.6\mathrm{m}$ \end{tabular} \\
                                                                                            \hline     
                                 Kernel-based \cite{r21} & AP coverage & \begin{tabular}[x]{@{}c@{}} Bhattacharyya distance  \\  Information potential \end{tabular}& Kernel density estimation  & $\sim 1.8\mathrm{m}$ & \begin{tabular}[x]{@{}c@{}}$66$ RPs, $2\mathrm{m}$ apart,\\ $4-200$ samples/RP \\   area: $36\mathrm{m} \times 42\mathrm{m}$ \end{tabular} \\
                                                                                                                   \hline    
                                 CS \cite{r30} & Affinity propagation & \begin{tabular}[x]{@{}c@{}} Strongest APs  \\  Fisher criterion \\ Random combination \end{tabular}& Compressive sensing  & $\sim 1.5\mathrm{m}$ & \begin{tabular}[x]{@{}c@{}}$72$ RPs, $1.5\mathrm{m}$ apart,\\ $50$ samples/RP \\   area: $30\mathrm{m} \times 46\mathrm{m}$ \end{tabular} \\
                                 \hline
                                  ACS-based \cite{r130} & Splitting-based & \begin{tabular}[x]{@{}c@{}} Joint selection   \end{tabular}& ML  & $\sim 0.8\mathrm{m}$ & \begin{tabular}[x]{@{}c@{}}$16384$ RPs, $1.56\mathrm{m}$ apart,\\ $100$ samples/RP \\   area: $200\mathrm{m} \times 200\mathrm{m}$ \end{tabular} \\
                                    \hline 
                                  CaDet \cite{r35} & K-means & \begin{tabular}[x]{@{}c@{}} InfoGain  \end{tabular}& Decision tree & $\sim 0.8\mathrm{m}$ & \begin{tabular}[x]{@{}c@{}}$99$ RPs, $1.5\mathrm{m}$ apart,\\ $100$ samples/RP \\   area: N/A \end{tabular} \\
                                   \hline  
                                   LASSO \cite{r139} & Weighted clustering & \begin{tabular}[x]{@{}c@{}} Fisher Criterion   \end{tabular}& LASSO sparse recovery  & $\sim 0.52\mathrm{m}$ & \begin{tabular}[x]{@{}c@{}}$192$ RPs, $0.91\mathrm{m}$ apart,\\ $100$ samples/RP \\   area: $300\mathrm{m} \times 35\mathrm{m}$ \end{tabular} \\
                                                                \hline
                                   GLMNET \cite{r139} & Weighted clustering & \begin{tabular}[x]{@{}c@{}} Fisher Criterion   \end{tabular}& Elastic net sparse recovery  & $\sim 0.96\mathrm{m}$ & \begin{tabular}[x]{@{}c@{}}$192$ RPs, $0.91\mathrm{m}$ apart,\\ $100$ samples/RP \\   area: $300\mathrm{m} \times 35\mathrm{m}$ \end{tabular} \\
                                                                                               \hline 
                                    GS \cite{r140} & Layered clustering & \begin{tabular}[x]{@{}c@{}} Fisher Criterion   \end{tabular}& GS sparse recovery  & $\sim 1.24\mathrm{m}$ & \begin{tabular}[x]{@{}c@{}}$192$ RPs, $0.91\mathrm{m}$ apart,\\ $100$ samples/RP \\   area: $300\mathrm{m} \times 35\mathrm{m}$ \end{tabular} \\
                                                                                                                              \hline  
                                \end{tabular}
                                \end{center}
                             \end{table*}
      \begin{table*}
                                  \label{proscons}                                
                                 \caption{Overview of Strengths and Weaknesses of Representative Fingerprinting Localization Approaches }
                                   \begin{center}
                                      \begin{tabular}{ |c|c|c| }                                                  
                                      \hline
                                      \textbf{Scheme} & \begin{tabular}[x]{@{}c@{}}\textbf{Strengths}\end{tabular}  & \begin{tabular}[x]{@{}c@{}}\textbf{Weaknesses} \end{tabular} \\
                                      \hline
                                      RADAR \cite{r15} & \begin{tabular}[x]{@{}l@{}}  Ease of implementation \\ \end{tabular}  & \begin{tabular}[x]{@{}c@{}}  No efficient AP selection; \\  Low localization accuracy \end{tabular}   \\
                                      \hline 
                                     \begin{tabular}[x]{@{}c@{}} Cosine  Similarity \cite{r233}  \end{tabular}    & \begin{tabular}[x]{@{}c@{}}  Enhanced metric between fingerprints and online measurements  \end{tabular} & \begin{tabular}[x]{@{}c@{}}  No coarse localization; \\ No AP selection \end{tabular} \\
                                      \hline
                                     Horus \cite{r232} & \begin{tabular}[x]{@{}c@{}}  High localization accuracy \end{tabular} &  Time-consuming implementation  \\
                                       \hline  
                                     Tilejunction \cite{r232} &  \begin{tabular}[x]{@{}c@{}}  Accounts for the constraints  \end{tabular} & \begin{tabular}[x]{@{}c@{}}  Complex implementation; \\  Needs model-based parameter estimation  \end{tabular} \\
                                                                     \hline     
                                     PCA \cite{r103} &  Suitable feature extraction &  Complex decomposition implementation \\
                                                                                                 \hline     
                                      Kernel-based \cite{r21} &  Enhanced metric between fingerprints and online measurements & \begin{tabular}[x]{@{}c@{}}  Complex kernel implementation  \end{tabular} \\
                                                                                                                        \hline    
                                      CS \cite{r30} &  High localization accuracy  & \begin{tabular}[x]{@{}c@{}}  Optimization's equality constraint; \\  Needs to satisfy special properties \end{tabular} \\
                                      \hline
                                       ACS-based \cite{r130} & Area-based AP selection & \begin{tabular}[x]{@{}c@{}}  Low accurate metric  \\  \end{tabular}\\
                                         \hline 
                                       CaDet \cite{r35} &  Enhanced AP selection technique & \begin{tabular}[x]{@{}c@{}} Complex Probabilistic AP selection  \end{tabular} \\
                                        \hline  
                                        LASSO \cite{r139} & \begin{tabular}[x]{@{}c@{}} High localization accuracy; \\  Enhanced optimality condition   \end{tabular}  & \begin{tabular}[x]{@{}c@{}} Needs parameter tuning  \end{tabular} \\
                                                                     \hline
                                        GLMNET \cite{r139} & \begin{tabular}[x]{@{}c@{}} High localization accuracy; \\  Enhanced optimality condition   \end{tabular} & \begin{tabular}[x]{@{}c@{}} Needs parameter tuning \end{tabular}  \\
                                                                                                    \hline 
                                         GS \cite{r140} & \begin{tabular}[x]{@{}c@{}}  High localization accuracy with low number of APs; \\  Integrates  coarse and fine localization in a \\ multicomponent optimization problem  \end{tabular} & \begin{tabular}[x]{@{}c@{}}  Needs parameter tuning   \end{tabular} \\
                                                                                                                                   \hline  
                                     \end{tabular}
                                     \end{center}
                                  \end{table*}

\section{Critical Summary and Recommendations for Future Work}
\label{Critical Summary and Outlook}
\subsection{Critical Summary}
The WLAN indoor localization has attracted great attention due to the low cost deployment, existing infrastructure, and ease of implementation. The WLAN fingerprinting approach became very popular as proven performance was in real environments. As indoor propagation is a very complex phenomenon distorted by multipath and signal blockages, traditional techniques such as trilateration do not show good performance. The research has become very broad and extensively branched due to necessity to address various issues. This paper attempts to systematize various aspects of the state-of-the-art.  
\par First, the paper categorizes conventional localization approaches at early stages. Then, the challenges that are associated with the fingerprinting approaches and conventional problems are enumerated. The state-of-the-art solutions to these challenges are categorized and the related works for each category has been overviewed. A key issue was to unify the misleading concepts and notations that varied among approaches and introduce them in a single trackable package. Recent approaches enhance the conventional methods, utilize the peculiarities of available environments and sensors, and leverage sparse recovery methods. 
\par Since localization approaches in the literature have been evaluated in different settings, representative fingerprinting approaches are implemented in a typical office environment for illustration purposes. In parallel, the details of some of the fingerprinting approaches are listed in Table V. A qualitative comparison over these methods is also included in Table VI. Table V shows the RP clustering method, AP selection method, fine localization technique, reported accuracy and details about the implemented setting. The comparison over the reported accuracies is difficult as the methods have been implemented in different testbeds which differ in the size of the area, number of RPs, granularity of RPs, and number of training samples. It is also commonly understood that the RP clustering and AP selection schemes have great impact on improving the accuracy. 
\par In addition, if one compares the accuracy of approaches with coarser granularity, such as Tilejunction \cite{r232}, the accuracy seems to be degraded compared to approaches with finer granularity. However, all methods should be implemented in a comparable granularity in order to extract safe conclusions. Therefore, comparison of many diverse localization techniques is hindered by the lack of standardized representative data that can be used for fair comparisons. To this end, we plan to create an open repository of our data that can be used by the community for comparative studies. 
\subsection{Recommendations for Future Work }
Emerging fields of Wi-Fi fingerprinting-based localization includes the following directions:
\begin{itemize}
\item  The future practical localization approaches should greatly care about the multipath effects of the indoor fingerprints. The fingerprinting profile may include a multipath profile of fingerprints instead of time collection of single fingerprints. This needs the access to the physical layer of the wireless front-ends. As far as the authors know, the smart devices do not yet allow to this access due to security issues. The team is working on a software defined radio implementation that can provide such capability.
\item The fingerprinting profile of an RP may also include the fingerprints of the user along with his trajectory. This associates a vector of the RSS to one RP and improves the available information in the system.
\item Localization approaches should care about the real environments performance when the infrastructure experiences intentional faults or in emergency scenarios when the navigation of people is of great importance.  
\end{itemize}

\bibliographystyle{IEEEtran}
\bibliography{mybib}

\begin{thebibliography}{100}
\providecommand{\url}[1]{#1}
\csname url@samestyle\endcsname
\providecommand{\newblock}{\relax}
\providecommand{\bibinfo}[2]{#2}
\providecommand{\BIBentrySTDinterwordspacing}{\spaceskip=0pt\relax}
\providecommand{\BIBentryALTinterwordstretchfactor}{4}
\providecommand{\BIBentryALTinterwordspacing}{\spaceskip=\fontdimen2\font plus
\BIBentryALTinterwordstretchfactor\fontdimen3\font minus
  \fontdimen4\font\relax}
\providecommand{\BIBforeignlanguage}[2]{{%
\expandafter\ifx\csname l@#1\endcsname\relax
\typeout{** WARNING: IEEEtran.bst: No hyphenation pattern has been}%
\typeout{** loaded for the language `#1'. Using the pattern for}%
\typeout{** the default language instead.}%
\else
\language=\csname l@#1\endcsname
\fi
#2}}
\providecommand{\BIBdecl}{\relax}
\BIBdecl

\bibitem{r158}
\BIBentryALTinterwordspacing
``Indoor location in retail: Where is the money?'' \emph{ABI Research: Location
  Technologies Market Research}, May 2015. [Online]. Available:
  \url{https://www.abiresearch.com/market-research/product/1013925-indoor-location-in-retail-where-is-the-mon/}
\BIBentrySTDinterwordspacing

\bibitem{r55}
``Wi-{F}i based real-time location tracking: Solutions and technology,'' CISCO
  Sytems, Tech. Rep., 2006.

\bibitem{r56}
``Ekahau,'' http://www.ekahau.com, 2006.

\bibitem{r60}
A.~Bandyopadhyay, D.~Hakim, B.~Funk, E.~Kohn, C.~Teolis, and G.~Blankenship,
  ``System and method for locating, tracking, and/or monitoring the status of
  personnel and/or assets both indoors and outdoors,'' Apr. 2014, {US} Patent
  8,712,686.

\bibitem{r123}
R.~Giuliano, F.~Mazzenga, M.~Petracca, and M.~Vari, ``Indoor localization
  system for first responders in emergency scenario,'' in \emph{Proceedings of
  the 9th International Wireless Communications and Mobile Computing
  Conference}, July 2013, pp. 1821--1826.

\bibitem{r57}
M.~Rodriguez, J.~Favela, E.~Martinez, and M.~Munoz, ``Location-aware access to
  hospital information and services,'' \emph{IEEE Transactions on Information
  Technology in Biomedicine}, vol.~8, no.~4, pp. 448--455, Dec. 2004.

\bibitem{r58}
H.~Harroud, M.~Ahmed, and A.~Karmouch, ``Policy-driven personalized multimedia
  services for mobile users,'' \emph{IEEE Transactions on Mobile Computing},
  vol.~2, no.~1, pp. 16--24, Jan. 2003.

\bibitem{r59}
R.~Muntz and C.~Pancake, ``Challenges in location-aware computing,'' \emph{IEEE
  Pervasive Computing}, vol.~2, no.~2, pp. 80--89, Apr. 2003.

\bibitem{r1}
P.~Misra and P.~Enge, \emph{{Global Positioning System: Signals, Measurements,
  and Performance}}, 2nd~ed.\hskip 1em plus 0.5em minus 0.4em\relax
  Ganga-Jamuna Press, Lincoln MA, 2006.

\bibitem{r145}
``Galileo,'' http://www.gsa.europa.eu/galileo/why-galileo.

\bibitem{r146}
``Beidou,'' http://en.beidou.gov.cn/.

\bibitem{r90}
R.~Mannings, \emph{Ubiquitous Positioning}.\hskip 1em plus 0.5em minus
  0.4em\relax Artech House, 2008.

\bibitem{r91}
A.~Bensky, \emph{Wireless Positioning Technologies and Applications}.\hskip 1em
  plus 0.5em minus 0.4em\relax Artech House, Inc., 2007.

\bibitem{r15}
P.~Bahl and V.~Padmanabhan, ``{RADAR}: an in-building {RF}-based user location
  and tracking system,'' in \emph{Proceedings of the 19th Annual Joint
  Conference of the IEEE Computer and Communications Societies}, vol.~2, 2000,
  pp. 775--784.

\bibitem{r83}
L.~M. Ni, Y.~Liu, Y.~C. Lau, and A.~P. Patil, ``{LANDMARC}: indoor location
  sensing using active {RFID},'' in \emph{Proceedings of the 1st IEEE
  International Conference on Pervasive Computing and Communications}, Mar.
  2003, pp. 407--415.

\bibitem{r163}
J.~Wang and D.~Katabi, ``Dude, where's my card?: {RFID} positioning that works
  with multipath and non-line of sight,'' in \emph{Proceedings of the ACM
  Conference on SIGCOMM}, Aug. 2013, pp. 51--62.

\bibitem{r164}
L.~Yang, Y.~Chen, X.-Y. Li, C.~Xiao, M.~Li, and Y.~Liu, ``Tagoram: Real-time
  tracking of mobile {RFID} tags to high precision using {COTS} devices,'' in
  \emph{Proceedings of the 20th Annual International Conference on Mobile
  Computing and Networking}, Sep. 2014, pp. 237--248.

\bibitem{r165}
W.~Zhuo, B.~Zhang, S.~H.~G. Chan, and E.~Y. Chang, ``Error modeling and
  estimation fusion for indoor localization,'' in \emph{Proceedings of the IEEE
  International Conference on Multimedia and Expo}, July 2012, pp. 741--746.

\bibitem{r92}
N.~A. Alsindi, B.~Alavi, and K.~Pahlavan, ``Measurement and modeling of
  ultrawideband {TOA}-based ranging in indoor multipath environments,''
  \emph{IEEE Transactions on Vehicular Technology}, vol.~58, no.~3, pp.
  1046--1058, Mar. 2009.

\bibitem{r93}
N.~Alsindi and K.~Pahlavan, ``Cooperative localization bounds for indoor
  {U}ltra-wideband wireless sensor networks,'' \emph{EURASIP Journal on
  Advanves in Signal Processing}, vol. 2008, pp. 125:1--125:13, Jan. 2008.

\bibitem{r5}
U.~Bandara, M.~Hasegawa, M.~Inoue, H.~Morikawa, and T.~Aoyama, ``Design and
  implementation of a bluetooth signal strength based location sensing
  system,'' in \emph{Proceedings of IEEE Radio and Wireless Conference}, Sep.
  2004, pp. 319--322.

\bibitem{r160}
X.~Zhao, Z.~Xiao, A.~Markham, N.~Trigoni, and Y.~Ren, ``Does {BTLE} measure up
  against {W}i{F}i? {A} comparison of indoor location performance,'' in
  \emph{Proceedings of 20th European Wireless Conference}, May 2014, pp. 1--6.

\bibitem{r207}
S.~S. Chawathe, ``Beacon placement for indoor localization using bluetooth,''
  in \emph{Proceedings of the 11th International IEEE Conference on Intelligent
  Transportation Systems}, Oct. 2008, pp. 980--985.

\bibitem{r188}
S.~Hilsenbeck, D.~Bobkov, G.~Schroth, R.~Huitl, and E.~Steinbach, ``Graph-based
  data fusion of pedometer and {W}i{F}i measurements for mobile indoor
  positioning,'' in \emph{Proceedings of the 2014 ACM International Joint
  Conference on Pervasive and Ubiquitous Computing}, Sep. 2014, pp. 147--158.

\bibitem{r222}
J.~Chen, X.~J. Wu, P.~Z. Wen, F.~Ye, and J.~W. Liu, ``A new distributed
  localization algorithm for {Z}ig{B}ee wireless networks,'' in \emph{2009
  Chinese Control and Decision Conference}, June 2009, pp. 4451--4456.

\bibitem{r161}
Y.~Chen, D.~Lymberopoulos, J.~Liu, and B.~Priyantha, ``{FM}-based indoor
  localization,'' in \emph{Proceedings of the 10th International Conference on
  Mobile Systems, Applications, and Services}, June 2012, pp. 169--182.

\bibitem{r162}
S.~Yoon, K.~Lee, and I.~Rhee, ``{FM}-based indoor localization via automatic
  fingerprint {DB} construction and matching,'' in \emph{Proceeding of the 11th
  Annual International Conference on Mobile Systems, Applications, and
  Services}, June 2013, pp. 207--220.

\bibitem{r221}
R.~Want, A.~Hopper, V.~Falc\~{a}o, and J.~Gibbons, ``The active badge location
  system,'' \emph{ACM Transactions on Information Systems}, vol.~10, no.~1, pp.
  91--102, Jan. 1992.

\bibitem{r4}
N.~B. Priyantha, A.~Chakraborty, and H.~Balakrishnan, ``{The Cricket
  Location-Support System},'' in \emph{Proceedings of the 6th Annual Conference
  on Mobile Computing}, Aug. 2000.

\bibitem{r61}
F.~Ijaz, H.~K. Yang, A.~Ahmad, and C.~Lee, ``Indoor positioning: A review of
  indoor ultrasonic positioning systems,'' in \emph{Proceedings of the 15th
  International Conference on Advanced Communication Technology (ICACT)}, Jan.
  2013, pp. 1146--1150.

\bibitem{r166}
Z.~Sun, A.~Purohit, K.~Chen, S.~Pan, T.~Pering, and P.~Zhang, ``{PANDAA}:
  Physical arrangement detection of networked devices through ambient-sound
  awareness,'' in \emph{Proceedings of the 13th International Conference on
  Ubiquitous Computing}, Sep. 2011, pp. 425--434.

\bibitem{r176}
H.~Liu, Y.~Gan, J.~Yang, S.~Sidhom, Y.~Wang, Y.~Chen, and F.~Ye, ``Push the
  limit of {W}i{F}i based localization for smartphones,'' in \emph{Proceedings
  of the 18th Annual International Conference on Mobile Computing and
  Networking}, Aug. 2012, pp. 305--316.

\bibitem{r206}
Q.~Zhang, Z.~Zhou, W.~Xu, J.~Qi, C.~Guo, P.~Yi, T.~Zhu, and S.~Xiao,
  ``Fingerprint-free tracking with dynamic enhanced field division,'' in
  \emph{Proceedings of the IEEE Conference on Computer Communications}, Apr.
  2015, pp. 2785--2793.

\bibitem{r194}
C.~Peng, G.~Shen, Y.~Zhang, Y.~Li, and K.~Tan, ``Beep{B}eep: A high accuracy
  acoustic ranging system using {COTS} mobile devices,'' in \emph{Proceedings
  of the 5th International Conference on Embedded Networked Sensor Systems},
  Nov. 2007, pp. 1--14.

\bibitem{r167}
Y.-S. Kuo, P.~Pannuto, K.-J. Hsiao, and P.~Dutta, ``Luxapose: {I}ndoor
  positioning with mobile phones and visible light,'' in \emph{Proceedings of
  the 20th Annual International Conference on Mobile Computing and Networking},
  Sep. 2014, pp. 447--458.

\bibitem{r168}
Z.~Yang, Z.~Wang, J.~Zhang, C.~Huang, and Q.~Zhang, ``Wearables can afford:
  {L}ight-weight indoor positioning with visible light,'' in \emph{Proceedings
  of the 13th Annual International Conference on Mobile Systems, Applications,
  and Services}, May 2015, pp. 317--330.

\bibitem{r142}
A.~T. Mariakakis, S.~Sen, J.~Lee, and K.-H. Kim, ``{SAIL}: Single access
  point-based indoor localization,'' in \emph{Proceedings of the 12th Annual
  International Conference on Mobile Systems, Applications, and Services}, June
  2014, pp. 315--328.

\bibitem{r192}
J.~Niu, B.~Lu, L.~Cheng, Y.~Gu, and L.~Shu, ``Zi{L}oc: Energy efficient
  {W}i{F}i fingerprint-based localization with low-power radio,'' in
  \emph{Proceedings of the IEEE Wireless Communications and Networking
  Conference (WCNC)}, Apr. 2013.

\bibitem{r169}
J.~Chung, M.~Donahoe, C.~Schmandt, I.-J. Kim, P.~Razavai, and M.~Wiseman,
  ``Indoor location sensing using {G}eo-magnetism,'' in \emph{Proceedings of
  the 9th International Conference on Mobile Systems, Applications, and
  Services}, June 2011, pp. 141--154.

\bibitem{r170}
H.~Xie, T.~Gu, X.~Tao, H.~Ye, and J.~Lv, ``Ma{L}oc: A practical magnetic
  fingerprinting approach to indoor localization using smartphones,'' in
  \emph{Proceedings of the 2014 ACM International Joint Conference on Pervasive
  and Ubiquitous Computing}, Sep. 2014, pp. 243--253.

\bibitem{r86}
E.~G. Villegas, E.~Lopez-Aguilera, R.~Vidal, and J.~Paradells, ``Effect of
  adjacent-channel interference in {IEEE} 802.11 {WLAN}s,'' in
  \emph{Proceedings of the 2nd International Conference on Cognitive Radio
  Oriented Wireless Networks and Communications,}, Aug. 2007, pp. 118--125.

\bibitem{r87}
W.~L. Tan, K.~Bialkowski, and M.~Portmann, ``Evaluating adjacent channel
  interference in {IEEE} 802.11 networks,'' in \emph{Proceedings of the IEEE
  71st Vehicular Technology Conference}, May 2010, pp. 1--5.

\bibitem{r6}
P.~Biswas, H.~Aghajan, and Y.~Ye, ``Integration of angle of arrival information
  for multimodal sensor network localization using semidefinite programming,''
  in \emph{Proceedings of the 39th Asilomar Conference on Signals, Systems and
  Computers}, Oct. 2005.

\bibitem{r7}
D.~Niculescu and B.~Nath, ``Ad hoc positioning system ({APS}) using {AOA},'' in
  \emph{Proceedings of the 22nd Annual Joint Conference of the IEEE Computer
  and Communications}, vol.~3, Mar. 2003, pp. 1734--1743.

\bibitem{r8}
R.~Peng and M.~Sichitiu, ``Angle of arrival localization for wireless sensor
  networks,'' in \emph{Proceedings of the 3rd Annual IEEE Communications
  Society on Sensor and Ad Hoc Communications and Networks,}, Sep. 2006, pp.
  374--382.

\bibitem{r10}
X.~Li and K.~Pahlavan, ``Super-resolution {TOA} estimation with diversity for
  indoor geolocation,'' \emph{IEEE Transactions on Wireless Communications},
  vol.~3, no.~1, pp. 224--234, Jan. 2004.

\bibitem{r11}
C.-R. Comsa, J.~Luo, A.~Haimovich, and S.~Schwartz, ``Wireless localization
  using time difference of arrival in narrow-band multipath systems,'' in
  \emph{International Symposium on Signals, Circuits and Systems}, vol.~2, July
  2007, pp. 1--4.

\bibitem{r12}
A.~Amar and G.~Leus, ``A reference-free time difference of arrival source
  localization using a passive sensor array,'' in \emph{IEEE Sensor Array and
  Multichannel Signal Processing Workshop (SAM)}, Oct. 2010, pp. 157--160.

\bibitem{r143}
Y.-T. Chan, W.-Y. Tsui, H.-C. So, and P.-C. Ching, ``Time-of-arrival based
  localization under {NLOS} conditions,'' \emph{IEEE Transactions on Vehicular
  Technology}, vol.~55, no.~1, pp. 17--24, Jan. 2006.

\bibitem{r144}
I.~G\"{u}ven\c{c}, C.-C. Chong, F.~Watanabe, and H.~Inamura, ``{NLOS}
  identification and weighted least-squares localization for {UWB} systems
  using multipath channel statistics,'' \emph{European Association for Signal
  Processing}, vol. 2008, Jan. 2008.

\bibitem{r73}
K.~H. Kim, J.~H. Kim, Y.~J. Yoon, J.~H. Seok, and J.~W. Lim, ``Propagation
  model for the {WLAN} service at the campus environments,'' in
  \emph{Proceedings of the 57th IEEE Semiannual Vehicular Technology
  Conference}, vol.~1, Apr. 2003, pp. 196--200.

\bibitem{r228}
T.~Hastie, R.~Tibshirani, and J.~Friedman, \emph{The Elements of Statistical
  Learning}, 2001.

\bibitem{r111}
A.~Bose and C.~H. Foh, ``A practical path loss model for indoor {W}i{F}i
  positioning enhancement,'' in \emph{Proceedings of 6th International
  Conference on Information, Communications Signal Processing}, Dec. 2007, pp.
  1--5.

\bibitem{r74}
M.~Brunato and R.~Battiti, ``Statistical learning theory for location
  fingerprinting in wireless {LAN}s,'' \emph{Computer Network}, vol.~47, no.~6,
  pp. 825--845, Apr. 2005.

\bibitem{r189}
S.~He, T.~Hu, and S.-H.~G. Chan, ``Contour-based trilateration for indoor
  fingerprinting localization,'' in \emph{Proceedings of the 13th ACM
  Conference on Embedded Networked Sensor Systems}, Nov. 2015, pp. 225--238.

\bibitem{r125}
M.~Seifeldin, A.~Saeed, A.~E. Kosba, A.~El-Keyi, and M.~Youssef, ``Nuzzer: A
  large-scale device-free passive localization system for wireless
  environments,'' \emph{IEEE Transactions on Mobile Computing}, vol.~12, no.~7,
  pp. 1321--1334, July 2013.

\bibitem{r126}
M.~Youssef, M.~Mah, and A.~Agrawala, ``Challenges: Device-free passive
  localization for wireless environments,'' in \emph{Proceedings of the 13th
  Annual ACM International Conference on Mobile Computing and Networking}, Sep.
  2007, pp. 222--229.

\bibitem{r129}
J.~Hong and T.~Ohtsuki, ``Signal eigenvector-based device-free passive
  localization using array sensor,'' \emph{IEEE Transactions on Vehicular
  Technology}, vol.~64, no.~4, pp. 1354--1363, Apr. 2015.

\bibitem{r131}
I.~Sabek, M.~Youssef, and A.~V. Vasilakos, ``{ACE}: An accurate and efficient
  multi-entity device-free {WLAN} localization system,'' \emph{IEEE
  Transactions on Mobile Computing}, vol.~14, no.~2, pp. 261--273, Feb. 2015.

\bibitem{r69}
S.-H. Fang, T.-N. Lin, and K.-C. Lee, ``A novel algorithm for multipath
  fingerprinting in indoor {WLAN} environments,'' \emph{IEEE Transactions on
  Wireless Communications}, vol.~7, no.~9, pp. 3579--3588, Sep. 2008.

\bibitem{r70}
E.~Kupershtein, M.~Wax, and I.~Cohen, ``Single-site emitter localization via
  multipath fingerprinting,'' \emph{IEEE Transactions on Signal Processing},
  vol.~61, no.~1, pp. 10--21, Jan. 2013.

\bibitem{r76}
Z.~Xiang, S.~Song, J.~Chen, H.~Wang, J.~Huang, and X.~Gao, ``A wireless
  {LAN}-based indoor positioning technology,'' \emph{IBM Journal of Research
  and Development}, vol.~48, pp. 617--626, Sep. 2004.

\bibitem{r107}
H.~Lim, L.-C. Kung, J.~C. Hou, and H.~Luo, ``Zero-configuration indoor
  localization over {IEEE} 802.11 wireless infrastructure,'' \emph{Wireless
  Network}, vol.~16, no.~2, pp. 405--420, Feb. 2010.

\bibitem{r72}
K.~Kaemarungsi and P.~Krishnamurthy, ``Properties of indoor received signal
  strength for {WLAN} location fingerprinting,'' in \emph{Proceedings of the
  1st Annual International Conference on Mobile and Ubiquitous Systems:
  Networking and Services}, Aug. 2004, pp. 14--23.

\bibitem{r94}
V.~Hodge and J.~Austin, ``A survey of outlier detection methodologies,''
  \emph{Artificial Intelligence Review}, vol.~22, no.~2, pp. 85--126, Oct.
  2004.

\bibitem{r99}
Y.~C. Chen, W.~C. Sun, and J.~C. Juang, ``Outlier detection technique for
  {RSS}-based localization problems in wireless sensor networks,'' in
  \emph{Proceedings of SICE Annual Conference}, Aug. 2010, pp. 657--662.

\bibitem{r180}
I.~Guvenc and C.~C. Chong, ``A survey on {TOA} based wireless localization and
  {NLOS} mitigation techniques,'' \emph{IEEE Communications Surveys and
  Tutorials}, vol.~11, no.~3, pp. 107--124, Third Quarter 2009.

\bibitem{r105}
H.~Liu, H.~Darabi, P.~Banerjee, and J.~Liu, ``Survey of wireless indoor
  positioning techniques and systems,'' \emph{IEEE Transactions on Systems,
  Man, and Cybernetics, Part C (Applications and Reviews)}, vol.~37, no.~6, pp.
  1067--1080, Nov. 2007.

\bibitem{r181}
F.~Seco, A.~R. Jimenez, C.~Prieto, J.~Roa, and K.~Koutsou, ``A survey of
  mathematical methods for indoor localization,'' in \emph{Proceedings of the
  IEEE International Symposium on Intelligent Signal Processing}, Aug. 2009,
  pp. 9--14.

\bibitem{r182}
V.~Honkavirta, T.~Perala, S.~Ali-Loytty, and R.~Piche, ``A comparative survey
  of {WLAN} location fingerprinting methods,'' in \emph{Proceedings of the 6th
  Workshop on Positioning, Navigation and Communication}, Mar. 2009, pp.
  243--251.

\bibitem{r183}
G.~Sun, J.~Chen, W.~Guo, and K.~J.~R. Liu, ``Signal processing techniques in
  network-aided positioning: a survey of state-of-the-art positioning
  designs,'' \emph{IEEE Signal Processing Magazine}, vol.~22, no.~4, pp.
  12--23, July 2005.

\bibitem{r184}
D.~Lymberopoulos, J.~Liu, X.~Yang, R.~R. Choudhury, V.~Handziski, and S.~Sen,
  ``A realistic evaluation and comparison of indoor location technologies:
  Experiences and lessons learned,'' in \emph{Proceedings of the 14th
  International Conference on Information Processing in Sensor Networks}, Apr.
  2015, pp. 178--189.

\bibitem{r185}
A.~M. Hossain and W.-S. Soh, ``A survey of calibration-free indoor positioning
  systems,'' \emph{Computer Communications}, vol.~66, pp. 1 -- 13, 2015.

\bibitem{r106}
A.~Roxin, J.~Gaber, M.~Wack, and A.~Nait-Sidi-Moh, ``Survey of wireless
  geolocation techniques,'' in \emph{IEEE Globecom Workshops}, Nov. 2007, pp.
  1--9.

\bibitem{r172}
D.~Lymberopoulos, J.~Liu, X.~Yang, R.~R. Choudhury, S.~Sen, and V.~Handziski,
  ``Microsoft indoor localization competition: Experiences and lessons
  learned,'' \emph{GetMobile: Mobile Computing and Communications}, vol.~18,
  no.~4, pp. 24--31, Jan. 2015.

\bibitem{r212}
S.~He and S.~H.~G. Chan, ``Wi-{F}i fingerprint-based indoor positioning: Recent
  advances and comparisons,'' \emph{IEEE Communications Surveys and Tutorials},
  vol.~18, no.~1, pp. 466--490, First Quarter 2016.

\bibitem{r82}
P.~Bahl and V.~N. Padmanabhan, ``Enhancements to the {RADAR} user location and
  tracking system,'' Microsoft Research, Tech. Rep., Feb. 2000.

\bibitem{r14}
S.~Saha, K.~Chaudhuri, D.~Sanghi, and P.~Bhagwat, ``Location determination of a
  mobile device using {IEEE} 802.11b access point signals,'' in
  \emph{Proceedings of the IEEE Wireless Communications and Networking},
  vol.~3, Mar. 2003, pp. 1987--1992 vol.3.

\bibitem{r18}
Q.~Tran, J.~Tantra, C.~H. Foh, A.-H. Tan, K.~C. Yow, and D.~Qiu, ``Wireless
  indoor positioning system with enhanced nearest neighbors in signal space
  algorithm,'' in \emph{Proceedings of the 64th IEEE Vehicular Technology
  Conference}, Sep. 2006, pp. 1--5.

\bibitem{r79}
P.~Prasithsangaree, P.~Krishnamurthy, and P.~Chrysanthis, ``On indoor position
  location with wireless {LAN}s,'' in \emph{Proceedings of the 13th IEEE
  International Symposium on Personal, Indoor and Mobile Radio Communications},
  vol.~2, Sep. 2002, pp. 720--724 vol.2.

\bibitem{r17}
Z.~Li, W.~Trappe, Y.~Zhang, and B.~Nath, ``Robust statistical methods for
  securing wireless localization in sensor networks,'' in \emph{Proceedings of
  the 4th International Symposium on Information Processing in Sensor
  Networks}, Apr. 2005, pp. 91--98.

\bibitem{r66}
A.~Haeberlen, E.~Flannery, A.~M. Ladd, A.~Rudys, D.~S. Wallach, and L.~E.
  Kavraki, ``Practical robust localization over large-scale 802.11 wireless
  networks,'' in \emph{Proceedings of the 10th Annual International Conference
  on Mobile Computing and Networking}, 2004.

\bibitem{r65}
A.~M. Ladd, K.~E. Bekris, A.~Rudys, G.~Marceau, L.~E. Kavraki, and D.~S.
  Wallach, ``Robotics-based location sensing using wireless ethernet,'' in
  \emph{Proceedings of the 8th Annual International Conference on Mobile
  Computing and Networking}, Sep. 2002, pp. 227--238.

\bibitem{r21}
A.~Kushki, K.~Plataniotis, and A.~Venetsanopoulos, ``Kernel-based positioning
  in wireless local area networks,'' \emph{IEEE Transactions on Mobile
  Computing}, vol.~6, no.~6, pp. 689--705, June 2007.

\bibitem{r64}
M.~Youssef and A.~Agrawala, ``Continuous space estimation for {WLAN} location
  determination systems,'' in \emph{Proceedings of the 13th International
  Conference on Computer Communications and Networks}, Oct. 2004, pp. 161--166.

\bibitem{r232}
------, ``The {H}orus {WLAN} location determination system,'' in
  \emph{Proceedings of the 3rd International Conference on Mobile Systems,
  Applications, and Services}, 2005, pp. 205--218.

\bibitem{r25}
J.~Pan, J.~Kwok, Q.~Yang, and Y.~Chen, ``Multidimensional vector regression for
  accurate and low-cost location estimation in pervasive computing,''
  \emph{IEEE Transactions on Knowledge and Data Engineering}, vol.~18, no.~9,
  pp. 1181--1193, Sep. 2006.

\bibitem{r26}
C.-L. Wu, L.-C. Fu, and F.-L. Lian, ``{WLAN} location determination in e-home
  via support vector classification,'' in \emph{Proceedings of the IEEE
  International Conference on Networking, Sensing and Control}, vol.~2, 2004,
  pp. 1026--1031.

\bibitem{r27}
T.-N. Lin, S.-H. Fang, W.-H. Tseng, C.-W. Lee, and J.-W. Hsieh, ``A
  group-discrimination-based access point selection for {WLAN} fingerprinting
  localization,'' \emph{IEEE Transactions on Vehicular Technology}, vol.~63,
  no.~8, pp. 3967--3976, Oct. 2014.

\bibitem{r220}
G.~Nu\~{n}o Barrau and J.~M. P\'{a}z-Borrallo, ``A new location estimation
  system for wireless networks based on linear discriminant functions and
  hidden markov models,'' \emph{EURASIP Journal on Applied Signal Processing},
  vol. 2006, pp. 159--159, Jan. 2006.

\bibitem{r175}
C.~Gentile, N.~Alsindi, R.~Raulefs, and C.~Teolis, \emph{Geolocation
  Techniques: Principles and Applications}.\hskip 1em plus 0.5em minus
  0.4em\relax Springer, 2012.

\bibitem{r179}
A.~Farshad, J.~Li, M.~K. Marina, and F.~J. Garcia, ``A microscopic look at
  {W}i{F}i fingerprinting for indoor mobile phone localization in diverse
  environments,'' in \emph{Proceedings of the International Conference on
  Indoor Positioning and Indoor Navigation (IPIN)}, Oct. 2013, pp. 1--10.

\bibitem{r233}
S.~Han, C.~Zhao, W.~Meng, and C.~Li, ``Cosine similarity based fingerprinting
  algorithm in {WLAN} indoor positioning against device diversity,'' in
  \emph{Proceedings of the IEEE International Conference on Communications
  (ICC)}, June 2015, pp. 2710--2714.

\bibitem{r153}
R.~Singh, L.~Macchi, and C.~S. Regazzoni, ``A statistical modelling based
  location determination method using fusion technique in {WLAN},'' in
  \emph{Proceedings on International Workshop on Wireless Ad-Hoc Networks},
  2005.

\bibitem{r154}
T.~Roos, P.~Myllym{\"a}ki, H.~Tirri, P.~Misikangas, and J.~Siev{\"a}nen, ``A
  probabilistic approach to {WLAN} user location estimation,''
  \emph{International Journal of Wireless Information Networks}, vol.~9, no.~3,
  pp. 155--164, July 2002.

\bibitem{r104}
M.~D. Redzic, C.~Brennan, and N.~E. O'Connor, ``{SEAMLOC}: Seamless indoor
  localization based on reduced number of calibration points,'' \emph{IEEE
  Transactions on Mobile Computing}, vol.~13, no.~6, pp. 1326--1337, June 2014.

\bibitem{r155}
M.~Youssef and A.~Agrawala, ``The {H}orus location determination system,''
  \emph{Wireless Networks}, vol.~14, no.~3, pp. 357--374, June 2008.

\bibitem{r156}
A.~Kushki, \emph{A Cognitive Radio Tracking System for Indoor Environments},
  2008.

\bibitem{r20}
A.~Goldsmith, \emph{Wireless Communications}.\hskip 1em plus 0.5em minus
  0.4em\relax Cambridge University Press, 2005.

\bibitem{r198}
D.~Madigan, E.~Einahrawy, R.~P. Martin, W.~H. Ju, P.~Krishnan, and A.~S.
  Krishnakumar, ``Bayesian indoor positioning systems,'' in \emph{Proceedings
  of the IEEE 24th Annual Joint Conference of the IEEE Computer and
  Communications Societies.}, vol.~2, Mar. 2005, pp. 1217--1227.

\bibitem{r19}
M.~Youssef, A.~Agrawala, and A.~Udaya~Shankar, ``{WLAN} location determination
  via clustering and probability distributions,'' in \emph{Proceedings of the
  1st IEEE International Conference on Pervasive Computing and Communications},
  Mar. 2003, pp. 143--150.

\bibitem{r67}
D.~W. Scott, \emph{Multivariate density estimation: theory, practice, and
  visualization}.\hskip 1em plus 0.5em minus 0.4em\relax Wiley-Interscience,
  1992.

\bibitem{r68}
B.~W. Silverman, \emph{Density Estimation for Statistics and Data
  Analysis}.\hskip 1em plus 0.5em minus 0.4em\relax London: Chapman \& Hall,
  1986.

\bibitem{r173}
D.~Han, S.~Jung, M.~Lee, and G.~Yoon, ``Building a practical {W}i-{F}i-based
  indoor navigation system,'' \emph{IEEE Pervasive Computing}, vol.~13, no.~2,
  pp. 72--79, Apr. 2014.

\bibitem{r174}
Y.~Liu and Z.~Yang, \emph{Location, Localization, and Localizability:
  Location-awareness Technology for Wireless Networks}.\hskip 1em plus 0.5em
  minus 0.4em\relax Springer Publishing Company, Incorporated, 2014.

\bibitem{r28}
A.~Kushki, K.~Plataniotis, and A.~Venetsanopoulos, ``Location tracking in
  wireless local area networks with adaptive radio maps,'' in \emph{Proceedings
  of the IEEE International Conference on Acoustics, Speech and Signal
  Processing}, vol.~5, May 2006.

\bibitem{r205}
M.~Azizyan, I.~Constandache, and R.~Roy~Choudhury, ``Surround{S}ense: Mobile
  phone localization via ambience fingerprinting,'' in \emph{Proceedings of the
  15th Annual International Conference on Mobile Computing and Networking},
  Sep. 2009, pp. 261--272.

\bibitem{r31}
Q.~Zhao, S.~Zhang, X.~Liu, and X.~Lin, ``An effective preprocessing scheme for
  {WLAN}-based fingerprint positioning systems,'' in \emph{Proceedings of the
  12th IEEE International Conference on Communication Technology (ICCT)}, Nov.
  2010, pp. 592--595.

\bibitem{r139}
A.~Khalajmehrabadi, N.~Gatsis, and D.~Akopian, ``A joint indoor {WLAN}
  localization and outlier detection scheme using {LASSO} and elastic-net
  optimization techniques,'' \emph{Submitted for publication in IEEE
  Transactions on Mobile Computing}.

\bibitem{r140}
------, ``Structured group sparsity: Indoor {WLAN} localization, outlier
  detection, and radio map interpolation scheme,'' \emph{Accepted with minor
  revisions in IEEE Transactions on Vehcular Technology}.

\bibitem{r141}
------, ``Indoor {WLAN} localization using group sparsity optimization
  technique,'' in \emph{Proceedings of the IEEE/ION Position, Location and
  Navigation Symposium (PLANS)}, Apr. 2016, pp. 584--588.

\bibitem{r35}
Y.~Chen, Q.~Yang, J.~Yin, and X.~Chai, ``Power-efficient access-point selection
  for indoor location estimation,'' \emph{IEEE Transactions on Knowledge and
  Data Engineering}, vol.~18, no.~7, pp. 877--888, July 2006.

\bibitem{r80}
R.~O. Duda, P.~E. Hart, and D.~G. Stork, \emph{Pattern Classification (2nd
  Edition)}.\hskip 1em plus 0.5em minus 0.4em\relax Wiley-Interscience, 2000.

\bibitem{r84}
M.~Youssef and A.~Agrawala, ``Handling samples correlation in the system,'' in
  \emph{Proceedings of the Annual Joint Conference of the IEEE Computer and
  Communications Societies}, vol.~2, Mar. 2004, pp. 1023--1031.

\bibitem{r85}
S.~H. Fang and C.~H. Wang, ``A dynamic hybrid projection approach for improved
  {W}i-{F}i location fingerprinting,'' \emph{IEEE Transactions on Vehicular
  Technology}, vol.~60, no.~3, pp. 1037--1044, Mar. 2011.

\bibitem{r30}
C.~Feng, W.~Au, S.~Valaee, and Z.~Tan, ``Received-signal-strength-based indoor
  positioning using compressive sensing,'' \emph{IEEE Transactions on Mobile
  Computing}, vol.~11, no.~12, pp. 1983--1993, Dec. 2012.

\bibitem{r88}
S.~Z. J.~L. Y.~Zhou, X.~Chen and D.~Liang, ``{AP} selection algorithm in {WLAN}
  indoor localization,'' \emph{Information Technology Journal}, vol.~12,
  no.~16, pp. 3773--3776, 2013.

\bibitem{r130}
D.~Liang, Z.~Zhang, and M.~Peng, ``Access point reselection and adaptive
  cluster splitting-based indoor localization in wireless local area
  networks,'' \emph{IEEE Internet of Things Journal}, vol.~2, no.~6, pp.
  573--585, Dec. 2015.

\bibitem{r102}
B.~W. Silverman, \emph{Density Estimation for Statistics and Data
  Analysis}.\hskip 1em plus 0.5em minus 0.4em\relax London: Chapman \& Hall,
  1986.

\bibitem{r103}
S.~H. Fang and T.~Lin, ``Principal component localization in indoor {WLAN}
  environments,'' \emph{IEEE Transactions on Mobile Computing}, vol.~11, no.~1,
  pp. 100--110, Jan. 2012.

\bibitem{r121}
A.~Tabibiazar and O.~Basir, ``Compressive sensing indoor localization,'' in
  \emph{Proceedings of the IEEE International Conference on Systems, Man, and
  Cybernetics (SMC)}, Oct. 2011, pp. 1986--1991.

\bibitem{r208}
E.~Martin, O.~Vinyals, G.~Friedland, and R.~Bajcsy, ``Precise indoor
  localization using smart phones,'' in \emph{Proceedings of the 18th ACM
  International Conference on Multimedia}, Oct. 2010, pp. 787--790.

\bibitem{r215}
R.~Harle, ``A survey of indoor inertial positioning systems for pedestrians,''
  \emph{IEEE Communications Surveys and Tutorials}, vol.~15, no.~3, pp.
  1281--1293, Third Quarter 2013.

\bibitem{r216}
N.~Roy, H.~Wang, and R.~Roy~Choudhury, ``I am a smartphone and {I} can tell my
  user's walking direction,'' in \emph{Proceedings of the 12th Annual
  International Conference on Mobile Systems, Applications, and Services}, June
  2014, pp. 329--342.

\bibitem{r217}
A.~Brajdic and R.~Harle, ``Walk detection and step counting on unconstrained
  smartphones,'' in \emph{Proceedings of the 2013 ACM International Joint
  Conference on Pervasive and Ubiquitous Computing}, Sep. 2013, pp. 225--234.

\bibitem{r224}
H.~Wang, S.~Sen, A.~Elgohary, M.~Farid, M.~Youssef, and R.~R. Choudhury, ``No
  need to war-drive: Unsupervised indoor localization,'' in \emph{Proceedings
  of the 10th International Conference on Mobile Systems, Applications, and
  Services}, June 2012, pp. 197--210.

\bibitem{r225}
H.~Abdelnasser, R.~Mohamed, A.~Elgohary, M.~F. Alzantot, H.~Wang, S.~Sen, R.~R.
  Choudhury, and M.~Youssef, ``Semantic{SLAM}: Using environment landmarks for
  unsupervised indoor localization,'' \emph{IEEE Transactions on Mobile
  Computing}, vol.~15, no.~7, pp. 1770--1782, July 2016.

\bibitem{r219}
S.~He, S.~H.~G. Chan, L.~Yu, and N.~Liu, ``Fusing noisy fingerprints with
  distance bounds for indoor localization,'' in \emph{Proceedings of the IEEE
  Conference on Computer Communications (INFOCOM)}, Apr. 2015, pp. 2506--2514.

\bibitem{r108}
Y.~Ji, S.~Biaz, S.~Pandey, and P.~Agrawal, ``{ARIADNE}: A dynamic indoor signal
  map construction and localization system,'' in \emph{Proceedings of the 4th
  International Conference on Mobile Systems, Applications and Services}, June
  2006, pp. 151--164.

\bibitem{r109}
M.~Ocana, L.~M. Bergasa, M.~A. Sotelo, R.~Flores, E.~Lopez, and R.~Barea,
  ``Training method improvements of a {W}i{F}i navigation system based on
  {POMDP},'' in \emph{Proceedings of the IEEE/RSJ International Conference on
  Intelligent Robots and Systems}, Oct. 2006, pp. 5259--5264.

\bibitem{r135}
H.~Wang, L.~Ma, Y.~Xu, and Z.~Deng, ``Dynamic radio map construction for {WLAN}
  indoor location,'' in \emph{Proceedings of the International Conference on
  Intelligent Human-Machine Systems and Cybernetics (IHMSC)}, vol.~2, Aug.
  2011, pp. 162--165.

\bibitem{r171}
S.~Yang, P.~Dessai, M.~Verma, and M.~Gerla, ``Free{L}oc: Calibration-free
  crowdsourced indoor localization,'' in \emph{Proceedings of the IEEE
  INFOCOM}, Apr. 2013, pp. 2481--2489.

\bibitem{r193}
K.~Chang and D.~Han, ``Crowdsourcing-based radio map update automation for
  {W}i-{F}i positioning systems,'' in \emph{Proceedings of the 3rd ACM
  SIGSPATIAL International Workshop on Crowdsourced and Volunteered Geographic
  Information}, Nov. 2014, pp. 24--31.

\bibitem{r226}
R.~W. Ouyang, A.~K.~S. Wong, C.~T. Lea, and M.~Chiang, ``Indoor location
  estimation with reduced calibration exploiting unlabeled data via hybrid
  generative/discriminative learning,'' \emph{IEEE Transactions on Mobile
  Computing}, vol.~11, no.~11, pp. 1613--1626, Nov. 2012.

\bibitem{r110}
M.~M. Atia, A.~Noureldin, and M.~J. Korenberg, ``Dynamic online-calibrated
  radio maps for indoor positioning in wireless local area networks,''
  \emph{IEEE Transactions on Mobile Computing}, vol.~12, no.~9, pp. 1774--1787,
  2013.

\bibitem{r112}
C.~E. Rasmussen and C.~K.~I. Williams, \emph{Gaussian Processes for Machine
  Learning}.\hskip 1em plus 0.5em minus 0.4em\relax The MIT Press, 2005.

\bibitem{r113}
M.~M. Atia, A.~Noureldin, and M.~Korenberg, ``Gaussian process regression
  approach for bridging {GPS} outages in integrated navigation systems,''
  \emph{Electronics Letters}, vol.~47, no.~1, pp. 52--53, Jan. 2011.

\bibitem{r133}
C.~Koweerawong, K.~Wipusitwarakun, and K.~Kaemarungsi, ``Indoor localization
  improvement via adaptive {RSS} fingerprinting database,'' in \emph{The
  International Conference on Information Networking}, Jan. 2013, pp. 412--416.

\bibitem{r16}
B.~Li, Y.~Wang, H.~Lee, A.~Dempster, and C.~Rizos, ``Method for yielding a
  database of location fingerprints in {WLAN},'' \emph{IEE Communications
  Proceedings}, vol. 152, no.~5, pp. 580--586, Oct. 2005.

\bibitem{r136}
V.~Rakovic, M.~Angjelicinoski, V.~Atanasovski, and L.~Gavrilovska, ``Location
  estimation of radio transmitters based on spatial interpolation of {RSS}
  values,'' in \emph{Proceedings of the 7th International ICST Conference on
  Cognitive Radio Oriented Wireless Networks and Communications (CROWNCOM)},
  June 2012, pp. 242--247.

\bibitem{r132}
J.~Talvitie, M.~Renfors, and E.~S. Lohan, ``Distance-based interpolation and
  extrapolation methods for {RSS}-based localization with indoor wireless
  signals,'' \emph{IEEE Transactions on Vehicular Technology}, vol.~64, no.~4,
  pp. 1340--1353, Apr. 2015.

\bibitem{r134}
R.~Kubota, S.~Tagashira, Y.~Arakawa, T.~Kitasuka, and A.~Fukuda, ``Efficient
  survey database construction using location fingerprinting interpolation,''
  in \emph{Proceedings of the 27th International Conference on Advanced
  Information Networking and Applications (AINA)}, Mar. 2013, pp. 469--476.

\bibitem{r204}
Y.~Tsuda, Q.~Kong, and T.~Maekawa, ``Detecting and correcting {W}i{F}i
  positioning errors,'' in \emph{Proceedings of the 2013 ACM International
  Joint Conference on Pervasive and Ubiquitous Computing}, Sep. 2013, pp.
  777--786.

\bibitem{r40}
S.~Mazuelas, A.~Bahillo, R.~Lorenzo, P.~Fernandez, F.~Lago, E.~Garcia, J.~Blas,
  and E.~Abril, ``Robust indoor positioning provided by real-time {RSSI} values
  in unmodified {WLAN} networks,'' \emph{IEEE Journal of Selected Topics in
  Signal Processing}, vol.~3, no.~5, pp. 821--831, Oct. 2009.

\bibitem{r95}
R.~Pearson, ``Exploring process data,'' \emph{Journal of Process Control},
  vol.~11, no.~2, pp. 179 -- 194, Apr. 2001.

\bibitem{r96}
U.~G. Laurie~Davies, ``The identification of multiple outliers,'' \emph{Journal
  of the American Statistical Association}, vol.~88, no. 423, pp. 782--792,
  Sep. 1993.

\bibitem{r97}
P.~Huber, \emph{Robust statistics}.\hskip 1em plus 0.5em minus 0.4em\relax
  Wiley New York, 1981.

\bibitem{r98}
R.~K. Pearson, ``Outliers in process modeling and identification,'' \emph{IEEE
  Transactions on Control Systems Technology}, vol.~10, pp. 55--63, Jan. 2002.

\bibitem{r36}
C.~Laoudias, M.~P. Michaelides, and C.~G. Panayiotou, ``Fault detection and
  mitigation in {WLAN RSS} fingerprint-based positioning,'' in
  \emph{Proceedings of the International Conference on Indoor Positioning and
  Indoor Navigation (IPIN)}, Sep. 2011, pp. 1--7.

\bibitem{r101}
J.~A. Morales, D.~Akopian, and S.~Agaian, ``Faulty measurements impact on
  wireless local area network positioning performance,'' \emph{IET Radar, Sonar
  Navigation}, vol.~9, no.~5, pp. 501--508, Aug. 2015.

\bibitem{r151}
J.-J. Fuchs, ``An inverse problem approach to robust regression,'' in
  \emph{IEEE International Conference on Acoustics, Speech, and Signal
  Processing}, vol.~4, Mar. 1999, pp. 1809--1812 vol.4.

\bibitem{r152}
V.~Kekatos and G.~Giannakis, ``From sparse signals to sparse residuals for
  robust sensing,'' \emph{IEEE Transactions on Signal Processing}, vol.~59,
  no.~7, pp. 3355--3368, July 2011.

\bibitem{r227}
A.~W. Tsui, Y.-H. Chuang, and H.-H. Chu, ``Unsupervised learning for solving
  {RSS} hardware variance problem in {W}i{F}i localization,'' \emph{Mobile
  Networks and Applications}, vol.~14, no.~5, pp. 677--691, Oct. 2009.

\bibitem{r229}
P.~Mirowski, P.~Whiting, H.~Steck, R.~Palaniappan, M.~MacDonald, D.~Hartmann,
  and T.~Ho, ``Probability kernel regression for {W}i{F}i localisation,''
  \emph{Journal of Location Based Services}, vol.~6, no.~2, pp. 81--100, 2012.

\bibitem{r186}
A.~Goswami, L.~E. Ortiz, and S.~R. Das, ``Wi{GEM}: A learning-based approach
  for indoor localization,'' in \emph{Proceedings of the 7th Conference on
  Emerging Networking Experiments and Technologies}, Dec. 2011, pp. 3:1--3:12.

\bibitem{r190}
A.~K. M.~M. Hossain, Y.~Jin, W.~S. Soh, and H.~N. Van, ``{SSD}: A robust {RF}
  location fingerprint addressing mobile devices' heterogeneity,'' \emph{IEEE
  Transactions on Mobile Computing}, vol.~12, pp. 65--77, Jan. 2013.

\bibitem{r137}
A.~Arya, P.~Godlewski, M.~Campedel, and G.~du~Chéné, ``Radio database
  compression for accurate energy-efficient localization in fingerprinting
  systems,'' \emph{IEEE Transactions on Knowledge and Data Engineering},
  vol.~25, no.~6, pp. 1368--1379, June 2013.

\bibitem{r23}
M.~Lin, X.~Yubin, and Z.~Mu, ``Accuracy enhancement for fingerprint-based
  {WLAN} indoor probability positioning algorithm,'' in \emph{Proceedings of
  the 1st International Conference on Pervasive Computing Signal Processing and
  Applications (PCSPA)}, Sep. 2010, pp. 167--170.

\bibitem{r29}
A.~Kushki, K.~Plataniotis, A.~Venetsanopoulos, and C.~Regazzoni, ``Radio map
  fusion for indoor positioning in wireless local area networks,'' in
  \emph{Proceedings of the 8th International Conference on Information Fusion},
  vol.~2, July 2005, pp. 8 pp.--.

\bibitem{r230}
P.~Mirowski, H.~Steck, P.~Whiting, R.~Palaniappan, M.~MacDonald, and T.~K. Ho,
  ``{KL}-divergence kernel regression for non-gaussian fingerprint based
  localization,'' in \emph{Proceedings of the International Conference on
  Indoor Positioning and Indoor Navigation (IPIN)}, Sep. 2011, pp. 1--10.

\bibitem{r234}
S.~He and S.~H.~G. Chan, ``Tilejunction: Mitigating signal noise for
  fingerprint-based indoor localization,'' \emph{IEEE Transactions on Mobile
  Computing}, vol.~15, no.~6, pp. 1554--1568, June 2016.

\bibitem{r235}
------, ``Sectjunction: Wi-fi indoor localization based on junction of signal
  sectors,'' in \emph{Proceedings of the IEEE International Conference on
  Communications (ICC)}, June 2014, pp. 2605--2610.

\bibitem{r237}
S.~He, T.~Hu, and S.-H.~G. Chan, ``Contour-based trilateration for indoor
  fingerprinting localization,'' in \emph{Proceedings of the 13th ACM
  Conference on Embedded Networked Sensor Systems}, 2015, pp. 225--238.

\bibitem{r50}
Y.~C. Pati, R.~Rezaiifar, Y.~C. P.~R. Rezaiifar, and P.~S. Krishnaprasad,
  ``Orthogonal matching pursuit: Recursive function approximation with
  applications to wavelet decomposition,'' in \emph{Proceedings of the 27 th
  Annual Asilomar Conference on Signals, Systems, and Computers}, Nov. 1993,
  pp. 40--44.

\bibitem{r52}
R.~Chartrand and W.~Yin, ``Iteratively reweighted algorithms for compressive
  sensing,'' in \emph{Proceedings of the IEEE International Conference on
  Acoustics, Speech and Signal Processing}, Mar. 2008, pp. 3869--3872.

\bibitem{r120}
S.~S. Chen, D.~L. Donoho, and M.~A. Saunders, ``Atomic decomposition by basis
  pursuit,'' \emph{SIAM Journal on Scientific Computing}, vol.~20, pp. 33--61,
  1998.

\bibitem{r157}
E.~Cand{\`e}s and M.~Wakin, ``{An introduction to compressive sampling},''
  \emph{IEEE Signal Processing Magazine}, vol.~25, no.~2, pp. 21--30, 2008.

\bibitem{r115}
E.~Candes and J.~Romberg, ``Sparsity and incoherence in compressive sampling,''
  2006.

\bibitem{r147}
R.~Tibshirani, ``Regression shrinkage and selection via the {L}asso,''
  \emph{Journal of the Royal Statistical Society, Series B}, vol.~58, pp.
  267--288, 1994.

\bibitem{r148}
J.~Friedman, T.~Hastie, and R.~Tibshirani, ``Regularization paths for
  generalized linear models via coordinate descent,'' \emph{Journal of
  Statistical Software}, vol.~33, no.~1, pp. 1--22, Jan. 2010.

\bibitem{r149}
N.~Simon, J.~Friedman, T.~Hastie, and R.~Tibshirani, ``A sparse-group
  {lasso},'' \emph{Journal of Computational and Graphical Statistics}, 2013.

\bibitem{r150}
J.~Liu, S.~Ji, and J.~Ye, \emph{{SLEP}: Sparse Learning with Efficient
  Projections}, Arizona State University, 2009.

\bibitem{r177}
W.~Sun, J.~Liu, C.~Wu, Z.~Yang, X.~Zhang, and Y.~Liu, ``Mo{L}oc: On
  distinguishing fingerprint twins,'' in \emph{Proceedings of the 33rd IEEE
  International Conference on Distributed Computing Systems (ICDCS)}, July
  2013, pp. 226--235.

\bibitem{r195}
Z.~Xiao, H.~Wen, A.~Markham, and N.~Trigoni, ``Lightweight map matching for
  indoor localisation using conditional random fields,'' in \emph{Proceedings
  of the 13th International Symposium on Information Processing in Sensor
  Networks}, Apr. 2014, pp. 131--142.

\bibitem{r197}
H.~Wen, Z.~Xiao, N.~Trigoni, and P.~Blunsom, ``On assessing the accuracy of
  positioning systems in indoor environments,'' in \emph{Proceedings of 10th
  European Conference on Wireless Sensor Networks}.\hskip 1em plus 0.5em minus
  0.4em\relax Springer Berlin Heidelberg, 2013, pp. 1--17.

\bibitem{r231}
Z.~Xiao, H.~Wen, A.~Markham, and N.~Trigoni, ``Indoor tracking using undirected
  graphical models,'' \emph{IEEE Transactions on Mobile Computing}, vol.~14,
  no.~11, pp. 2286--2301, Nov. 2015.

\bibitem{r38}
H.~Wang, A.~Szabo, J.~Bamberger, D.~Brunn, and U.~Hanebeck, ``Performance
  comparison of nonlinear filters for indoor {WLAN} positioning,'' in
  \emph{Proceedings of the 11th International Conference on Information
  Fusion}, June 2008, pp. 1--7.

\bibitem{r213}
R.~Nandakumar, K.~K. Chintalapudi, and V.~N. Padmanabhan, ``Centaur: Locating
  devices in an office environment,'' in \emph{Proceedings of the 18th Annual
  International Conference on Mobile Computing and Networking}, Aug. 2012, pp.
  281--292.

\bibitem{r159}
S.~Liu, Y.~Jiang, and A.~Striegel, ``Face-to-face proximity estimation using
  bluetooth on smartphones,'' \emph{IEEE Transactions on Mobile Computing},
  vol.~13, no.~4, pp. 811--823, Apr. 2014.

\bibitem{r127}
Y.~Chen, W.~Trappe, and R.~P. Martin, ``{ADLS}: Attack detection for wireless
  localization using least squares,'' in \emph{Proceedings of the 5th Annual
  IEEE International Conference on Pervasive Computing and Communications
  Workshops}, Mar. 2007, pp. 610--613.

\bibitem{r128}
D.~Liu, P.~Ning, and W.~K. Du, ``Attack-resistant location estimation in sensor
  networks,'' in \emph{Proceedings of the International Symposium on
  Information Processing in Sensor Networks}, Apr. 2005, pp. 99--106.

\bibitem{cvx}
I.~CVX~Research, ``{CVX}: Matlab software for disciplined convex programming,
  version 2.0,'' \url{http://cvxr.com/cvx}, Aug. 2012.

\bibitem{gb08}
M.~Grant and S.~Boyd, ``Graph implementations for nonsmooth convex programs,''
  in \emph{Recent Advances in Learning and Control}, V.~Blondel, S.~Boyd, and
  H.~Kimura, Eds.\hskip 1em plus 0.5em minus 0.4em\relax Springer-Verlag
  Limited, 2008, pp. 95--110.

\end{thebibliography}

\newpage
\begin{IEEEbiography}[
{\includegraphics[width=1in,height=1.25in,clip,keepaspectratio, angle =0 ]{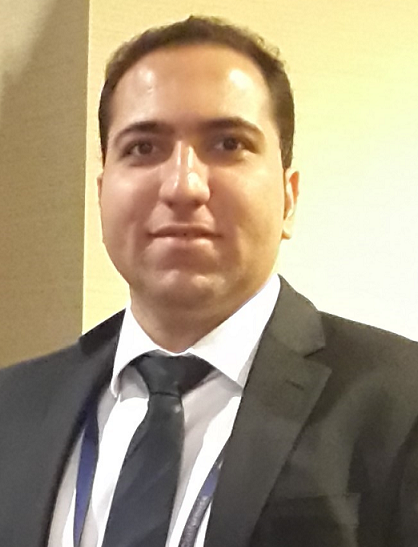}}
]{Ali Khalajmehrabadi} received his B.Sc. from Babol Noshirvani University of Technology, Iran, in 2010 and M.Sc. from University Technology Malaysia (UTM), Malaysia, in 2012 with the best graduate student award. He is currently a Ph.D. candidate (with specialization in Communications) in the Department of Electrical and Computer Engineering, the University of Texas at San Antonio (UTSA). His research interests include collaborative localization, indoor localization and navigation systems, and Global Positioning System (GPS). He is a student member of IEEE. 
\end{IEEEbiography}
\vspace{-20mm}
\begin{IEEEbiography}[
{\includegraphics[width=1in,height=1.25in,clip,keepaspectratio, angle =0 ]{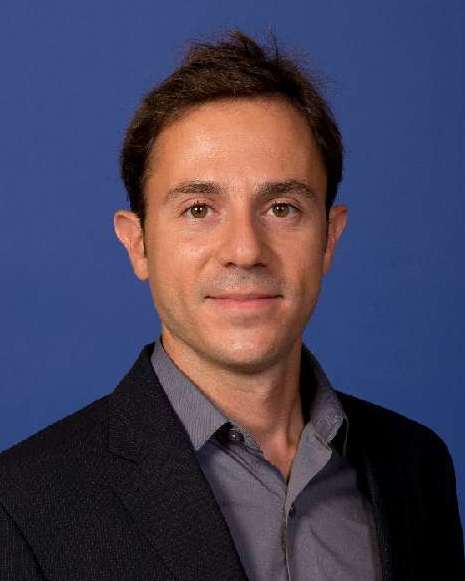}}
]{Nikolaos Gatsis}
 received the Diploma degree in Electrical and Computer Engineering from the University of Patras, Greece, in 2005 with honors. He received the M.Sc. degree in Electrical Engineering in 2010, and the Ph.D. degree in Electrical Engineering with minor in Mathematics in 2012, both from the University of Minnesota. He is currently an Assistant Professor with the Department of Electrical and Computer Engineering at the University of Texas at San Antonio. His research interests lie in the areas of smart power grids, communication networks, and cyberphysical systems, with an emphasis on optimal resource management. Prof. Gatsis co-organized symposia in the area of Smart Grids in IEEE GlobalSIP 2015 and IEEE GlobalSIP 2016. He also served as a Technical Program Committee member for symposia in IEEE SmartGridComm from to 2013 through 2016, and in GLOBECOM 2015.
\end{IEEEbiography}
\vspace{-20mm}

\begin{IEEEbiography} [

{\includegraphics[width=1in,height=1.25in,clip,keepaspectratio, angle =0 ]{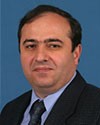}}
]{David Akopian} (M’02-SM’04) is a Professor at the University of Texas at San Antonio (UTSA). Prior to joining UTSA he was a Senior Research Engineer and Specialist with Nokia Corporation from 1999 to 2003. From 1993 to 1999 he was a researcher and instructor at the Tampere University of Technology, Finland, where he received his Ph.D. degree in electrical engineering in 1997. Dr. Akopian’s current research interests include digital signal processing algorithms for communication and navigation receivers, positioning, dedicated hardware architectures and platforms for software defined radio and communication technologies for healthcare applications. He authored and co-authored more than 30 patents and 140 publications. He served in organizing and program committees of many IEEE conferences and co-chairs annual SPIE Multimedia on Mobile Devices conference. His research has been supported by National Science Foundation, National Institutes of Health, USAF, US Navy and Texas foundations.
\end{IEEEbiography}

\end{document}